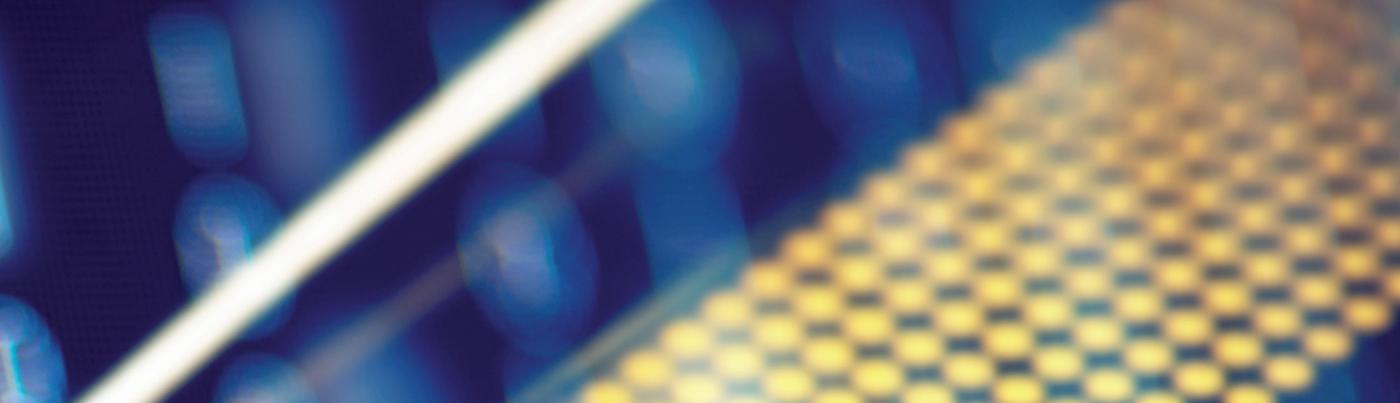

# Computer and Network Security

*Edited by Jaydip Sen*

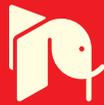

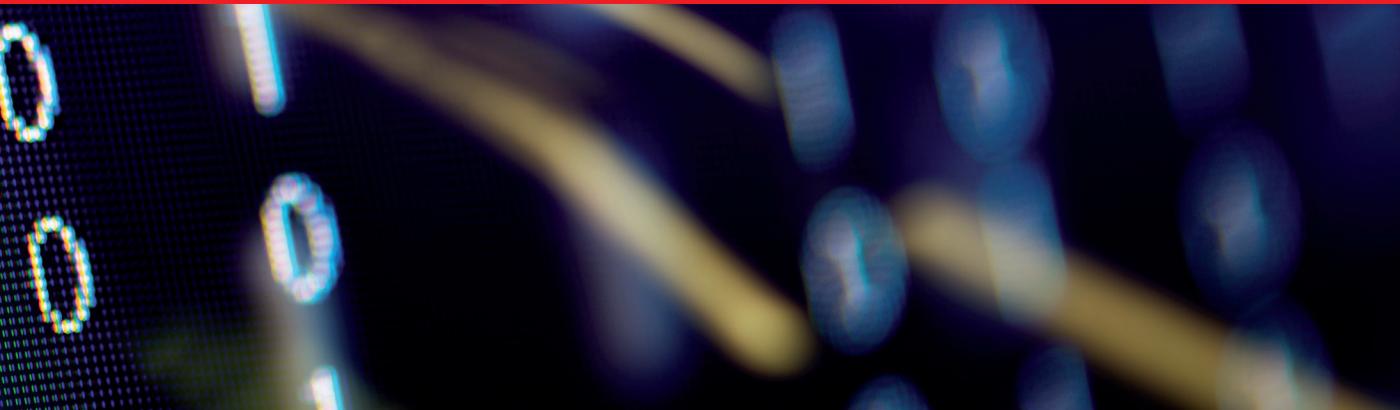

# Computer and Network Security

*Edited by Jaydip Sen*

Notice
Statements and opinions expressed in the chapters are these of the individual contributors and not necessarily those of the editors or publisher. No responsibility is accepted for the accuracy of information contained in the published chapters. The publisher assumes no responsibility for any damage or injury to persons or property arising out of the use of any materials, instructions, methods or ideas contained in the book.

# Meet the editor

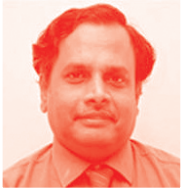 Professor Jaydip Sen has worked for many reputed organizations, including Oil and Natural Gas Corporation Ltd., India; Oracle India Pvt. Ltd.; Akamai Technology Pvt. Ltd.; Tata Consultancy Services Ltd.; and National Institute of Science and Technology, India. Prior to joining the NSHM Knowledge Campus, India, in December 2018 as the head of the Data Science and Computing School, Prof. Sen worked with Praxis Business School as a Professor of Business Analytics. Prior to that, he was a lead scientist in the Innovation Lab of Tata Consultancy Services, Kolkata, India, where he was involved in research and development in security and privacy aspects in wireless networks and Internet of Things. His research areas include security in wired and wireless networks, intrusion detection systems, secure routing protocols in wireless ad hoc and sensor networks, secure multicast and broadcast communication in next-generation broadband wireless networks, trust- and reputation-based systems, quality of service in multimedia communication in wireless networks and cross-layer optimization-based resource allocation algorithms in next-generation wireless networks, sensor networks, and privacy issues in ubiquitous and pervasive communication, big data analytics, R, Python, Spark, Hadoop and MapReduce programming. Currently, he is active in the fields of applied statistical modelling, data mining and machine learning, data warehousing and multi-dimensional modelling, social media and mobile analytics, Artificial Intelligence, and Deep Learning. He has more than 200 publications in reputed international journals and referred conference proceedings (IEEE Xplore, ACM Digital Library, Springer LNCS etc.), and eight chapters in books published by internationally renowned publishing houses. He has delivered expert talks and keynote lectures at various international conferences and symposia. He is a senior member of ACM, a member of IEEE, and a lifetime member of Indian Society of Technical Education (ISTE). He was also an active member of the security group of IEEE 802.16 standard body and European Telecommunication Standards Institute (ETSI). His biography has been listed in Marquis Who's Who in the World annually since 2008. He has delivered invited talks at many prestigious international conferences both in India and abroad and has conducted a number of training programs for teachers of Computer Science and Engineering and Data Science at various universities.

# Contents

# Preface

In an age of explosive worldwide growth of electronic data storage and communications, effective protection of information has become a critical requirement. Especially when used in coordination with other tools for information security, cryptography in all of its applications, including data confidentiality, data integrity, and user authentication, is the most powerful tool for protecting information. While the importance of cryptographic technique (i.e., encryption) in protecting sensitive and critical information and resources cannot be overemphasized, an examination of technical evolution within several industries reveals an approaching precipice of scientific change. The glacially paced but inevitable convergence of quantum mechanics, nanotechnology, computer science, and applied mathematics will revolutionize modern technology. The implications of such changes will be far reaching, with one of its greatest impacts affecting information security. More specifically, that of modern cryptography.

With the exponential growth of wireless communications, the Internet of Things, and Cloud Computing, along with the increasingly dominant roles played by electronic commerce in every major industry, safeguarding the information in storage and the information travelling over communication networks is a critical challenge for technology innovators. The key prerequisite for the sustained development and successful exploitation of information technology and other related industries is the notion of information security and assurance that includes operations for protecting information systems by ensuring their availability, integrity, authentication, non-repudiation, information confidentiality, and privacy.

While it is true that cryptography has failed to provide its users the real security it promised, the reasons for its failure do not have much to do with cryptography as a mathematical science. Rather, poor implementation of protocols and algorithms has been the major source of the problem. Cryptography will continue to play lead roles in developing new security solutions that will be in great demand with the increasing bandwidth and data rate of next-generation communication systems and networks. New cryptographic algorithms, protocols, and tools must follow up in order to adapt to the new communication and computing technologies. Computing systems and communication protocols, like IEEE 802.11 and IEEE 802.15, have become targets of attacks because their underlying radio communication medium contains security loopholes. New security mechanisms should be designed to defend against the increasingly complex and sophisticated attacks launched on networks and web-based applications. In addition to classical cryptographic algorithms, approaches like chaos-based cryptography, DNA-based cryptography, and quantum cryptography will increasingly play important roles.

However, one must not forget that the fundamental problems in security are not new. What has changed over the decades, however, is the exponential growth in the number of connected devices, evolution of networks with data communication speeds as high as terabits/second, massive increase in the volume of data communication, availability of high-performance hardware, and massively parallel architecture for computing and intelligent software. As security systems design becomes more and more complex to meet these challenges, a common mistake made most often by security specialists is not comprehensively analyzing the system to be secured before making a choice about which security mechanism to deploy. On many occasions, the security mechanism chosen turns out to be either incompatible with or inadequate for handling the complexities of the system. This, however, does not vitiate the ideas, algorithms, and protocols of the security mechanisms. While the same old security mechanism, even with appropriate extensions and enhancements, may not be strong enough to secure the multiplicity of complex systems today, the underlying principles will continue to work on the next-generation systems, and indeed, for next era of computing and communications.

## About the book

This book discusses some of the critical security challenges being faced by today's computing world, as well as mechanisms to defend against them using classical and modern techniques of cryptography. With this goal, the book presents a collection of research work from experts in the field of cryptography and network security.

The book is organized into two sections. Section 1 contains six chapters that focus on various aspects of network security. Section 2 consists of three chapters dealing with various mechanisms of cryptography. In Chapter 1, "Introductory Chapter: Machine Learning in Misuse and Anomaly Detection," Sen and Mehtab examine how various machine learning approaches can be gainfully utilized in network security and intrusion detection systems. They illustrate systems exploiting supervised learning, unsupervised learning, and hybrid learning, and discuss their relative advantages and disadvantages. In Chapter 2, "A New Cross-Layer FPGA-Based Security Scheme for Wireless Networks," Ekonde proposes a scheme for enforcing security in a cross-layer mode by using a coding technique in the physical layer in the communication protocol stack in a wireless environment. The coding scheme is implemented using *residue number system* (RNS) and non-linear convolution coding at the physical layer, and the RSA security protocol in the higher layer of the protocol stack, to achieve security in communication. The error correction ability is achieved using a non-linear convolution code. The chapter also presents details about an FPGA implementation of the proposed scheme. In Chapter 3, "Anomaly-Based Intrusion Detection System," Jyothsna emphasizes the need for efficient and effective intrusion detection systems for defending against anomaly-based attacks, and then proposes some approach to detect anomaly-based intrusions in a network using unsupervised learning methods. The chapter also presents a brief review of various intrusion detection approaches, such as statistical approaches, knowledge-based techniques, data mining-based methods, and approaches based on machine learning. The author discusses two specific methods of anomaly detection—*feature correlation analysis and association impact scale* (FCAAIS) and *feature association impact scale* (FAIS)—and analyzes their performance. In Chapter 4, "Security in Wireless Local Area Networks (WLANs) ," Singh and Sharma discuss various currently available security and authentication mechanisms for handoff, and confidentiality of messages in a *wireless local area network* (WLAN) environment. The authors argue that security protocols in a WLAN environment should be lightweight in computation and should not also involve heavy message communication. In this context, the authors propose two protocols—*control and provisioning of wireless access points* (CAPWAP) and *hand over keying* (HOKEY)—for secure handoff in a WLAN. In Chapter 5, "Analysis of Network Protocols: The Ability of Concealing the Information," Noskov discusses methods of hiding data in a network communication using various approaches of steganography. The author compares performances of various schemes including *transcoding steganography* (TransSteg), *lost audio packets* (LACK), *hidden communication system for corrupted networks* (HICCUPS), *retransmission steganography* (RSTEG), modifications in the headers of TCP and IP packets, modification in the data blocks in *stream control transmission protocol* (SCTP) protocols, hybrid SCTP protocol, and SCTP multi-homing method. In Chapter 6, "Multifactor Authentication Methods: A Framework for Their Comparison and Selection," Velasquez et al. propose a detailed guideline for choosing various multi-factor authentication systems for secure access to information. The guidelines have been designed using an extensive action-research methodology in collaboration with experts in the field of secure information system design. In Chapter 7, "Secure Communication Using Cryptography and Covert Channel," Fatayer proposes a scheme of merging encryption, authentication, and covert channel to realize a covert channel of communication that ensures integrity and confidentiality of information communicated over the channel. This covert channel is also used for generating keys for encrypting messages. The chapter also presents results of the performance of the proposed scheme. In Chapter 8, "High-Speed Area-Efficient Implementation of AES Algorithm on Reconfigurable Platform," Mane and Mulani present a scheme of implementation of the *advanced encryption standard* (AES) of a symmetric key encryption algorithm. The scheme uses Xilinx SysGen on Nexys4 and simulates the encryption environment using Simulink. The experimental results show that the scheme produces an overall data throughput of 14.1125 GBPS while consuming 121 slice registers. In Chapter 9, "Hybrid Approaches to Block Cipher," Chitrakar et al. propose two schemes of DNA cryptography. The authors first present a *DNA hybridization scheme* (DHES) in which DNA cryptography is used for encryption and decryption using a *onetime password* (OTP)–based approach for key generation. In the second scheme the authors propose a *hybrid graphical encryption algorithm* (HGEA) utilizing pattern recognition and transformation using mono-alphabetic or poly-alphabetic substitution with a range of characters consisting of 256 possible values. The chapter presensts results of the performance of both schemes with plaintexts of different length and content using different key size.

We hope that the volume will be useful for researchers, engineers, graduate and doctoral students, and faculty members of graduate schools and universities who work in the field of cryptography and network security. However, since it is not an introductory book on the subject, the subject matter does not deal with any basic information. Rather, the chapters in the book present some in-depth cryptography and network security-related theories, as well as some of the latest updates that might be of interest to advanced students and researchers in identifying their research problems and focussing on their solutions. It is assumed that readers have knowledge of mathematical and theoretical backgrounds of cryptography and network security algorithms and protocols.

We express our sincere thanks to all authors for their valuable contributions. Without their cooperation and eagerness to contribute, this project would never have been successfully completed. All the authors have been extremely cooperative and punctual during the submission, editing, and publication processes. We express our heartfelt thanks to Ms. Kristina Kardum, Publishing Process Manager at IntechOpen Publishing, London, for support, encouragement, patience, and cooperation during the period of publishing the volume. Our sincere thanks also go to Ms. Ana Pantar, Senior Commissioning Editor at IntechOpen, for having faith in us and delegating to us the critical responsibility of editing such a prestigious academic volume. We would surely be failing in our duty if we do not acknowledge the encouragement, motivation, and assistance that we received from graduate students of the School of Computing and Analytics of NSHM Knowledge Campus, Kolkata, India. While there are too many to name, the contributions of Abhishek Dutta, Manjari Mukherjee, Saikat Mondal, and Ashmita Paul stand out as being invaluable in ensuring this volume is as error-free as possible. Last but not the least, we would like to thank all members of our respective families for being the major sources of our motivation, inspiration, and strength.

**Jaydip Sen & Sidra Mehtab (Editors)**
Department of Computing and Analytics,
NSHM Knowledge Campus,
Kolkata, India

# Network Security

# Introductory Chapter: Machine Learning in Misuse and Anomaly Detection

*Jaydip Sen and Sidra Mehtab*

## 1. Introduction

Over the last 30 years, ubiquitous and networked computing has increasingly gained importance in our life. With the increase in complexity of computer networks, cybersecurity threats have also manifested in a variety of which was unimaginable even a decade back. While the rule-based intrusion detection systems (IDSs) can accurately detect already known attacks on a cyberinfrastructure, these systems are not capable to detect novel, unknown, and polymorphic cyber threats. Moreover, the computational overheads including CPU cycles and memory overheads are unacceptably high for most of the detection systems. Hence, it has been a constant challenge for security researchers to design automated, fast, and yet accurate IDSs for deployment in real-world cyberinfrastructures. From expert-crafted rules to sophisticated machine learning and deep learning algorithms, researchers have explored and attempted to push the boundary of the detection accuracy while minimizing the false alarm rates.

Applications of machine learning and data mining algorithms in both signature and anomaly detection systems have been widely proposed in the literature. In misuse detection systems, following approaches of machine learning are quite popular: (1) classification using association rules [1–3], (2) artificial neural networks [4], (3) support vector machines [5], (4) classification and regression trees [6, 7], (5) Bayesian network classifier [8–10], and (6) naïve Bayes method [11]. While the signature detection systems require labeled training data in order to learn the features of the attack and the normal traffic, anomaly detection systems are based on identifying any significant changes in the system from its normal state. Various approaches to machine learning in anomaly detection have been proposed in the literature. Some of these approaches are as follows: (1) association rule mining [12–14], (2) fuzzy association rule mining [15], (3) artificial neural network [16–18], (4) support vector machines [19, 20], (5) nearest neighbor [21], (6) hidden Markov model [22–24], (7) Kalman filter [25], (8) clustering [26], and (9) random forest [27, 28]. Other machine learning methods have been proposed for learning the probability distribution of data and in applying statistical tests to detect outliers [29–35].

The hybrid detection approach combines the adaptability and the powerful detection ability of an anomaly detection system with the higher accuracy and reliability of the misuse detection approach [28, 36–43]. The selection of misuse and anomaly detection systems for designing a hybrid detection system is dependent on the application in which the detection system is to be deployed. Following a combinational approach, the integration of an anomaly detection system with a misuse detection counterpart has been classified into four categories [28, 36]. These types are: (1) anomaly–misuse sequence detection, (2) misuse–anomaly sequence detection, (3) parallel detection, and (4) complex mixture detection. The complex mixture model is highly application-specific. **Figure 1** depicts the first three categories of hybrid detection systems. Complex detection systems are application-specific, and these systems cannot be represented by any generic architecture.

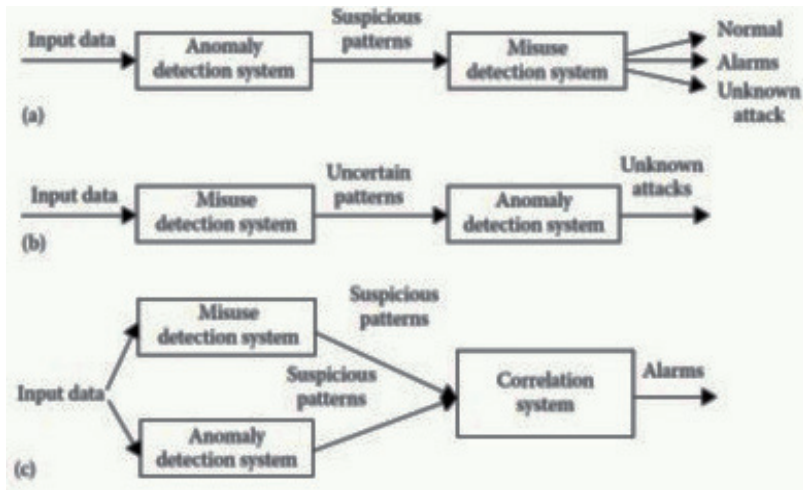

**Figure 1.**
*Three categories of hybrid detection systems. (a) Anomaly–misuse sequence, (b) misuse–anomaly sequence, and (c) parallel detection system.*

## 2. Conclusion

A fundamental challenge in designing an intrusion detection system is the limited availability of appropriate data for model building and testing. Generating data for intrusion detection is an extremely painstaking and complex task that mandates the generation of normal system data as well as anomalous and attack data. If a real-world network environment, generating normal traffic data is not a problem. However, the data may too privacy-sensitive to be made available for public research.

Classification-based methods require training data to be well balanced with normal traffic data and attack traffic data. Although it is desirable to have a good mix of a large variety of attack traffic data (including some novel attacks), it may not be feasible in practice. Moreover, the labeling of data is mandatory with attack and normal traffic data clearly distinguished by their respective labels.

Unlike classification-based approaches that are mostly used in misuse detection, unsupervised anomaly detection-based approaches do not require any prior labeling of the training data. In most of the cases, the attack traffic constitutes the sparse class, and hence, the smaller clusters are most likely to correspond to the attack traffic data. Although unsupervised anomaly detection is a very interesting approach, the results produced by this method are unacceptably low in terms of their detection accuracies.

In a pure anomaly detection approach, the training data are assumed to be consisting of only normal traffic. By training the detection model only on the normal traffic data, the detection accuracy of the system can be significantly improved. Anomalous states are indicated by only a significant state change from the normal state of the system.

In a real-world network that is connected to the Internet, an assumption of attack free traffic is utopian. A pure anomaly detection system can still be trained on a training data that include attack traffic. In that case, those attack traffic data will be considered as normal traffic, and the detection system will not raise an alert when such traffic is encountered in real-world operations. Hence, in order to increase the detection accuracy, attack traffic should be removed from the training data as much as possible. The removal of attack traffic from the training data can be done using updated misuse detection systems or by deploying multiple anomaly detection systems and combining their results by a voting mechanism.

For an intrusion detection system that is deployed in a real-world network, it is mandatory to have a real-time detection capability under a high-speed, high-volume data environment. However, most of the cluster techniques used in unsupervised detection require quadratic time. This renders their deployment infeasible in practical applications. Moreover, the cluster algorithms are not scalable, and they need the entire training data to reside in the memory during the training process. This requirement puts a restriction on the model size. The future direction of research may include studies on scalability and performance of anomaly detection algorithms in conjunction with the detection rate and false positive rate. Most of the currently existing propositions on intrusion detection have not paid adequate attention to these critical issues.

In this book, the following chapters deal with various aspects of network security and cryptography. While the chapters belonging to the network security section broadly discuss different aspects of applications and deployment of security protocols and secure system architecture, the cryptography section discusses various theoretical algorithms and their complexities.

## Author details

Jaydip Sen* and Sidra Mehtab
School of Computing and Analytics, NSHM Knowledge Campus, Kolkata, India

*Address all correspondence to: jaydip.sen@acm.org

IntechOpen

# A New Cross-Layer FPGA-Based Security Scheme for Wireless Networks

*Michael Ekonde Sone*

## Abstract

This chapter presents a new cross-layer security scheme which deploys efficient coding techniques in the physical layer in an upper layer classical cryptographic protocol system. The rationale in designing the new scheme is to enhance security-throughput trade-off in wireless networks which is in contrast to existing schemes which either enhances security at the detriment of data throughput or vice versa. The new scheme is implemented using the residue number system (RNS), non-linear convolutional coding and subband coding at the physical layer and RSA cryptography at the upper layers. The RNS reduces the huge data obtained from RSA cryptography into small parallel data. To increase the security level, iterated wavelet-based subband coding splits the ciphertext into different levels of decomposition. At subsequent levels of decomposition, the ciphertext from the preceding level serves as data for encryption using convolutional codes. In addition, throughput is enhanced by transmitting small parallel data and the bit error correction capability of non-linear convolutional code. It is shown that, various passive and active attacks common to wireless networks could be circumvented. An FPGA implementation applied to CDMA could fit into a single Virtex-4 FPGA due to small parallel data sizes employed.

**Keywords:** residue number system (RNS), RSA cryptography, field programmable gate array (FPGA), subband coding, convolutional coding, CDMA

## 1. Introduction

Generally, wireless networks consist of low capacity links with nodes that rely on batteries. An efficient communication scheme for such networks should minimize both congestion in the links and control information in the nodes. Security is a critical parameter in wireless applications and any efficient communication scheme has to integrate security vulnerabilities of the system in its implementation. Unfortunately, existing schemes have network security implemented at the upper layer such as the application layer; meanwhile parameters such as congestion, which affect data throughput, are the physical layer. Hence, any attempt to increase the security level in a communication system greatly compromises data throughput. In [1], the authors developed a metric to estimate a timeframe for cyberattacks using the RSA public key cryptography. In the analysis, the authors estimated the attacker's human time in carrying out a successful attack based on the key length.

Such an implementation at the upper layer will curb any security attack at the prescribed time but will greatly compromise data throughput at the physical layer due to the huge modular exponentiation involved in its implementation. In [2], the authors developed a secure and efficient method for mutual authentication and key agreement protocol with smart cards. The implementation, which is based on the constant updating of the password, will involve a considerable amount of control information, which is detrimental to the optimum functioning of the nodes. Research work using different information-theoretic models to develop physical layer security based on the characteristics of the wireless links has been carried out [3, 4]. However, the existing methods for implementing physical layer security under the different information-theoretic security models is expensive and requires assumptions about the communication channels that may not be accurate in practice [5]. Hence any deployment of the physical-layer security protocol to supplement a well-established upper layer security scheme will be a pragmatic approach for robust data transmission and confidentiality [6]. It is in this light that, a new cross-layer approach is presented in this research. Major research efforts have targeted cross-layer implementation of security schemes in wireless networks [7–9]. In this research, the proposed cross-layer security scheme uses signal processing techniques as well as efficient coding and well-established cryptographic algorithm to implement a security scheme, which greatly enhances security-throughput trade-off, and curb many security threats common to wireless networks.

The rest of the chapter is organized in eight subsequent sections. In Section 2, we present the background knowledge required for the design and implementation of the new cross layer security scheme. This will involve a review of the different techniques used in the development of the new security scheme. The first subsection presents the implementation of the multi-level convolutional cryptosystem. This implementation involves the combination of subband coding and a new non-linear convolutional coding. Next, a review of the residue number system (RNS) with brief description of the Chinese remainder theorem (CRT) is presented. We conclude the section with an overview of RSA public-key cryptography. Section 3 presents the protocol for the implementation of the new cross layer security scheme. The FPGA-based implementation applied to CDMA using the new layered security scheme will be presented in Section 4. In Section 5, cryptanalysis of the cross-layer security scheme will be carried out in order to quantify the security. Quantification of data throughput is performed in section 6 while different security threats which could be circumvented by the cross-layer security scheme are presented in Section 7. We end the chapter in Section 8 with conclusions of our work.

## 2. Background

This section presents the background knowledge required for the design and implementation of the new cross layer security scheme. It involves a review of the different techniques used in the development of the new security scheme.

### 2.1 Multi-level convolutional cryptosystem

The multi-level convolutional cryptosystem constitutes the second stage of implementation at the physical layer. It receives integers from the RNS implemented at the first stage. The multi-level cryptosystem is implemented using subband coding and non-linear convolutional cryptosystem.

### 2.1.1 Subband coding

The integers from the RNS block are split into different levels of decomposition based on subband coding. Subband coding is implemented using integer wavelet lifting scheme [10, 11]. It is shown in [12] that, a judicious choice of filter banks could result into an integer transform despite the fact that, wavelet transform is an approximation process. A four-tap Daubechies polyphase matrix, which results into integer transforms, is given as follows [12]:

$$P^{new}(z) = \begin{bmatrix} h_e(z) & g_e^{new}(z) \\ h_o(z) & g_o^{new}(z) \end{bmatrix} \tag{1}$$

where h and g are filter coefficients with suffix e and o denoting even and odd coefficients. The factorization of the polyphase matrix is as follows [12–14].

$$P(z) = \tilde{P}(z) = \begin{bmatrix} 1 & -\sqrt{3} \\ 0 & 1 \end{bmatrix} \begin{bmatrix} 1 & 0 \\ \frac{\sqrt{3}}{4} + \frac{\sqrt{3}-2}{4}z^{-1} & 1 \end{bmatrix} \begin{bmatrix} 1 & z \\ 0 & 1 \end{bmatrix} \begin{bmatrix} \frac{\sqrt{3}+1}{\sqrt{2}} & 0 \\ 0 & \frac{\sqrt{3}-1}{\sqrt{2}} \end{bmatrix} \tag{2}$$

(Eq. (2)) forms the basis of integer to integer wavelet transform which in effect is progressive transmission. The factored coefficients $h_1 = -\sqrt{3}$, $h_2 = \frac{\sqrt{3}}{4}$, $h_3 = \frac{\sqrt{3}-2}{4}$, $h_4 = \frac{\sqrt{3}+1}{\sqrt{2}}$, $h_5 = \frac{\sqrt{3}-1}{\sqrt{2}}$ are used to compute the approximate and detail sequences. In such transmission, the original data is split into two portions, namely approximate and detail sequences. The detail sequence is transmitted while the approximate sequence is further split into two halves. The process is repeated until the final data point. Using (Eq. (2)), the approximate, $a_n$ and the detail, $d_n$ sequences could be computed as follows [12, 15]:

$$
\begin{aligned}
|a_n|_{m_j} &= \left| \left| \left| |x_{2l}\, h_1|_{m_j} + x_{2l\text{-}1} \right|_{m_j} \cdot h_2 \right|_{m_j} + x_{2l\text{-}1} \right|_{m_j} \\
|d_n|_{m_j} &= \left| \left| |x_{2l}\, h_1|_{m_j} + x_{2l\text{-}1} \right|_{m_j} + a_{n\text{-}1} \right|_{m_j}
\end{aligned}
\tag{3}
$$

where $a_n$ and $d_n$ are the approximation and detail sequences of wavelet coefficients of the nth sample. The subsequent transmissions are fed into the non-linear convolutional coding block as depicted in **Figure 1** for the first level kth detail and approximation sequences [12, 15]. The processing blocks (PEs) shown in the figure depicts the computations of (Eq. (3)).

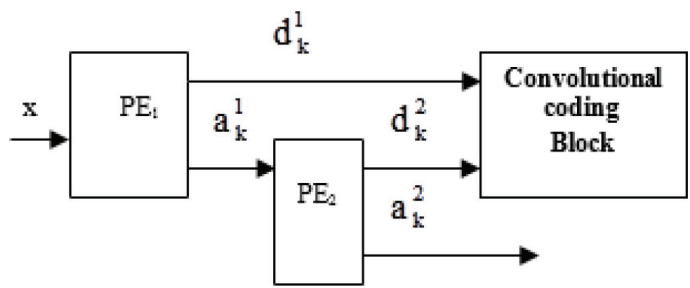

**Figure 1.**
*Synopsis for the computation of the kth coefficients for the 1st and 2nd levels of decomposition.*

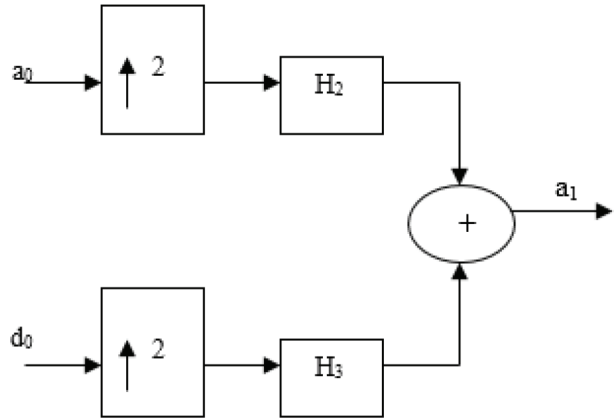

**Figure 2.**
*The first stage of the inverse transform.*

The final decomposition level which comprises one data point will give one approximation coefficient and one detail coefficient.

At the destination, the inverse wavelet transform is performed to obtain the successive approximation sequences. (Eq. (3)) is used to perform the inverse transform by reversing the operations for the forward transform and flipping signs [12, 15]. The process starts with the approximation and detail coefficients $a_0$ and $d_0$ respectively obtained at the final decomposition level of the forward transform. The first stage of the inverse transform is shown in **Figure 2** [12, 15].

The ($\uparrow 2$) symbol represents upsampling by 2, which means that zeros are inserted between samples while $H_2$ and $H_3$ are the filter coefficients used in (Eq. (3)).

### 2.1.2 Nonlinear convolutional cryptosystem

A major advantage of symmetric cryptography is the ability of composing primitives to produce stronger ciphers although on their own the primitives will be weak. Hence, the vulnerable convolutional code block will be cascaded into different stages using the product ciphers obtained from the S-box and P-box to form a non-linear convolutional cryptosystem.

- **Key generation:** The specifications of the private keys used in the implementation of the cascaded convolutional cryptosystem are as follows [12, 15, 16]:

    a. States of each transducer or convolutional code block in the cascade given by the contents of the sub-matrices in the generator matrix;

    b. The transition functions. These are mappings used to compare the input bits and present state and switches to the appropriate next state;

    c. n-bit S-boxes. They are used to shuffle the output bits.

    d. n-bit P-boxes. They are used for the different permutations per level of decomposition.

For illustrative purposes, an (8, 8, 2) convolutional encoder will be considered to demonstrate the keys generation process.

### 2.1.2.1 States of convolutional code block or transducer

For an $8 \times 8$ matrix, there are at least $2^{16}$ ways or keys in which the connections of a register to the modulo-2 adder could be made. A possible key which gives the contents of the $8 \times 8$ matrix are shown in (Eq. (4))

$$g_{m1,0}^1 = \begin{bmatrix} 1 & 1 & 0 & 0 & 0 & 0 & 0 & 0 \\ 0 & 1 & 0 & 0 & 0 & 0 & 0 & 0 \\ 0 & 0 & 1 & 0 & 0 & 0 & 0 & 0 \\ 0 & 0 & 0 & 1 & 0 & 0 & 0 & 0 \\ 0 & 0 & 0 & 0 & 1 & 0 & 0 & 0 \\ 0 & 0 & 0 & 0 & 0 & 1 & 0 & 0 \\ 0 & 0 & 0 & 0 & 0 & 0 & 1 & 0 \\ 0 & 0 & 0 & 0 & 0 & 0 & 0 & 1 \end{bmatrix}, g_{m1,1}^1 = \begin{bmatrix} 0 & 0 & 0 & 0 & 0 & 0 & 0 & 0 \\ 0 & 0 & 0 & 0 & 0 & 0 & 0 & 0 \\ 0 & 0 & 0 & 0 & 0 & 0 & 0 & 0 \\ 0 & 0 & 0 & 0 & 0 & 0 & 0 & 0 \\ 0 & 0 & 0 & 0 & 0 & 0 & 0 & 0 \\ 0 & 0 & 0 & 0 & 0 & 0 & 0 & 0 \\ 0 & 0 & 0 & 0 & 0 & 0 & 0 & 0 \\ 0 & 0 & 0 & 0 & 0 & 0 & 0 & 0 \end{bmatrix}, g_{m1,2}^1 = \begin{bmatrix} 1 & 0 & 0 & 0 & 0 & 0 & 0 & 0 \\ 0 & 1 & 1 & 0 & 0 & 0 & 0 & 0 \\ 0 & 0 & 1 & 1 & 0 & 0 & 0 & 0 \\ 0 & 0 & 0 & 1 & 1 & 0 & 0 & 0 \\ 0 & 0 & 0 & 0 & 1 & 1 & 0 & 0 \\ 0 & 0 & 0 & 0 & 0 & 1 & 0 & 0 \\ 0 & 0 & 0 & 0 & 0 & 0 & 1 & 0 \\ 0 & 0 & 0 & 0 & 0 & 0 & 0 & 1 \end{bmatrix} \quad (4)$$

The generator matrix is used to specify the following set of 8 vectors.

$X(7) := A1(7) \oplus A3(7);$ $\qquad$ $X(6) := A1(7) \oplus A1(6) \oplus A3(6).$
$X(5) := A1(5) \oplus A3(5) \oplus A3(6);$ $\qquad$ $X(4) := A1(4) \oplus A3(4) \oplus A3(5).$
$X(3) := A1(3) \oplus A3(3) \oplus A3(4);$ $\qquad$ $X(2) := A1(2) \oplus A3(2) \oplus A3(3).$
$X(1) := A1(1) \oplus A3(1);$ $\qquad$ $X(0) := A1(0) \oplus A3(0).$

It should be recalled that, for an (n,k,L) convolutional encoder, each vector has Lk dimensions and contains the connections of the encoder to the modulo-2 adder.

The structure of the (8, 8, 2) convolutional encoder is shown in **Figure 3** with A2, A3 representing the registers $M_1^1, M_2^1$ while A1 represents the input data.

### 2.1.2.2 The transition functions

For an (8,8,2) convolutional code, there are $2^8 = 256$ mappings or keys. There are two sets of transition functions denoted as $f_1$ for the two possible states.

For example, a transition function that compares input data to state 1 and remains in state 1 is given as follows:

$$f_1(1, \{[00000000], [00000001], [00000010], [00000011], [00000100],$$
$$[00000101], [00000110], [00000111]\}) = \{1\} \quad (5)$$

In (Eq. (5)), if the input data is any of the sequences $\{[00000000], [00000001], [00000010], [00000011], [00000100], [00000101], [00000110], [00000111]\}$, the present state of the transducer which is state 1 is retained.

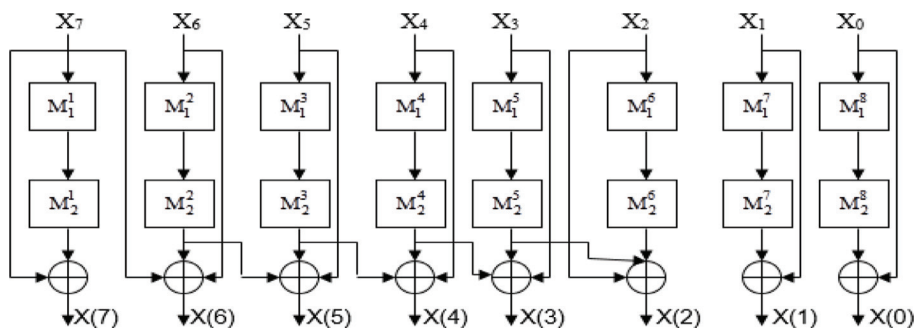

**Figure 3.**
*Structure of an (8, 8, 2) convolutional encoder.*

A transition function that compares input data to state 1 and switches to state 2 is given as follows:

$$f_1(1, \{[00001000], [00001001], [00001010], [00001011], [00001100],$$
$$[00001101], [00001110], [00001111]\}) = \{2\}, \tag{6}$$

In (Eq. (6)), if the input data is any of the sequences $\{[00001000], [00001001], [00001010], [00001011], [00001100], [00001101], [00001110], [00001111]\}$, the present state of the transducer which is state 1 is changed to state 2.

At the destination, the transition functions are similar to those for the encoder at the source but change roles. The transition functions are very critical in the implementation of convolutional cryptosystem since it accounts for its dynamic nature, hence an increase in security level.

### 2.1.2.3 S-box entries

For an (8,8,2) convolutional code, using 2-bit shuffling boxes, there are 16 S-boxes or keys. For higher n-bit shuffling boxes, the number of keys increases, for example 8-bit shuffling boxes will give $2^8$ keys. The four 2-bit S-boxes used to illustrate the scheme are shown in **Table 1**. Given an 8-bit data sequence as $[A_7, A_6, A_5, A_4, A_3, A_2, A_1, A_0]$, the look-up S-box, $Sub_{1,1}$ is used to shuffle the first pair of bits, $[A_7, A_6]$, $Sub_{1,2}$ is used to shuffle the second pair $[A_5, A_4]$, $Sub_{1,3}$ is used to shuffle the third pair $[A_3, A_2]$, and $Sub_{1,4}$ is used to shuffle the last pair $[A_1, A_0]$.

### 2.1.2.4 P-box entries

The interconnections between inputs and outputs are implemented using a permutation set look-up table. For an (8,8,2) code, the eight (08) inputs and outputs could be permuted or interconnected in at least $7^7 = 823,543$ ways. A permissible permutation is shown in **Table 2** [12, 15].

| Input | 0 0 | 0 1 | 1 0 | 1 1 |
|---|---|---|---|---|
| output | 0 0 | 1 1 | 1 0 | 0 1 |

$Sub_{1,1}$

| Input | 0 0 | 0 1 | 1 0 | 1 1 |
|---|---|---|---|---|
| output | 0 1 | 0 0 | 1 1 | 1 0 |

$Sub_{1,2}$

| Input | 0 0 | 0 1 | 1 0 | 1 1 |
|---|---|---|---|---|
| output | 1 0 | 0 1 | 0 0 | 1 1 |

$Sub_{1,3}$

| Input | 0 0 | 0 1 | 1 0 | 1 1 |
|---|---|---|---|---|
| output | 1 1 | 1 0 | 0 1 | 0 0 |

$Sub_{1,4}$

**Table 1.**
*2-bit shuffle look-up-table.*

| Input | 1 | 2 | 3 | 4 | 5 | 6 | 7 | 8 |
|---|---|---|---|---|---|---|---|---|
| Output | 8 | 7 | 6 | 5 | 4 | 3 | 2 | 1 |

**Table 2.**
*Input–output interconnect look-up-table for encoder.*

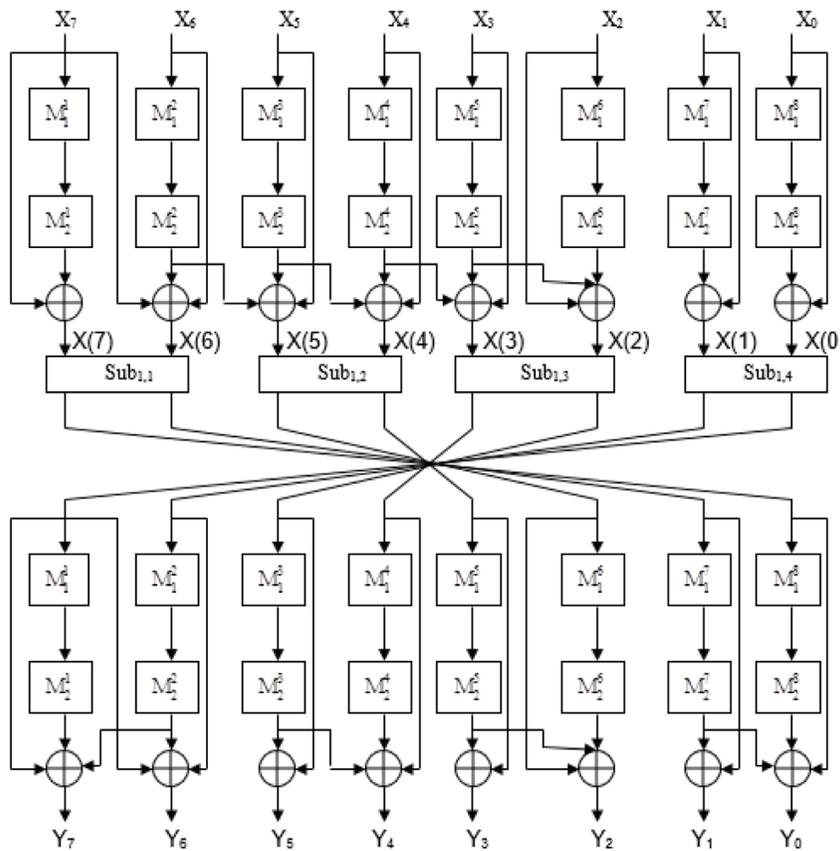

**Figure 4.**
*Initial structure of the cascade before encoding.*

After the specification of the keys, the vulnerable convolutional code block will be cascaded into different stages using the product ciphers obtained from the S-box and P-box. Using two (02) stages, a non-linear (8, 8, 2) 2-cascaded is as shown in **Figure 4**.

In **Figure 4**, $Sub_{1,1}$, $Sub_{1,2}$, $Sub_{1,3}$ and $Sub_{1,4}$ are S-boxes used for pairwise bit shuffling and the input vector to the first transducer, $\{X_7, X_6, X_5, X_4, X_3, X_2, X_1, X_0\}$ is the output set from the subband encoding block while the output vector $\{Y_7, Y_6, Y_5, Y_4, Y_3, Y_2, Y_1, Y_0\}$ is the ciphertext from the second transducer stage.

It is worth noting that the security level could be greatly increased by increasing the number of stages to be cascaded.

## 2.2 Residue number system (RNS)

The residue number system uses the Chinese remainder theorem (CRT) to compute unknown values from the remainders left or residues when unknown values are divided by known numbers.

The modular Chinese remainder theorem states that [17, 18]:

Assume $m_1, m_2, ..., m_N$ are positive integers, relatively prime pairs: $(m_i, m_k) = 1$ if $i \neq k$. Let $\{b_1, b_2, ..., b_N\}$ be arbitrary integers, then the system of simultaneous linear congruence

$$x_1 = b_1 \ (mod \ m_1) - - - x_N = b_N \ (mod \ m_N) \tag{7}$$

| Authentication and non-repudiation | Confidentiality |
|---|---|
| Choose plaintext X. | Choose plaintext X. |
| Compute $X_s \equiv X^d \pmod{m}$ | Use Bob's public key (m, e) to |
| Send (X, $X_s$) to Alice. $X_s$ is the RSA digital signature of | compute |
| message, X | $C \equiv X^e \pmod{m}$. |
| | Send ciphertext, C to Bob |

**Table 3.**
*Summary of security services of RSA public-key cryptography.*

has exactly one solution modulo the product $m_1$, $m_2$, ..., $m_N$. The solution to the simultaneous linear congruence is formally given as [15, 17, 18].

$$|x|_M \;=\; \left| \sum_{j=1}^{L} \hat{m}_j \left| \frac{x_j}{\hat{m}_j} \right|_{m_j} . \; x_j \right|_M \qquad (8)$$

where $\hat{m}_j = \frac{M}{m_j}$ and $M = \prod_{j=1}^{N} m_j$.

(Eq. (8)) establishes the uniqueness of the solution. In this research, the integers $T_j = |\frac{x_j}{\hat{m}_j}|$ often referred to as the multiplicative inverses of $\hat{m}_j$ are computed a priori by solving the linear congruence in (Eq. (7)). In other words, if $M_j = M/m_j$, then the multiplicative inverses, $T_j$ are computed by solving the congruence $T_j M_j = 1 \pmod{m_j}$.

## 2.3 RSA public-key cryptography

- Key creation
  - Choose secrete primes p and q and compute m = p.q
  - Choose encryption exponent, e
  - Compute d satisfying $e.d \equiv 1 \mod ((p-1). (q-1))$.
  - Public key: (m, e) and Private key: d

- Security services achieved using RSA cryptography are authentication and non-repudiation based on digital signatures and confidentiality based on encryption. Implementation of these services are summarized in **Table 3**

## 3. Protocol of implementation of cross-layer security scheme

The new scheme is implemented at the application and physical layers. The detail operations of the application and physical layers at the source and destination are as follows:

- Source:
  - Application layer:

    a. Traditional RSA encryption

  - Physical layer:

    a. Step 1: Residue number system (RNS) converts the message points into residues based on the moduli set;

    b. Step 2: RNS-based RSA ciphertext is converted into different levels of decomposition using subband coding;

    c. Step 3: Symmetric encryption using Convolutional cryptosystem;

    At the destination, the entire process is reversed starting with convolutional decoding at the physical layer and ending with RSA decryption at the application layer.

## 3.1 Illustrative example

    Consider an arbitrary array of integers for plaintext as follows {398, 453, 876, 200, 356, 165, 265, 897}.

- Source

  ∘ Application layer: Traditional RSA encryption

    a. Primes, $p = 13$; $q = 37 \Rightarrow n = p \cdot q = 481$

    b. Encryption key, $e = 5$

    c. Decryption key: $e \cdot d \equiv 1 \bmod (432) \Rightarrow d = 173$

Array due to RSA encryption is given as {151, 293, 252, 135, 304, 315, 265, 182}

  ∘ Physical layer:

    a. Step 1: Moduli set of {107, 109, 113} is used to convert the RSA ciphertext into 8-bit data point arrays. The residue set, $r_1$ for $m_1 = 107$ is as follows:

$r_1 = \{44, 79, 38, 28, 90, 101, 51, 75\}$

    b. Step 2: Subband coding is performed to split residues obtained using moduli set into three levels of decomposition since $m = 8 = 2^3$ data points are used. Subband coding is basically down sampling by 2 using (Eq. (3)). The corresponding arrays for the first level of decomposition are as follows:

$r_{11} = \{-9, -48, -79, -27\}; - r12 = \{-9, -42, -75, -21\}; - r13 = \{-9, -30, -67, -9\}$

    Note that $r_{11}$ refers to first level of decomposition array for modulus, $m_1 = 107$ and the first element is obtained using (Eq. (3)) with integer lifting filter coefficients set, $h = \{2, 0, 0\}$ as follows: $r_1(1) - 2 \times r_1(0) = 79 - 2 \times 44 = -9$.

    The same procedure is performed for the second and third levels of decomposition.

    c. Step 3: **Table 4** summarizes the manual computation of the encryption and decryption process of the convolutional cryptosystem for the data $r_{11}(0) = -9$ from the subband encoding stage based on the entries of the product cipher and combinational logic of the non-linear (8,8,2) 2-cascaded convolutional transducer in **Figure 4**.

- Destination
  - Physical layer:

The entire process is reversed starting with convolutional decoding at the physical layer which was illustrated in **Table 4**.

      a. Subband decoding: It involves upsampling and the use of quadrature mirror filter (QMF) bank as depicted in **Figure 5** [12, 13]

The ($\uparrow 2$) symbol represents upsampling by 2, which means that zeros are inserted between samples. The QMF bank are the coefficients derived from the 4-tap Daubechies filter bank [12, 13, 19]. Hence, using upsampling and the QMF bank coefficients the residue sets $r_1$, $r_2$ and $r_3$ are retrieved.

**Figure 5** will be used to perform a numerical illustration of subband decoding for the first level of decomposition of the array of modulus $m_1 = 107$ to obtain approximation coefficients $a_1$.

The moduli sets obtained from subband encoding for the three levels of decomposition for $m_1 = 107$ are as follows:

- Level 3: $r_{31} = \{2, 44\}$; − Level 2: $r_{21} = \{−50, −20\}$; − Level 1: $r_{11} = \{−9, −48, −79, −27\}$.

For subband decoding, the entire process is reversed with level 3 of encoding becoming level 1 for decoding.

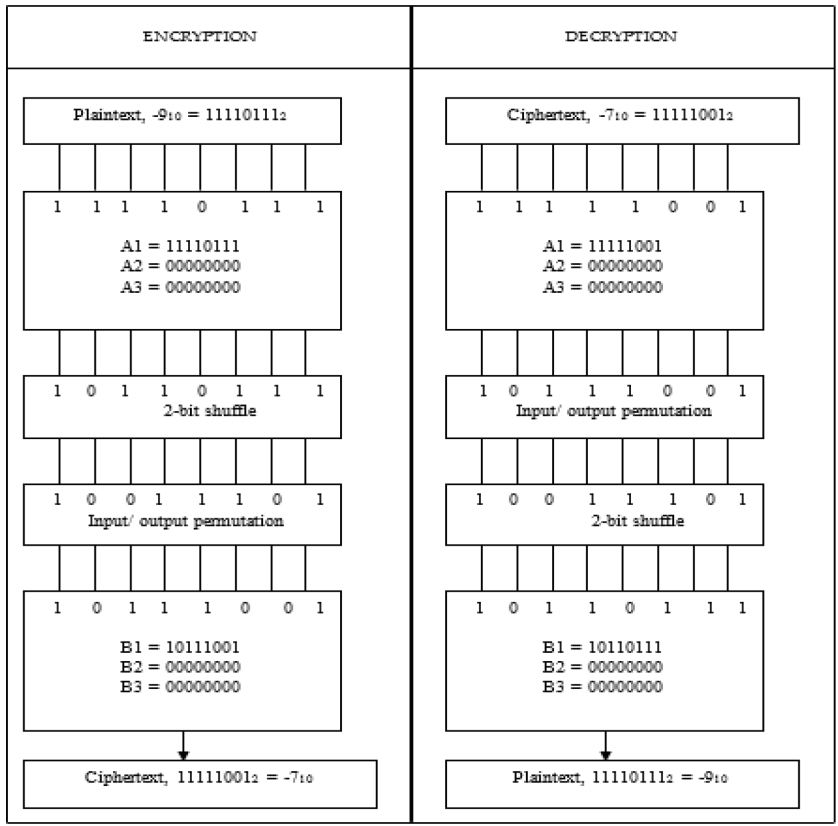

**Table 4.**
*Manual computation for the first sample of first level of decomposition for modulus 107.*

$r_{31}$ from subband encoding namely $a_0$ = 2 and $d_0$ = 44 is used. From **Figure 5**, upsampling performed on the approximation, $a_0$ and detail, $d_0$ data points gives the sets y_1 = {2, 0} and z_1 = {44, 0} respectively. Using 4-tap Daubechies integer lifting filter coefficients set, h = {2, 0, 0} we have.

$$w\_1(0) = z\_1(0) - h(2)^*y\_1(0) - h(3)^*y\_1(1) = 44 - 0 - 0 = 44$$
$$w\_1(1) = z\_1(1) - h(2)^*y\_1(0) - h(3)^*y\_1(1) = 0 - 0 - 0 = 0$$
$$a\_1(0) = y\_1(0) - h(1)^*w\_1(0) = 2 - 2^*44 = -86$$
$$a\_1(1) = y\_1(1) - h(1)^*w\_1(1) = 0 - 2^*0 = 0$$

Hence the first level approximation data points, $a_1$ are obtained as follows.
$a_1(0)$ = w_1(0) = 44 and $a_1(1)$ = w_1(1) + z_1(0) = 0 - 86 = -86
$\Rightarrow a_1$ = {44, -86}.

The process is repeated to obtain the second and third levels approximation data points. The third level approximation data points, $a_3$ should be equal to residue set, $r_1$ obtained from the RNS-based RSA ciphertext using the modulus, $m_1$ = 107.

b. RNS-based Chinese Remainder Theorem (CRT): It is applied to the residue sets $r_1$, $r_2$ and $r_3$.

(Eq. (8)) will be used to retrieve the RSA ciphertext from the residue sets. Using (Eq. (8)) and moduli set, m = {107, 109, 113} to compute the first data point of the ciphertext set we have.

$$\begin{aligned} X \quad &\equiv (r_{11} \times \hat{m}_1 \times T_1) \, (r_{21} \times \hat{m}_2 \times T_2) \, (r_{31} \times \hat{m}_3 \times T_3) (\mathrm{mod} \, M) \\ &\equiv ((44 \times 12,317 \times 9) \quad + \quad (42 \times 12,091 \times 68) \quad + \quad (38 \times 11,663 \times 33)) (\mathrm{mod} \, 1,317,919) \\ &\Rightarrow X = 151 \end{aligned}$$

The same process is repeated to obtain all the other data points of the RSA ciphertext set. The RSA ciphertext set is fed to the RSA decryption block at the application layer.

○ Application layer: RSA decryption

Decryption of the first data point is given as M = $151^d$ mod n = $151^{173}$ mod 481 = 398 which represents the original data which was sent at the source.

The same process is repeated to obtain all the other data points of the plaintext set.

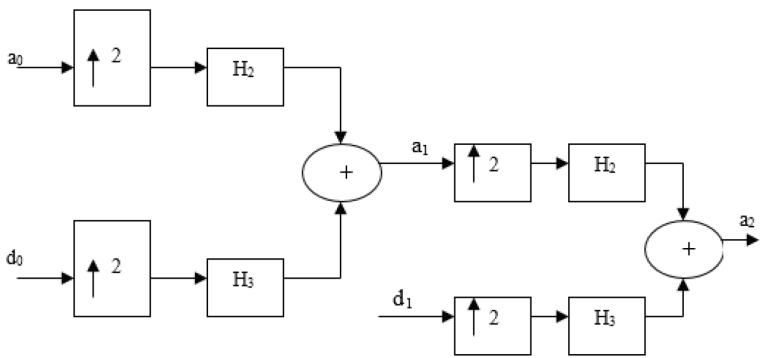

**Figure 5.**
*The first two stages of the wavelet inverse transform.*

## 4. FPGA implementation of security scheme applied to CDMA

In this section, FPGA implementation of new scheme applied to CDMA, the new VHDL code package to implement A mod n operations, synthesis report and behavioral simulation results will be presented.

### 4.1 FPGA implementation of new scheme applied to CDMA

Code division multiple access (CDMA) enables several users to transmit messages simultaneously over the same channel bandwidth in such a way that each transmitter/ receiver user pair has its own distinct signature code for transmitting over the common channel bandwidth. This distinct signature is ensured by using spread spectrum techniques whereby the message from each user is transmitted using orthogonal waveforms. In orthogonal signaling, the residues are mapped to orthogonal waveforms which constitute the CDMA signal [20]. The orthogonal waveforms used in this research are Walsh functions.

Considering an (8, 8, 2) multi-level cryptosystem for illustrative purposes, the mapping process will involve $M = 2^8 = 256$ orthogonal waveforms. Using the dynamic range of $(-128, 126)$, a set of $M = 2^8 = 256$ orthogonal waveforms is required to completely represent all the integers or symbols. Based on this, the corresponding Hadamard matrix obtained from the procedure elaborated in [21] is as follows:

$$
\begin{aligned}
H_{256} \ &= \ H_{128} \ \otimes \ H_{128} \\
&= \ \begin{pmatrix} H_{128} & H_{128} \\ H_{128} & \overline{H}_{128} \end{pmatrix}
\end{aligned}
$$

The $H_{256}$ matrix is a large matrix comprising of 256 rows and 256 columns. The Hadamard matrix results into a multi–dimensional array. Multi–dimensional arrays are arrays with more than one index. Multi-dimensional arrays are not allowed for hardware synthesis. One way around this is to declare two one–dimensional array types. This approach is easier to use and more representative of actual hardware. The VHDL code used to declare the two one–dimensional array types is shown in **Figure 6** [22].

The other operations in the hardware Walsh function generator implementation are trivial since they involve modulo–2 addition with built–in operators in VHDL code to handle such operations.

### 4.2 New VHDL code package

In this research, a new algorithm is presented which implements modular exponentiation without the use of the Montgomery algorithm. A package is developed in the VHDL code to extract residues similarly to the X mod N operation for any randomly generated data. Meanwhile, the large operand lengths which resulted

```
Subtype Depth_Typ is Integer range 0 to 255;
Subtype Width_Typ is Integer range 255 downto 0;
Subtype Data_Typ is Bit_vector (Width_Typ);
Type Memory_Typ is array (Depth_Typ) of Data_Typ;
```

**Figure 6.**
*VHDL code for synthesizable 256 × 256 Hadamard matrix.*

from the modular exponentiation are reduced using binary exponentiation and the RNS.

The principle used to develop the package is as follows:

To perform the x = X mod N calculation where X has a large operand length of b bits say b = 1024 bits and N is modulus of small operand length of $b_1$ bits say $b_1$ = 8 bits, the following steps are used:

1. X is converted to binary equivalent;

2. The $b_1$ least significant bits of the b bits of X are chosen;

3. The integer equivalent, $x_1$ of the chosen $b_1$ bits is determined;

4. The residue, x = X mod N is obtained from the following equation;

$$x = x_1 + \left(2^{b_1-1} - N\right) \times \frac{X}{2^{b_1-1}} \qquad (9)$$

5. If the residue, x is greater than the modulus, N the process is iterated until the residue is less than the modulus.

(Eq. (9)) forms the basis for the x = X mod N calculation.

Based on this new algorithm which implements modular exponentiation without the use of the Montgomery algorithm, the entire physical layer security scheme could fit into a single FPGA chip.

## 4.3 Behavioral simulation and synthesis report

In order to verify the performance of the proposed architecture, a VHDL programme was written and implemented on a Xilinx Virtex-4 FPGA chip (device: xc4vlx 200, package: ff 1513, speed grade − 11) [22]. Sixteen (16) randomly generated integers were fed into the FPGA. For this value, the number of bonded IOBs is 760 out of 960 resulting to 79% resource used. The behavioral simulation results for array {39,870, 45,378, 87,654, 20,087, 35,689, 16,592, 564, 276,509, 89,732, 56,287, 4527, 89,065, 4321, 7654, 5489, 512} using moduli set {111, 115, 119} are displayed in [15].

In [15], the complete synthesis report showing device utilization summary is presented. Due to the additional implementation of orthogonal signaling compared to the implementation in [15], the following parameters are different compared to results displayed in [15]:

The device utilization summary is as follows:

• Number of slices: 5411 out of 89,088 6%

• Number of slice flip flops: 60 out of 178,176 0%

• Number of four input LUTs: 7452 out of 178,176 4%

• Total REAL time to Router completion: 24 min 7 s.

• Total REAL time to place and route (PAR) completion: 24 min 41 s.

• Pin delays less than 1.00 ns: 21928 out of 30,651 71.5%.

## 5. Cryptanalysis of cross-layer scheme

The cryptanalysis will be performed separately at the application and physical layers and later combined in the cross-layer scheme to demonstrate the high security level of the new scheme compared to separate implementations.

### 5.1 Cryptanalysis at the application layer

The RSA public key cryptography is implemented at the application layer. The most successful method to break the RSA cryptosystem is the Number Field Sieve (NFS) method used for partial key exposure attacks. The NFS is based on a method known as "Fermat Factorization": one tries to find integers x, y, such that $x^2 \equiv y^2$ mod n but $x \neq \pm y$ mod n [12]. We assume that the two primes p and q should be close and approximately equal to the square root of n, where n = p.q. If one of the integers could be written as x = (p + q)/2 then number of steps, $S_1$ required to determine the other integer, y could be computed as follows [23].

$$S_1 = \frac{p + q}{2} - \sqrt{n} = \frac{\left(\sqrt{q} - \sqrt{p}\right)^2}{2} = \frac{(\sqrt{n} - p)^2}{2\,p} \tag{10}$$

It is partial key exposure attack since the number of steps, $S_1$ required for the attack depends on one of the primes.

**Table 5** gives a summary of the number of steps required to break the traditional RSA cryptography implemented at the application layer using Fermat Factorization.

### 5.2 Cryptanalysis at the physical layer

Security at the physical layer is ensured by the multi-level convolutional cryptosystem which encrypts already encrypted data emanating from the RNS-based RSA. The cryptanalysis of the multi-level convolutional cryptosystem will be based on the ciphertext-only attack whereby, it is assumed that the attacker knows ciphertext of several messages encrypted with the same key and/or several keys. The keys used in the encryption are those mentioned in Section 2.1.2 for the nonlinear (8, 8, 2) 2-cascaded convolutional cryptosystem.

It is shown in [12] that, for an (n, k, L) convolutional code, each generator matrix reveals at most p – k – 1 values of a private parameter, using Gaussian elimination for p blocks of input data. Hence, if q is the number of states, then to completely break the (k, k, L) N-cascaded cryptosystem, the minimum number of plaintext-ciphertext pairs (u, v) required is [12].

| Operand key length | Total number of steps |
| --- | --- |
| 16-bit | 1 |
| 32-bit | 1 |
| 64-bit | $1.8 \times 10^8$ |
| 128-bit | $8.0 \times 10^{17}$ |
| 256-bit | $1.26 \times 10^{25}$ |
| 512-bit | $2.53 \times 10^{63}$ |
| 1024-bit | $3.3 \times 10^{140}$ |

**Table 5.**
*Number of steps required to break the traditional RSA.*

$$S_2 = \left\{ \left[ p . \left[ \frac{q.2^k}{p-k-1} \times \frac{q.2^k}{p} \times \frac{k!}{p} \times \frac{1}{p} \times \left( \frac{k}{2} \right)^2 \times 2^2 \right] \right]^N \right\} \quad (11)$$

For an (8, 8, 2) 2-cascaded cryptosystem, k = 8 and the least number of plaintext-ciphertext blocks required is p = 10 due to the number of rows and columns in the generator matrix. Assuming q = 2 states, $S_2$ could be as

$$S_2 = \left\{ \left[ 10 . \left[ \frac{2.2^8}{10-8-1} \times \frac{2.2^8}{10} \times \frac{8!}{10} \times \frac{1}{10} \times \left( \frac{8}{2} \right)^2 \times 2^2 \right] \right]^2 \right\} = 7.8 \times 10^{34} \quad (12)$$

**Table 6** gives a summary of the number of steps required to break the (8, 8, 2) 2-cascaded cryptosystem using ciphertext-only attack.

### 5.3 Cryptanalysis of the new cross-layer security scheme

At the upper layer, huge key lengths such as 1024 bits and 2048 bits are used to implement the RSA. Such implementations will greatly compromise throughput at the physical layer due to modular exponentiation. Hence, the main objective of the new cross-layer security scheme is to increase security level at the physical layer despite the small valued data points transmitted derived from the RNS-based RSA in order to enhance throughput. Cryptanalysis is performed on the small residue RSA encrypted values. The analysis will be based on partial key exposure and ciphertext-only attacks at the physical layer for eavesdropper who could wiretap the transmitted data. The number of steps, S required to break the new cross-layer security scheme should be a product of $S_1$ and $S_2$ given as [12].

$$S = S_1 . \left\{ \left[ p . \left[ \frac{q.2^k}{p-k-1} \times \frac{q.2^k}{p} \times \frac{k!}{p} \times \frac{1}{p} \times \left( \frac{k}{2} \right)^2 \times 2^2 \right] \right]^N \right\} \quad (13)$$

**Table 7** gives a summary of the number of steps required to break the new cross-layer security scheme by using partial key exposure attack and ciphertext-only attacks for different cascaded stages.

Comparing **Tables 5–7**, it can be seen that high security levels comparable to the traditional 1024-bit RSA implemented at the upper layer could be attained using short operand key lengths of 128 bits and 256 bits for cross-layer security implemented at the physical layer. It is worth noting that, the security level could be much higher compared to the values displayed in **Table 7** if the S-boxes were implemented using 4-bit and 8-bit shuffling instead of the aforementioned 2-bit shuffling.

| Operand key length | Total number of steps |
|---|---|
| 8-bit | $7.8 \times 10^{34}$ |
| 16-bit | $2.1 \times 10^{57}$ |
| 32-bit | $1.4 \times 10^{57}$ |
| 64-bit | $1.5 \times 10^{62}$ |
| 128-bit | $6.1 \times 10^{71}$ |
| 256-bit | $9.87 \times 10^{87}$ |
| 512-bit | $2.68 \times 10^{112}$ |
| 1024-bit | $1.42 \times 10^{134}$ |

**Table 6.**
*Number of steps required to break the (8, 8, 2) 2-cascaded cryptosystem.*

| Operand key length | Total number of steps | | |
|---|---|---|---|
| | N = 2 | N = 3 | N = 4 |
| 16-bit | $2.1 \times 10^{57}$ | $9.6 \times 10^{85}$ | $4.4 \times 10^{114}$ |
| 32-bit | $1.4 \times 10^{57}$ | $5.2 \times 10^{85}$ | $1.9 \times 10^{114}$ |
| 64-bit | $1.5 \times 10^{62}$ | $1.8 \times 10^{93}$ | $2.3 \times 10^{124}$ |
| 128-bit | $6.1 \times 10^{71}$ | $4.8 \times 10^{107}$ | $3.7 \times 10^{143}$ |
| 256-bit | $9.87 \times 10^{87}$ | $9.8 \times 10^{131}$ | $9.7 \times 10^{175}$ |
| 512-bit | $2.68 \times 10^{112}$ | $4.4 \times 10^{168}$ | $7.2 \times 10^{224}$ |

**Table 7.**
*Number of steps required to break the cross-layer security scheme.*

## 6. Data throughput quantification

The data throughput, T could be given as [24].

$$T = R(1–P_e)^N \qquad (14)$$

where $P_e$ is the bit error probability, N is the number of bits in the block length and R is a fixed transmission rate for the frames. For $P_e << 1$, the throughput could approximate to

$$T \cong R(1–NP_e) \qquad (15)$$

From (Eq. (15)) it could be seen that, for a fixed transmission rate, R the throughput, T could be increased by either minimizing N or $P_e$. In this section, it will be shown how convolutional coding could be used to achieve both conditions through orthogonal signaling and forward error correction respectively.

### 6.1 Coded orthogonal signaling

It is shown in [16, 25] that, for coded orthogonal signaling, the bit error probability to transmit k-bit symbols is as follows:

$$P_b \leq \left( \frac{2^{k-1}}{2^k - 1} \right) \sum_{3}^{2^k} a_d \left[ \frac{4 \left( 1 + \frac{k}{L} \overline{\gamma}_b \right)}{\left( 2 + \frac{k}{L} \overline{\gamma}_b \right)^2} \right]^L \qquad (16)$$

where $a_d$ denotes the number of paths of distance d from the all-zero path which merge with the all-zero path for the first time and $d_{free} = 3$ in this case, is the minimum distance of the code. $d_{free}$ is also equal to the diversity, L. $\overline{\gamma}_b$ is the average signal-to-noise ratio (SNR) per bit [25]. For each convolutional code, the transfer function is obtained and the sum of the coefficients $\{a_d\}$ calculated.

For illustrative purposes, the transfer function, T(D) of smaller convolutional codes such as (2, 2, 2) and (4, 4, 2) will be used.

The transfer function T(D) for the (2, 2, 2) code is given as follows [16]:

$$T(D) = D^3 + 2\,D^4 + - - - \tag{17}$$

The transfer function for the (4, 4, 2) code is given as follows [16]:

$$T(D) = D^3 + 2D^4 + 3D^5 + 5D^6 + 9D^7 + 16D^8 + 28D^9 + 49D^{10} + 85D^{11} + - - - \tag{18}$$

For both the (2, 2, 2) code and the (4, 4, 2) code, $d_{free} = L = 3$. Using the values of L and $\{a_d\}$, the probability of a binary digit error, $P_b$ as a function of the SNR per bit, $\overline{\gamma}_b$ is shown in **Figure 7** for k = 2 and 4 [16].

The curves illustrate that, the error probability increases with an increase in k for the same value of SNR. Hence, better performance for wireless transmission should involve lower order codes and many independent parallel channels rather than higher order codes with fewer independent parallel channels. Hence, high data throughput could be attained by using small number of bits in the block length, N.

## 6.2 Forward error correction (FEC) code

The Viterbi algorithm [25] is the most extensively decoding algorithm for Convolutional codes and has been widely deployed for forward error correction in wireless communication systems. In this sub-section Viterbi algorithm will be applied to the non-linear convolutional code. The constraint length, L for a (n,k,m) convolutional code is given as L = k(m-1). The constraint length is very essential in convolutional encoding since a Trellis diagram which gives the best encoding representation populates after L bits. Hence to encode blocks of n bits, each block has to be terminated by L zeros (0 s) before encoding.

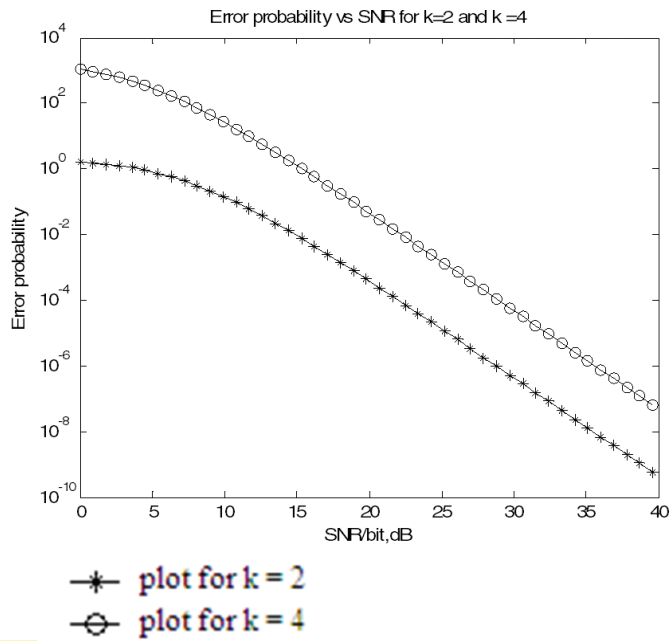

**Figure 7.**
*Performance of coded orthogonal signaling for k = 2 and k = 4.*

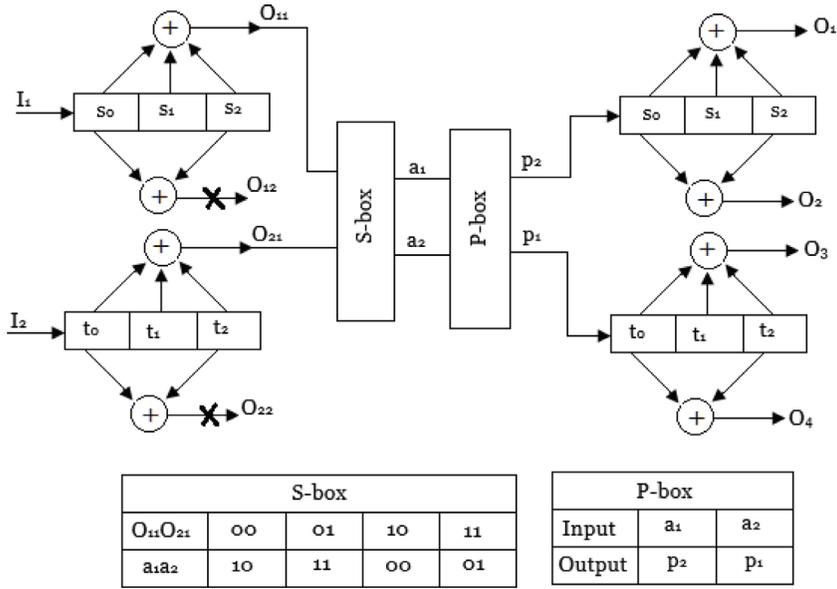

**Figure 8.**
*2-stage non-linear (4,2,3) convolutional code.*

For illustrative purposes, a non-linear (4,2,3) convolutional code will be used to demonstrate encoding and Viterbi decoding. A possible non-linear (4,2,3) convolutional code showing mod-2 connections and the product cipher is shown in **Figure 8**.

### 6.2.1 Example: encode/decode the message M = 10,011

• Encoding process

The constraint length, L = k(m-1) = 2(3–1) = 4.

Hence 4 zeros will be appended to message M before encoding. The modified message becomes M' = 10110000. Transition tables in appendix are used to encode the modified message.

   a. Using transition tables in appendix, the transmitted sequence from the 1st stage is given as $T_{in}$ = 10 01 01 11

   b. S-box output is given as S = 00 11 11 01

   c. P-box output is given as P = 00 11 11 10

   d. Transmitted sequence into the 2nd stage is given as P = 00 11 11 10

   e. Using transition tables in appendix, the final transmitted sequence which is the output bits from the 2nd stage is given as $T_{out}$ = 0000 1111 0101 1001

• Viterbi decoding process

In performing the Viterbi algorithm, a bit in the sequence $T_{out}$ will be altered. Let the received sequence be $T_R$ = 1000 1111 0101 1001 instead of $T_{out}$ = 0000 1111 0101 1001. The Viterbi algorithm applied to the 2nd stage is summarized in **Table 8**.

| Incoming bits | Input State | Output bits | Output state | Present metric | Cumulative metric |
|---|---|---|---|---|---|
| 1000 | 0000 | 0000 00 | 0000 | 3 | 3 |
| | 0000 | 0011 | 0010 | 1 | 1 |
| | 0000 | 1100 | 1000 | 3 | 3 |
| | 0000 | 1111 | 1010 | 1 | 1 |
| 1111 | 0000 | 0000 | 0000 | 0 | 3 |
| | 0000 | 0011 | 0010 | 2 | 5 |
| | 0000 | 1100 | 1000 | 2 | 5 |
| | 0000 | 1111 11 | 1010 | 4 | 7 |
| | 1000 | 1000 | 0100 | 1 | 4 |
| | 1000 | 1011 | 0110 | 3 | 6 |
| | 1000 | 0100 | 1100 | 1 | 4 |
| | 1000 | 0111 | 1110 | 3 | 6 |
| 0101 | 1010 | 1010 | 0101 | 0 | 7 |
| | 1010 | 1001 | 0111 | 2 | 9 |
| | 1010 | 0110 | 1101 | 2 | 9 |
| | 1010 | 0101 11 | 1111 | 4 | 11 |
| 1001 | 1111 | 0101 | 0101 | 2 | 13 |
| | 1111 | 0110 | 0111 | 0 | 11 |
| | 1111 | 1001 10 | 1101 | 4 | 15 |
| | 1111 | 1010 | 1111 | 2 | 13 |

**Table 8.**
*Viterbi algorithm applied to 2nd stage of (4,2,3) code.*

| Incoming bits | Input State | Output bits | Output state | Present metric | Cumulative metric |
|---|---|---|---|---|---|
| 10 | 0000 | 00 | 0000 | 1 | 1 |
| | 0000 | 01 | 0010 | 0 | 0 |
| | 0000 | 10 10 | 1000 | 2 | 2 |
| | 0000 | 11 | 1010 | 1 | 1 |
| 01 | 1000 | 10 | 0100 | 0 | 2 |
| | 1000 | 11 | 0110 | 1 | 2 |
| | 1000 | 00 | 1100 | 1 | 3 |
| | 1000 | 11 01 | 1110 | 2 | 4 |
| 01 | 1110 | 00 01 | 0101 | 2 | 6 |
| | 1110 | 00 | 0111 | 1 | 5 |
| | 1110 | 11 | 1101 | 1 | 5 |
| | 1110 | 10 | 1111 | 0 | 4 |
| 11 | 0101 | 00 11 | 0000 | 2 | 8 |
| | 0101 | 10 | 0010 | 1 | 7 |
| | 0101 | 01 | 1000 | 1 | 7 |
| | 0101 | 00 | 1010 | 0 | 6 |

**Table 9.**
*Viterbi algorithm applied to 1st stage of (4,2,3) code.*

The bits above the arrows will constitute the retrieved sequence from the 2nd stage. Hence, the retrieved sequence is given as, $R_1$ = 00 11 11 10. This sequence is fed to the P-box.

- P-box output is given as P1 = 00 11 11 01. Sequence, P1 is fed to the S-box

- S-box output is given as S1 = 10 01 01 11

Sequence, S1 is fed into the 1st stage to retrieve the final correct message. The Viterbi algorithm applied to the 1st stage is summarized in **Table 9**.

For a good trellis, the final state is the all-zero state as seen in the winning path in **Table 9**. The final received sequence is identical to the original transmitted message of M' = $R_{final}$ = 10110000 despite the first bit error. Hence, using the non-linear convolutional code, the error bit was identified and corrected. The forward error correction capability will therefore enhance throughput, since the bit error rate, $P_e$ is reduced.

## 7. Security attacks in wireless networks

Most attacks in wireless networks are classified into two categories: passive and active. Passive attacks such as eavesdropping and traffic analysis do not interfere with normal network operations as opposed to active attacks. Some of the attacks could be circumvented by the cross-layer security scheme presented in this research due to the following characteristics inherent in its implementation:

- RSA cryptographic algorithm at the upper layer: **Table 3** summarized the security services, which could be achieved by implementing RSA cryptography such as authentication, non-repudiation and confidentiality. These services are essential network security requirements which are vital in curbing attacks such as eavesdropping, masquerade attack and information disclosure since there will be a possibility of not attaining the final all-zero state if message is modified.

- Convolutional cryptosystem at the physical layer: The different keys generated are essential in ensuring confidentiality while the forward error correction capability is essential in curbing message modification attack.

Other attacks such as denial of service and replay attack could be circumvented if the cross-layer security scheme is associated with Transmission Control Protocol (TCP).

## 8. Conclusions

In this chapter, we have described a new cross-layer security scheme which has the advantage of enhancing both security and throughput as opposed to existing schemes which either enhances security or throughput but not both. The new scheme is implemented using the residue number system (RNS), non-linear convolutional coding and subband coding at the physical layer and RSA cryptography at the upper layers. By using RSA cryptography, the scheme could be used in encryption, authentication and non-repudiation with efficient key management as opposed to existing schemes, which had poor key management for large wireless networks since their implementation, was based on symmetric encryption techniques. Results show that, the new algorithm exhibits high security level for key sizes of 64, 128 and 256 bits when using three or more convolutional-cascaded stages. The security level is far above the traditional 1024-bit RSA which is already vulnerable. The vulnerability of 1024-bit RSA has led to the proposal of implementing higher levels such as 2048-bit and 4096-bit. These high level RSA schemes when implemented will greatly compromise throughput due to modular exponentiation. Hence the usefulness of a scheme such as the one presented in this chapter. In addition, Viterbi algorithm was performed for the new non-linear convolutional code in order to highlight the error correction capability. It was shown that, by using error correction codes on many small block lengths compared to one huge block length, throughput increases. Hence non-linear convolutional code is very critical in the implementation of the new scheme, since it contributes in enhancing both security and throughput. The entire scheme could be implemented at different access points in a wireless network since it fits in a single FPGA. Finally, the new cross-layer security scheme is essential in circumventing some attacks in wireless and computer networks.

## A. Appendix: transition tables

### Stage 1

| Input Bits (stage 1) $I_1$ $I_2$ | Input State (Stage 1) $S_1 S_2 I_1 I_2$ | Output Bits (Stage 1) $O_{11}$ $O_{12}$ | Output State (Stage 1) $S_1 S_2 I_1 I_2$ |
|---|---|---|---|
| 0 0 | 0000 | 0 0 | 0000 |
| 0 1 | 0000 | 0 1 | 0010 |
| 1 0 | 0000 | 1 0 | 1000 |
| 1 1 | 0000 | 1 1 | 1010 |
| 0 0 | 0001 | 0 1 | 0000 |
| 0 1 | 0001 | 0 0 | 0010 |
| 1 0 | 0001 | 1 1 | 1000 |
| 1 1 | 0001 | 1 0 | 1010 |
| 0 0 | 0010 | 0 1 | 0001 |
| 0 1 | 0010 | 0 0 | 0011 |
| 1 0 | 0010 | 1 1 | 1011 |
| 1 1 | 0010 | 1 0 | 1011 |
| 0 0 | 0011 | 0 0 | 0001 |
| 0 1 | 0011 | 0 1 | 0011 |
| 1 0 | 0011 | 1 0 | 1001 |
| 1 1 | 0011 | 1 1 | 1011 |
| 0 0 | 0100 | 1 0 | 0000 |
| 0 1 | 0100 | 1 1 | 0010 |
| 1 0 | 0100 | 0 0 | 1000 |
| 1 1 | 0100 | 0 1 | 1010 |
| 0 0 | 0101 | 1 1 | 0000 |
| 0 1 | 0101 | 1 0 | 0010 |
| 1 0 | 0101 | 0 1 | 1000 |
| 1 1 | 0101 | 0 0 | 1010 |
| 0 0 | 0110 | 1 1 | 0001 |
| 0 1 | 0110 | 1 0 | 0011 |
| 1 0 | 0110 | 0 1 | 1001 |
| 1 1 | 0110 | 0 0 | 1011 |
| 0 0 | 0111 | 1 0 | 0001 |
| 0 1 | 0111 | 1 1 | 0011 |
| 1 0 | 0111 | 0 0 | 1001 |
| 1 1 | 0111 | 0 1 | 1011 |
| 0 0 | 1000 | 1 0 | 0100 |
| 0 1 | 1000 | 1 1 | 0110 |
| 1 0 | 1000 | 0 0 | 1100 |
| 1 1 | 1000 | 0 1 | 1110 |
| 0 0 | 1001 | 1 1 | 0100 |
| 0 1 | 1001 | 1 0 | 0110 |
| 1 0 | 1001 | 0 1 | 1100 |
| 1 1 | 1001 | 0 0 | 1110 |
| 0 0 | 1010 | 1 1 | 0101 |
| 0 1 | 1010 | 1 0 | 0111 |
| 1 0 | 1010 | 0 1 | 1101 |
| 1 1 | 1010 | 0 0 | 1111 |
| 0 0 | 1011 | 1 0 | 0101 |
| 0 1 | 1011 | 1 1 | 0111 |
| 1 0 | 1011 | 0 0 | 1101 |
| 1 1 | 1011 | 0 1 | 1111 |
| 0 0 | 1100 | 0 0 | 0100 |
| 0 1 | 1100 | 0 1 | 0110 |
| 1 0 | 1100 | 1 0 | 1100 |
| 1 1 | 1100 | 1 1 | 1110 |
| 0 0 | 1101 | 0 1 | 0100 |
| 0 1 | 1101 | 0 0 | 0110 |
| 1 0 | 1101 | 1 1 | 1100 |
| 1 1 | 1101 | 1 0 | 1110 |
| 0 0 | 1110 | 0 1 | 0101 |
| 0 1 | 1110 | 0 0 | 0111 |
| 1 0 | 1110 | 1 1 | 1101 |
| 1 1 | 1110 | 1 0 | 1111 |
| 0 0 | 1111 | 0 0 | 0101 |
| 0 1 | 1111 | 0 1 | 0111 |
| 1 0 | 1111 | 1 0 | 1101 |
| 1 1 | 1111 | 1 1 | 1111 |

### Stage 2

| Input Bits (Stage 2) $P_1$ $P_2$ | Input State (Stage 2) $S_1 S_2 I_1 I_2$ | Output Bits (Stage 2) $O_1$ $O_2$ $O_3$ $O_4$ | Output State (Stage 2) $S_1 S_2 I_1 I_2$ |
|---|---|---|---|
| 0 0 | 0000 | 0 0 0 0 | 0000 |
| 0 1 | 0000 | 0 0 1 1 | 0010 |
| 1 0 | 0000 | 1 1 0 0 | 1000 |
| 1 1 | 0000 | 1 1 1 1 | 1010 |
| 0 0 | 0001 | 0 0 1 1 | 0000 |
| 0 1 | 0001 | 0 0 0 0 | 0010 |
| 1 0 | 0001 | 1 1 1 1 | 1000 |
| 1 1 | 0001 | 1 1 0 0 | 1010 |
| 0 0 | 0010 | 0 0 1 0 | 0001 |
| 0 1 | 0010 | 0 0 0 1 | 0011 |
| 1 0 | 0010 | 1 1 1 0 | 1011 |
| 1 1 | 0010 | 1 1 0 1 | 1011 |
| 0 0 | 0011 | 1 1 0 1 | 0001 |
| 0 1 | 0011 | 1 1 1 0 | 0011 |
| 1 0 | 0011 | 1 1 1 1 | 1001 |
| 1 1 | 0011 | 1 1 1 0 | 1011 |
| 0 0 | 0100 | 1 1 0 0 | 0000 |
| 0 1 | 0100 | 1 1 1 1 | 0010 |
| 1 0 | 0100 | 0 0 0 0 | 1000 |
| 1 1 | 0100 | 0 0 1 1 | 1010 |
| 0 0 | 0101 | 1 1 1 1 | 0000 |
| 0 1 | 0101 | 1 1 0 0 | 0010 |
| 1 0 | 0101 | 0 0 1 1 | 1000 |
| 1 1 | 0101 | 0 0 0 0 | 1010 |
| 0 0 | 0110 | 1 1 1 0 | 0001 |
| 0 1 | 0110 | 1 1 0 1 | 0011 |
| 1 0 | 0110 | 0 0 1 0 | 1001 |
| 1 1 | 0110 | 0 0 0 1 | 1011 |
| 0 0 | 0111 | 1 1 0 1 | 0001 |
| 0 1 | 0111 | 1 1 1 0 | 0011 |
| 1 0 | 0111 | 0 0 0 1 | 1001 |
| 1 1 | 0111 | 0 0 1 0 | 1011 |
| 0 0 | 1000 | 1 0 0 0 | 0100 |
| 0 1 | 1000 | 1 0 1 1 | 0110 |
| 1 0 | 1000 | 0 1 0 0 | 1100 |
| 1 1 | 1000 | 0 1 1 1 | 1110 |
| 0 0 | 1001 | 1 0 1 1 | 0100 |
| 0 1 | 1001 | 1 0 0 0 | 0110 |
| 1 0 | 1001 | 0 1 1 1 | 1100 |
| 1 1 | 1001 | 0 1 0 0 | 1110 |
| 0 0 | 1010 | 1 0 1 0 | 0101 |
| 0 1 | 1010 | 1 0 0 1 | 0111 |
| 1 0 | 1010 | 0 1 1 0 | 1101 |
| 1 1 | 1010 | 0 1 0 1 | 1111 |
| 0 0 | 1011 | 1 0 0 1 | 0101 |
| 0 1 | 1011 | 1 0 1 0 | 0111 |
| 1 0 | 1011 | 0 1 0 1 | 1101 |
| 1 1 | 1011 | 0 1 1 0 | 1111 |
| 0 0 | 1100 | 0 1 0 0 | 0100 |
| 0 1 | 1100 | 0 1 1 1 | 0110 |
| 1 0 | 1100 | 1 0 0 0 | 1100 |
| 1 1 | 1100 | 1 0 1 1 | 1110 |
| 0 0 | 1101 | 0 1 1 1 | 0100 |
| 0 1 | 1101 | 0 1 0 0 | 0110 |
| 1 0 | 1101 | 1 0 1 1 | 1100 |
| 1 1 | 1101 | 1 0 0 0 | 1110 |
| 0 0 | 1110 | 0 1 1 0 | 0101 |
| 0 1 | 1110 | 0 1 0 1 | 0111 |
| 1 0 | 1110 | 1 0 1 0 | 1101 |
| 1 1 | 1110 | 1 0 0 1 | 1111 |
| 0 0 | 1111 | 0 1 0 1 | 0101 |
| 0 1 | 1111 | 0 1 1 0 | 0111 |
| 1 0 | 1111 | 1 0 0 1 | 1101 |
| 1 1 | 1111 | 1 0 1 0 | 1111 |

## Author details

Michael Ekonde Sone
College of Technology, University of Buea, Buea, Cameroon

*Address all correspondence to: michael.sone@ubuea.cm

IntechOpen

# Anomaly-Based Intrusion Detection System

*Veeramreddy Jyothsna and Koneti Munivara Prasad*

## Abstract

Anomaly-based network intrusion detection plays a vital role in protecting networks against malicious activities. In recent years, data mining techniques have gained importance in addressing security issues in network. Intrusion detection systems (IDS) aim to identify intrusions with a low false alarm rate and a high detection rate. Although classification-based data mining techniques are popular, they are not effective to detect unknown attacks. Unsupervised learning methods have been given a closer look for network IDS, which are insignificant to detect dynamic intrusion activities. The recent contributions in literature focus on machine learning techniques to build anomaly-based intrusion detection systems, which extract the knowledge from training phase. Though existing intrusion detection techniques address the latest types of attacks like DoS, Probe, U2R, and R2L, reducing false alarm rate is a challenging issue. Most network IDS depend on the deployed environment. Hence, developing a system which is independent of the deployed environment with fast and appropriate feature selection method is a challenging issue. The exponential growth of zero-day attacks emphasizing the need of security mechanisms which can accurately detect previously unknown attacks is another challenging task. In this work, an attempt is made to develop generic meta-heuristic scale for both known and unknown attacks with a high detection rate and low false alarm rate by adopting efficient feature optimization techniques.

**Keywords:** intrusion detection, data mining, classification based, DoS, Probe, U2R, R2L, false alarm rate, zero-day attacks

## 1. Introduction

### 1.1 Internet security

Today, the world has numerous inventions and technological developments with proliferation of the Internet. Advances in business forced the organizations and governments worldwide to invent and use sophisticated and modern networks. These networks mix a variety of security aspects such as encryption, data integrity, authentication, and technologies like distributed storage systems, voice over Internet protocol (VoIP), wireless access, and web services.

Enterprises are more available to these systems. For instance, numerous business associations enable access to their administration on the system through intranet and web to their partners; endeavors empower clients to connect with the systems by means of web-based business exchanges that enable representatives to get to data by methods for virtual private systems. This usage makes it more vulnerable to attacks and intrusions. A security threat comes not only from the external intruders but also from internal user in the form of abuse and misuse. A firewall simply blocks the network but cannot protect against intrusion attempts. In contrast, intrusion detection system (IDS) can monitor the abnormal activities on the network.

## 1.2 Intrusion detection systems (IDS)

Intrusion detection systems play a vital role in research and development with an increase in attacks on computers and networks [1]. Intrusion detection systems monitor the events occurring in a computer system or networks for analyzing the patterns of intrusions. IDS examine a host or network to spot the potential intrusions. Host-based systems explore the system calls and process identifiers mainly related to the operating system data. On the other hand, network-based systems analyze network-related events like traffic volume, IP address, service ports, and protocol used. Intrusion detection systems will

i.  analyze and monitor the system and user activities;

ii.  assess the integrity of critical system and data files; and

iii.  provide statistical analysis of activity patterns.

## 1.3 Taxonomy of intrusion detection systems

The intrusion detection systems are broadly classified as

i.  misuse detection systems and

ii.  anomaly-based detection systems.

### 1.3.1 Misuse detection systems

A misuse detection system is also called as signature-based detection that uses recognized patterns [2]. These patterns describe suspect, collection of sequences of activities or operations that can be possibly be harmful and stored in database. It uses well-defined patterns of the attack that exploits the weaknesses in system. The time taken to match with the patterns stored in the database is minimal. A key benefit of these systems is that the patterns or signatures can easily develop and understand the network behavior if familiar. It is more efficient to handle the attacks whose patterns are already maintained in the database.

The major restriction of these signature-based approaches is that they can only detect the intrusions whose attack patterns are already stored in the database. For every attack, its signature is to be created. Attacks whose patterns are not present in the database cannot be detected. Such technique can be easily deceived as they are dependent on a specific set of expressions and string matching. In addition, the signature works well only against fixed behavioral patterns; they fail to handle the attacks with human interference or attacks with inherent self-modifying behavioral characteristics.

These detection systems are also ineffective in cases where client works on new technology platforms such as no operation (NoP) generators, encoding, and decoding payloads. The efficiency of the signature-based systems decreases due to the need of creating dynamic signatures for different variations. With growing volume of signatures, the performance of the engine also might lose the momentum. Because of this, intrusion detection frameworks are conducted on multiprocessors and Gigabit cards. IDS developers develop new signatures before the attackers develop solutions, in order to prevent any new kind of attacks on the system.

### 1.3.2 Anomaly-based detection systems

Network behavior is the major parameter on which the anomaly detection systems rely upon. If the network behavior is within the predefined behavior, then the network transaction is accepted or else it triggers the alert in the anomaly detection system [3]. Acceptable network performance can be either predetermined or learned through specifications or conditions defined by the network administrator.

The crucial stage of behavior determination is regarding the ability of detection system engine toward multiple protocols at each level. The IDS engine must be able to understand the process of protocols and its goal. Despite the fact that the protocol analysis is very expensive in terms of computation, the benefits like increasing rule set assist in lesser levels of false-positive alarms.

Defining the rule sets is one of the key drawbacks of anomaly-based detection. The efficiency of the system depends on the effective implementation and testing of rule sets on all the protocols. In addition, a variety of protocols that are used by different vendors impact the rule defining the process.

In addition to the aforesaid, custom protocols also add complexity to the process of rule defining. For accurate detection, the administration should clearly understand the acceptable network behavior. However, with strong incorporation of rules and protocol, the anomaly detection procedure would likely to perform more efficiently.

However, if the malicious behavior falls under the accepted behavior, in such conditions it might get unnoticed. The major benefit of the anomaly-based detection system is about the scope for detection of novel attacks. This type of intrusion detection approach could also be feasible, even if the lack of signature patterns matches and also works in the condition that is beyond regular patterns of traffic.

## 2. Network intrusion detection systems framework

In **Figure 1**, common intrusion detection framework (CIDF) integrated with Internet Engineering Tasks Force (IETF) and Intrusion Detection Working Group (IDWG) has successfully achieved efficient performance in representing the framework. This group defines a basic IDS structural design based on four functional modules.

*Event modules (E-Modules)* are defined as a combination of sensing elements and are engaged in continuous monitoring of the end system. In addition, these modules are also involved in processing the information events to the bottom three modules for further analysis.

*Analysis modules (A-Modules)* analyze the events and detect probable aggressive behavior, in order to ensure that some kind of alarm generated in essential conditions.

*Data storage modules (D-modules)* store the data from the E-Modules for further processing by the other modules.

*Response modules (R-Modules)* are used to provide the response to the transactions based on the information obtained from the analysis module.

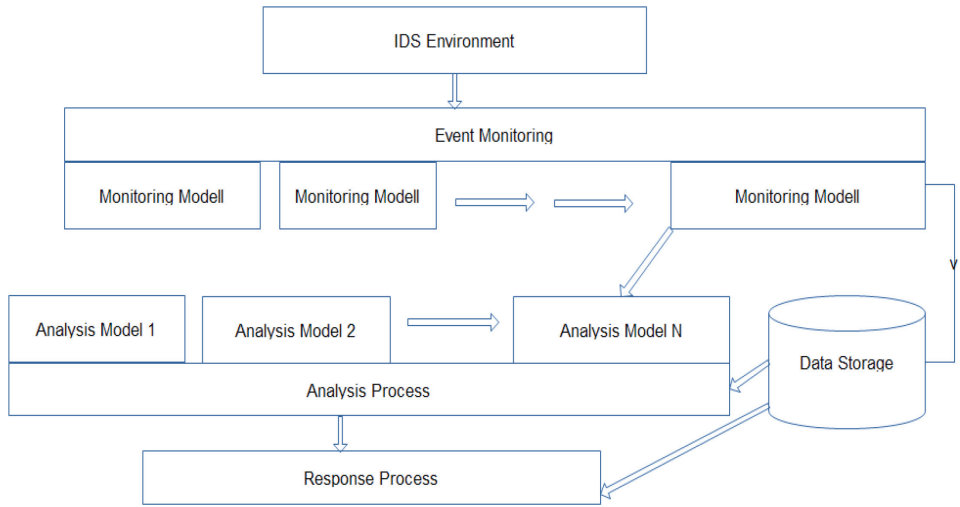

**Figure 1.**
*Common intrusion detection framework architecture.*

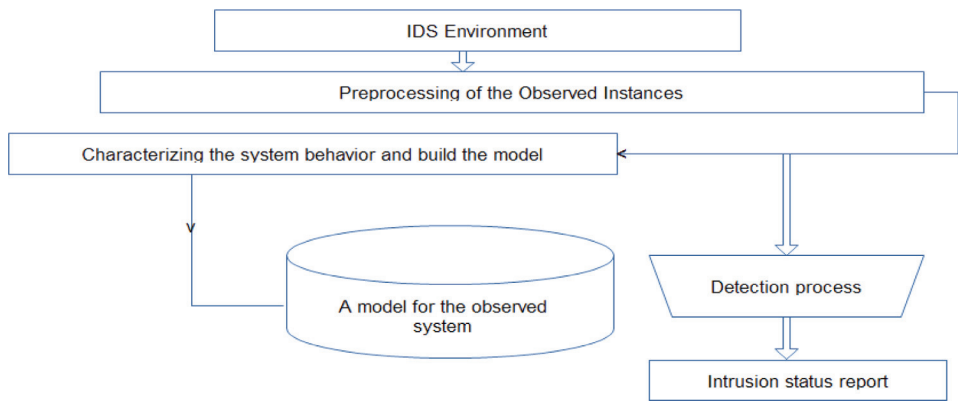

**Figure 2.**
*Common anomaly-based network IDS.*

**Figure 2** represent the Common anomaly-based network IDS. The functional stages normally adopted in the anomaly-based network intrusion detection systems (ANIDS) are as follows:

*Formation of attributes:* In this stage, preprocessing of the attributes is done based on the target system.

*Observation stage:* A model that is built on the basis of behavioral features of the specified system where observations of intrusions can be carried out either through automatically or by manual detection procedure.

*Functional stage*: It is also called as detection stage. If the characterizing system model is available, it will match with the observed traffic.

## 3. Anomaly-based intrusion detection techniques

**Figure 3** represents the taxonomy of anomaly-based intrusion detection techniques. They are statistical based, cognitive based or knowledge based, machine learning or soft

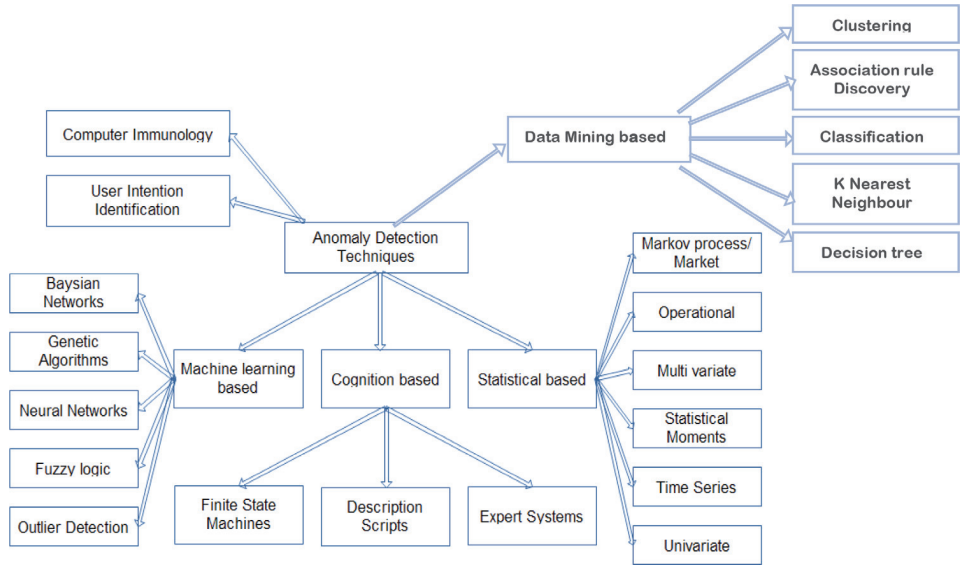

**Figure 3.**
*Classification of anomaly-based intrusion detection techniques.*

computing based, data mining based, user intention identification, and computer immunology.

### 3.1 Statistical-based techniques

Statistical-based techniques use statistical properties such as mean and variance on normal transaction to build the normal profile [4]. The statistical tests are employed to determine whether the observed transaction deviates from the normal profile. The IDS assigns a score to the transactions whose profile deviates from the normal. If the score reaches the threshold, alarm is raised. The threshold value is set based on count of events that occur over a period of time.

Statistical-based techniques are further classified into operational model or threshold metric, time series model, Markov process model or Marker model, parametric approaches, statistical moments or mean and standard deviation model, multivariate model, and nonparametric approaches.

The main advantages of statistical-based techniques are as follows:

i. They do not require any prior knowledge about the signatures of the attacks. So, they can detect zero-day attacks.

ii. As the system is not depended on any of the signatures, updating is not required. Hence it is easy to maintain.

iii. The intrusion activities that were occurred over extended period of time can be identified accurately and are good at detecting DoS attacks.

The disadvantages of statistical-based techniques are as follows:

i. They need accurate statistical distributions.

ii. The learning process of statistical-based techniques takes days or weeks to become accurate and effective.

### 3.2 Cognitive-based or knowledge-based techniques

Knowledge-based techniques are used to extract the knowledge from the specific attacks and system vulnerabilities. This knowledge can be further used to identify the intrusions or attacks happening in the network or system. They generate alarm as soon as an attack is detected. They can be used for both misuse and anomaly-based detection [5].

The knowledge-based techniques are broadly classified as state transition analysis, expert systems, and signature analysis.

The knowledge-based techniques possess good accuracy and very low false alarm rates. The knowledge gathered makes security analyst easier to take preventive or corrective action.

The knowledge-based techniques are maintaining the knowledge of each attack based on the careful and detailed analysis performed; it is a time-consuming task. A prior knowledge to update the each attack is a difficult task.

### 3.3 Data mining-based techniques

The knowledge-based IDS can detect the attacks whose patterns are known, but it is difficult to detect the inside attacks. One of the solutions is data mining techniques. The core idea is to extract the useful patterns and also the previously ignored patterns from the dataset [6].

The data mining-based techniques are further classified into clustering, association rule discovery, classification, K-nearest neighbor, and decision tree methods.

The key advantages of data mining-based techniques are as follows:

  i.  They can handle high dimensional data.

 ii.  As the precomputed models are designed in the training phase, comparing each instance at the testing phase can be done in faster way.

iii.  They can generate the patterns in unsupervised mode.

The key disadvantages of data mining-based techniques are as follows:

  i.  These methods identify abnormalities as a by-product of clustering and as are not optimized for anomaly detection.

 ii.  They require high storage and are slow in classifying due to high dimensionality.

### 3.4 Machine learning or soft computing-based techniques

Machine learning can be characterized as the capacity of a program or potentially a framework to learn and improve their performance on a specific task or group of tasks over a time [7]. Machine learning strategies emphasize on building a framework that enhances its execution based on previous results, that is, it can change their execution strategy based on recently acquired data.

Machine learning-based techniques are broadly classified as Bayesian approaches, support vector machines, neural networks, fuzzy logic, and genetic algorithms. Their key advantage is flexibility, adaptability, and capture of interdependencies. The disadvantage is high algorithmic complexity and long training times.

### 3.5 User intention identification

Intrusion detection system can be built based on the features that categorize the user or the system usage, to distinguish the abnormal activities from normal activities. During the early investigation of anomaly detection, the main emphasis was on profiling system or user behavior from monitored system log or accounting log data. The log data or system log may contain UNIX shell commands, system calls, key strokes, audit events, and network packages used.

### 3.6 Computer immunology

Computer immunology is a field of science that includes high-throughput genomic and bioinformatics approaches to immunology. The main objective is to convert immunological data into computational problems, solve these problems using statistical and computational approaches, and then convert the results into immunologically meaningful interpretations.

## 4. NSL-KDD dataset

The NSL-KDD [8] dataset is a refined version of its predecessor KDD99 dataset. NSL-KDD dataset comprises close to 4,900,000 unique connection vectors, where every connection vector consists of 41 features of which 34 are continuous features and 07 are discrete features. Each vector is labeled as either normal or attack. There are four major categories of attacks labeled in NSL-KDD: denial of service attack, probing attack, users-to-root attack, and remote-to-local attack.

i. ***Denial of service attack (DoS):*** Denial of service is an attack category, which exhausts the victim's assets, thereby making it unable to handle legitimate requests. Examples of DoS attacks are "teardrop," "neptune," "ping of death (pod)," "mail bomb," "back," "smurf," and "land."

ii. ***Probing attack (PROBE):*** Objective of surveillance and other probing attacks is to gain information about the remote victim. Examples of probing attacks are "nmap," "satan," "ipsweep," and "portsweep."

iii. ***Users-to-root attack (U2R):*** The attacker enters into the local system by using the authorized credentials of the victim user and tries to exploit the vulnerabilities to gain the administrator privileges. Examples of U2R attacks are "load module," "buffer overflow," "rootkit," and "perl."

iv. ***Remote-to-local attack (R2L):*** The attackers access the targeted system or network from the remote machine and try to gain the local access of the victim machine. Examples of R2L attacks are "phf," "warezmaster," "warezclient," "spy," "imap," "ftp write," "multihop," and "guess passwd."

## 5. Issues and challenges in anomaly-based intrusion detection systems

Although many methods and systems have been developed by the research community, there are still a number of open research issues and challenges. Some of the research issues and challenges of AIDS are as follows:

i. A network anomaly-based IDS should reduce the false alarm rate. But, totally mitigating the false alarm is not possible. Developing an intrusion detection system independent of the environment is another challenge task for the network anomaly-based intrusion detection system development community [9–13].

ii. Developing a general methodology or a set of parameters that can be used to evaluate the intrusion detection system is another challenging task [12, 13].

iii. When new patterns are identified in ANIDS, updating the database without compromise of performance is another challenging task [9, 13].

iv. Another task to be addressed is to reduce the computational complexities of data preprocessing in the training phase and also in the deployment phase [9, 10].

v. Developing a suitable method for selecting the attributes for each category of attack is another important task [9–11].

vi. Identifying a best classifier from a group of classifiers that is nonassociated and unbiased to build an effective ensemble approach for anomaly detection is another challenge [9–11].

## 6. Feature optimization using canonical correlation analysis

The preprocessed set of network transactions are partitioned based on its labeling ("normal" transactions as one set, "DoS" transactions as the other set and similar other range of sets). Unique values of each feature value set $f_i v(NTS)$ in the resultant normal transactions set (NTS) and its percentage of coverage are:

$$f_i v = \{f_i(v_1, c_1), f_i(v_2, c_2), f_i(v_3, c_3), f_i(v_4, c_4), \dots\dots\dots f_i(v_j, c_j)\} \tag{1}$$

The procedure for feature optimization for each attack $A_k$ is as follows:

i. Consider the transactions set $ts(A_k)$ denoting attack type $A_k$ (as an example considers DoS as an attack).

ii. For every feature $f_i(A_k)$, consider all the values as a set $f_i v(A_k)$. An empty set $\overline{f_i v}$ of size $|f_i v(A_k)|$ is created and fills it based on its coverage as $|f_i v(A_k)| \cong |\overline{f_i v}|$, in which $|f_i v(A_k)|$ denotes the size of the feature values set of $f_i(A_k)$.

iii. The process is used to generate the feature values vector $\overline{f_i v}$ of the NTS, such that $\overline{f_i v}$ is compatible to the "$f_i v(A_k)$" toward size and that also represents the coverage ratio of the values in $f_i v(NTS)$.

iv. The process is applied for all feature values set in network transactions of attack $A_k$.

v. Find the canonical correlation between $f_i v(A_k)$ and $\overline{f_i v}$. If the resultant canonical correlation is less than the threshold or zero, then the feature $f_i(A_k)$ can be considered as optimal toward assessing the scale of intrusion scope.

It is imperative from the implementation of the above procedure that optimal features of a specific attack $A_k$ can be identified. Further, the optimal features are ordered using the canonical correlation values. The values with lower than threshold are considered as optional set of features. Reducing the features leads to lesser computational complexities to the minimal level. The optimal features shall be used for further assessing the impact scale intrusion of type $A_k$.

## 7. Feature association impact scale (FAIS)

The approach for measuring the proposed feature association support (*fas*) metric considers the network transaction of the training dataset. The feature categorical values used in the network transactions are in the form of two independent sets. These values are used to develop a duplex graph between them.

### 7.1 Assumptions

Let $\{f1, f2, f3, \ldots\ldots fn \forall f_i = \{f_i v_1, f_i v_2, \ldots\ldots\ldots, f_i v_m\}\}$ be the set of categorical features values used for forming the set of network transactions $T$. Here $T$ is a set of network transaction records of the given training set such as:

$$T = \{t_1, t_2, t_3, \ldots\ldots t_n \forall t_i = \{val(f_1), val(f_2), \ldots\ldots val(f_i), val(f_{i+1}), \ldots\ldots val(f_n)\}\} \tag{2}$$

Categorical values of the set of features related to every network transaction shall be considered as transaction value set *tvs* and all transaction value sets are treated as "*STVS*."

In the description above in Eq. 2, $val(f_i)$ can be expressed as $val(f_i) \in \{f_i v_1, f_i v_2, \ldots\ldots, f_i v_m\}$. The term "feature" refers to the current categorical value of the feature. The two features "$val(f_i)$" and "$val(f_j)$," "$val(f_i)$" are connected with "$val(f_j)$" if and only if $\left(val(f_i), val(f_j)\right) \in tvs_k$.

### 7.2 Algorithm for FAIS technique

**Step 1:** The edge weight between the features $val(f_1)$ and $val(f_2)$ is estimated as:

$$w\left(val(f_1) \leftrightarrow val(f_2)\right) = \frac{ctvs}{|STVS|} \tag{3}$$

**Step 2:** The edge weight between transaction value sets and its corresponding set of feature categorical values can be measured as:

$$E = \left\{(tvs_i, val_j) : val_j \in tvs_i, tvs_i \in STVS, val_j \in v\right\} \tag{4}$$

**Step 3:** Further assuming the transaction value sets of the given duplex graph as pivots and the feature categorical values as pure prerogatives, the pivot and prerogative values are measured.

**Step 3.1:** Consider matrix u, which denotes pivot initial value as 1.

**Step 3.2:** Transpose the matrix A as $A'$.

**Step 3.3:** Calculate prerogative weights by multiplying $A'$ with u.

**Step 3.4:** Calculate original pivot weights using matrix multiplication between A and V.

**Step 4:** Calculate the feature categorical value *fas* of $f_i v_j$ as:

$$fas(f_i v_j) = \frac{\sum_{k=1}^{|STVS|} \{u(tvs_k) : (f_i v_j \rightarrow tvs_k) \neq 0\}}{\sum_{k=1}^{|STVS|} u(tvs_k)} \tag{5}$$

**Step 5:** the Feature Association Impact Scale *fais* for every transaction value set $tvs_i$ is estimated as:

$$fais(tvs_i) = 1 - \frac{\sum_{j=1}^{m} \{fas(\{val_j \exists val_j \in V\}) : (val_j \subset tvs_i)\}}{|tvs_i|} \tag{6}$$

**Step 6:** The Feature Association Impact Scale threshold *faist* can be measured as:

$$faist = \frac{\sum_{i=1}^{|STVS|} fais(tvs_i)}{|STVS|} \tag{7}$$

**Step 7:** Calculate the standard deviation as:

$$sdv_{faist} = \sqrt{\frac{\left(\sum_{i=1}^{|STVS|} fais(tvs_i) - faist^2\right)}{(|STVS| - 1)}} \tag{8}$$

**Step 8:** The Feature Association Impact Scale range can be explored as Step 8.1 and Step 8.2:

**Step 8.1:** Calculate lower threshold of *faist* as $faist_l = faist - sdv_{faist}$.

**Step 8.2:** Calculate higher threshold of *faist* as $faist_h = faist + sdv_{faist}$.

## 8. Analysis of experimental results

The total number of records chosen for the test is 25% of the actual dataset, that is, 34,361. The combination of test records chosen is from various categories such as Probe, DoS, U2R, R2L, and Normal. The difference between CC average and standard deviation of CC is called as lower bound of CC threshold. The sum of CC average and standard deviation of CC is called as upper bound of CC threshold.

The records that identified to be normal are 19.8% of the total test data records, with observations of 4.7% of it as "false negatives" and 15.1% of it as "true negatives." The cumulative number of records that are detected as "intruded transactions" is 80.2%, with 75.3% of them being "truly intruded transactions" of test data records and the "false positive" percentage of 4.9% of test data records.

As per the results obtained, the proposed model is found to be accurate up to 90.4%. The experiments are conducted on the same dataset using "anomaly-based network intrusion detection through assessing Feature Association Impact Scale (FAIS)" [14]. The results depict that the proposed model is also scalable and effective for detecting the scope of intrusion from a network transaction. Despite the fact that the FAIS model proposed shows 88% accuracy, the major limitation is process complexity in training the system. Such process complexities of designing the scale using FAIS are due to the number of features selected for assessing the scale. The issue of selecting the optimal features for training the Intrusion Detection System using Association Impact Scale is significantly addressed in the FCAAIS [15] model.

**Table 1** indicates the comparison of performance metrics such as precision, recall/sensitivity, specificity, accuracy, and F-measure of FCAAIS over FAIS. **Figure 4** indicates that the accuracy of FCAAIS with optimal features is 91%, whereas the FAIS accuracy with all features is 88%. The precision of the FCAAIS model with optimal features and FAIS with all features is 92%. The other performance metrics such as sensitivity, specificity, and F-measure is calculated on FCAAIS over FAIS. The sensitivity, specificity, and F-measure are 96, 49, and 95%, respectively, for FCAAIS, whereas sensitivity, specificity, and F-measure are 95, 46, and 91%, respectively, for FAIS.

| | | FCAAIS | FAIS |
|---|---|---|---|
| | Total number of records tested | 34,361 | 34,361 |
| TP (true positive) | The number of transactions identified as normal, which are actually normal | 29,379 | 27,889 |
| FP (false positive) | The number of transactions identified as normal, which are actually intruded | 1968 | 2752 |
| TN (true negative) | The number of transactions identified as intruded, which are actually intruded | 1901 | 2375 |
| FN (false negative) | The number of transactions identified as intruded, which are actually normal | 1113 | 1345 |
| Precision | TP/(TP + FP) | 0.937218873 | 0.910185699 |
| Recall/sensitivity | TP/(TP + FN) | 0.963498623 | 0.953991927 |
| Specificity | TN/(FP + TN) | 0.491341432 | 0.46323386 |
| Accuracy | (TP + TN)/(TP + TN + FP + FN) | 0.910334391 | 0.880765985 |
| F-measure | 2 × (PRECISION × RECALL)/ (PRECISION + RECALL) | 0.951646837 | 0.91131588 |

**Table 1.**
*Comparison of performance metrics of FCAAIS and FAIS.*

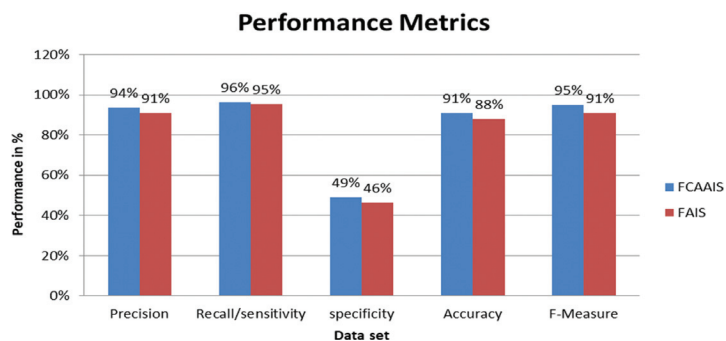

**Figure 4.**
*The performance metrics observed for FCAAIS over FAIS.*

According to the results, the accuracy of FCAAIS (selected feature set using canonical correlation) minimized the process complexity of designing the scale using FAIS (**Figure 5** and **Table 2**).

The observed time complexity is adaptable, as the completion time is not directly related to the ratio of features count, which is due to the higher CC threshold as shown in **Figure 6**. Hence it is obvious to conclude that the applying canonical correlation toward optimized attribute selection is significant improvement to the FAIS model (shown in **Figure 6**).

It is observed that applying canonical correlation toward optimized attribute selection results in 3% improvement in the accuracy of FAIS [14]. **Table 3** indicates precision, recall, and F-measure values calculated under divergent canonical correlation threshold values (**Figure 7**).

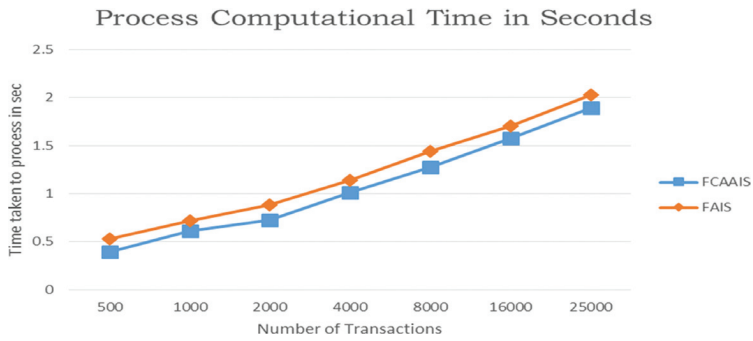

**Figure 5.**
*The process computational time observed for FCAAIS over FAIS.*

| Number of transactions | FCAAIS (s) | FAIS (s) |
| --- | --- | --- |
| 500 | 0.397 | 0.527 |
| 1000 | 0.611 | 0.714 |
| 2000 | 0.723 | 0.882 |
| 4000 | 1.012 | 1.139 |
| 8000 | 1.275 | 1.439 |
| 16,000 | 1.578 | 1.703 |
| 25,000 | 1.891 | 2.031 |

**Table 2.**
*Process computational time of FCAAIS and FAIS.*

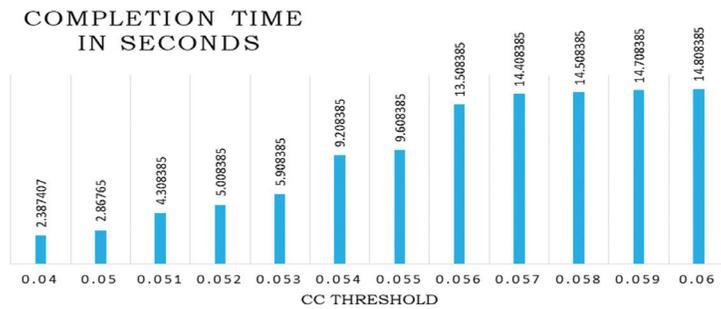

**Figure 6.**
*The FCAAIS consumption of time under divergent canonical correlation thresholds.*

|  | Precision | F-measure | Recall |
|---|---|---|---|
| Less than the upper bound of CC threshold | 0.989 | 0.987998988 | 0.987 |
| Less than the lower bound of CC threshold | 0.98 | 0.984974619 | 0.99 |
| Less than the CC threshold | 0.985 | 0.985 | 0.985 |

**Table 3.**
*Precision, recall, and F-measure values calculated under divergent canonical correlation threshold.*

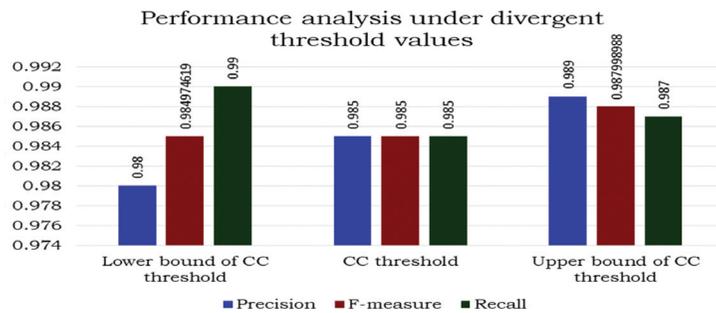

**Figure 7.**
*Performance analysis of the prediction accuracy of FCAAIS under divergent canonical correlation threshold value.*

## 9. Conclusion

It is desirable for anomaly-based network intrusion detection system to achieve high classification accuracy and reduce the process complexity of extracting the rules from training data. In this chapter, a canonical correlation analysis is proposed to optimize the features toward designing the scale to detect the intrusions. The selection of optimal features simplifies the process of FAIS. The experiments were conducted using a benchmark NSL-KDD dataset. The results indicate that the accuracy of FCAAIS with optimal features is 91%, whereas the FAIS accuracy with all features is 88%. The precision of the FCAAIS model with optimal features and FAIS with all features is almost close to 92%. It is observed that applying canonical correlation toward optimized attribute selection has 3% improvement in the accuracy of FAIS. The other performance metrics such as sensitivity, specificity, and F-measure is calculated on FCAAIS over FAIS. The sensitivity, specificity, and F-measure are 96, 49, and 95%, respectively, for FCAAIS, whereas they are 95, 46, and 91%, respectively, for FAIS.

## Author details

Veeramreddy Jyothsna[1*] and Koneti Munivara Prasad[2]

1 Sree Vidyanikethan Engineering College, Tirupati, India

2 Chadalawada Ramanamma Engineering College, Tirupati, India

*Address all correspondence to: jyothsna1684@gmail.com

IntechOpen

# Security in Wireless Local Area Networks (WLANs)

*Rajeev Singh and Teek Parval Sharma*

## Abstract

Major research domains in the WLAN security include: access control & data frame protection, lightweight authentication and secure handoff. Access control standard like IEEE 802.11i provides flexibility in user authentication but on the other hand fell prey to Denial of Service (DoS) attacks. For Protecting the data communication between two communicating devices—three standard protocols i.e., WEP (Wired Equivalent Privacy), TKIP (Temporal Key Integrity Protocol) and AES-CCMP (Advanced Encryption Standard—Counter mode with CBC-MAC protocol) are used. Out of these, AES-CCMP protocol is secure enough and mostly used in enterprises. In WLAN environment lightweight authentication is an asset, provided it also satisfies other security properties like protecting the authentication stream or token along with securing the transmitted message. CAPWAP (Control and Provisioning of Wireless Access Points), HOKEY (Hand Over Keying) and IEEE 802.11r are major protocols for executing the secure handoff. In WLANs, handoff should not only be performed within time limits as required by the real time applications but should also be used to transfer safely the keying material for further communication. In this chapter, a comparative study of the security mechanisms under the above-mentioned research domains is provided.

**Keywords:** WLAN security, WEP, WPA, 802.11i, denial of service (DoS), lightweight authentication, secure handoff

## 1. Introduction

Wireless Local Area Networks (WLANs) provide an extension to the wired network. The wireless stations (STAs) connect to an Access Point (AP) for communication. The messages involved in the communication between STA and AP are visible to other STAs lying in the communication range. This makes WLANs insecure and hence WLANs requires protection.

As with any other computer network, the major security goals in WLANs are: confidentiality, integrity and availability (termed as CIA triad). Prominent techniques that help in attaining these goals include: access control, authentication, encryption, message authentication codes (MAC). Under Access control domain, the entity authentication is performed initially. Depending upon the entity authentication results, access into the WLAN network is controlled. For controlling access into the WLANs IEEE 802.11i (WPA2) is the main standard [1]. This standard though provides flexibility in user authentication but has several issues under the Denial of Service (DoS) attacks [2]. For providing protection to individual WLAN data frames encryption mechanisms like WEP (Wired Equivalent Privacy),

TKIP (Temporal Key Integrity Protocol) and AES-CCMP (Advanced Encryption Standard—Counter mode with CBC-MAC protocol) are used. Lepaja et al. [3] have demonstrated through experiment that WPA with AES provides high TCP through-put. Also, AES-CCMP protocol provides strong security properties, and hence is mostly used in the enterprises [3]. In WLANs, sometimes handoff by the STA is required to maintain communication continuity. There exist several protocols like CAPWAP (Control and Provisioning of Wireless Access Points), HOKEY (Hand Over Keying) and IEEE 802.11r that claim safe and continuous handoff by the STAs [4]. These protocols transfer safely the keying material to STA for further communication. The time limit constraint is imposed on such handoff as the handoff should be performed within short interval required by the real time applications.

This chapter is further divided into four sections. Section2 discusses access control methodologies in WLANs while section3 provides understanding of frame authentication methodologies. Section 4 explains secure handoff methods along with the requirements of secure handoff in WLAN environment. Each of these sections also provides comparative analysis among various methodologies. Section5 provides conclusions and future directions.

## 2. Access control

Traditionally, the entity authentication and access control is provided by the legacy authentication standard i.e., WEP. It has proved insufficient [2] and is hence, deprecated. Currently, IEEE 802.11i (WPA2) [1] security standard is used as an entity authentication and access control mechanism. This security standard is used to secure data communication over 802.11 wireless LANs. The IEEE 802.11i authentication specifies 802.1X authentication mechanism for large networks. The 4-way handshake follows an 802.1X authentication process to confirm the shared keys on Wireless Station (STA) and AP, evolving alongside the Pairwise Transient Key (PTK). This key is used to secure the data sessions between STA and AP using either Temporal Key Integrity Protocol (TKIP) or Advanced Encryption Standard (AES) in counter mode with a Cipher Block Chaining Message Authentication Code (CBC-MAC) Protocol (CCMP). As per the findings of Asante and Akomea-Agyin, use of simple passwords/passphrases makes CCMP susceptible to dictionary attacks [5]. The authentication and 4-way handshake are performed sequentially in 802.11i. Once STAs are authenticated, the standard evolves fresh secret keys to secure data communication over 802.11 wireless LANs. A large numbers of packets are used in these processes [2], which results in an increased process length, communication overhead and network overhead. The authentication and 4-way handshake both are prone to Denial of Service (DoS) attacks. This is due to the lack of proper authentication and insecure message communications between wireless devices [2, 6].

In 802.11i based Networks, 4-way handshake is used for evolving and sharing the keys between the two communicating partners. This 4-way handshake is one of the major concerns in WPA2/802.11i because of Denial of Service (DoS) attacks and therefore researchers target to reduce the 4-way handshake latency. Some suggested to make it 3-way while other suggested to make it 2-way [7]. One such improvement is proposed by Singh and Sharma [7]. In their proposal, the authors try to eliminate the entire 4-way handshake while maintaining the security and key refreshing requirements. For their purpose, they have utilized frame sequence numbers and the striking feature of the proposal is that the key freshness is maintained for each communicating frame. The key refreshed is used for fulfilling the security aspects like frame encryption and integrity management. The overheads in the proposal are bare minimum and it is lightweight as no changes in the existing MAC frame are done. Also, no extra messages are required. Their improvement is more useful under frequent key refreshing situations where users are joining and leaving the wireless environment frequently like in a short duration conference/workshop or in lounge of railway station/airport. The improved technique provides a secure authentication mechanism and no explicit synchronization is required in case of loss of frames. The timings analysis done in the work shows that this technique is effective while security analysis shows that it enjoys almost equivalent security as compared with 4-way handshake of 802.11i. Removal of handshake ensures that the attacks conducted in the 4-way handshake are also removed.

Another improvement in the 802.11i standard is proposed by Singh and Sharma [8] wherein a novel sequence number based scheme is proposed to reduce the MIC field overhead in the WLANs. The existing security frameworks (WPA, 802.11i) provide MIC for maintaining the integrity and authentication for each data frame. MIC is kept in separate field in the frame, and hence adds to the communication overhead. The scheme of Singh and Sharma [8] introduces the notion of authentication token (AT). This AT is calculated based upon the existing sequence number of the WLAN frame. The AT serves both frame integrity and frame authentication purposes. After calculation, it is placed instead of sequence number in the sequence number field of the WLAN frame which means no extra bit or field overhead involvements. As MIC field is removed and AT placement requires no overheads, the scheme is effective as far as WLAN communication overheads and space managements are considered. In addition, the authors have shown that their method is resistant against replay attacks and also provided details on how to attain synchronization in case of frame loss.

In October 2017, a new and major weakness was documented in WPA2 WLAN standard termed as Key Reinstallation AttaCK or KRACK [9]. It was noted that this affected all kinds of WLAN security and hence the reputation of WPA2 got decreased. The WPA2 standard also suffered under DoS attacks. Hence, Wi-Fi Alliance comes up with the improvement. The improvement is termed as WPA3. Its main features involve: (1) ease of use (2) natural password selection (3) an improved and robust handshake and, (4) forward secrecy. The WPA3 is backward compatible with WPA2 which means the upgraded devices can work in WPA2 or WPA3 modes [10]. The market adoption of this standard is now picking and it will take some more time for getting stabilized. Thus, this work on WLAN security considers the present widespread standard i.e., WPA2.

Li et al. proposed an initial entity authentication scheme termed as fast WLAN initial access authentication protocol (FLAP) [11]. FLAP is targeted towards making access authentication faster by reducing the number of initial authentication messages. It is assumed in the protocol that STA and AS share common secret key which simplifies the entire mechanism. Overall, this method involves 6 messages (approx. Two round trip times, **Figure 1**), proves STA authentication at the AS via shared key, has key hierarchy equivalent to 802.11i and protects the messages by MIC. Through practical measurements it is shown that FLAP can improve the efficiency of EAP-TLS by 94.7 percent. It is suggested that this method is compatible with 802.11i and can coexist with existing 802.11i standard. Depending upon circumstances either 802.11i or FLAP can be chosen from suite selector. Like standard 802.11i security protocol, FLAP scheme also depends upon MIC for frame integrity and authentication despite of the fact that MIC verification is computation intensive. This protocol hence may fall an easy prey to Denial of Service (DoS) attacks wherein the attacker may send large number of frames having incorrect MICs. The successive MIC failures on the receiver results in a kind of DoS attack termed as computation DoS attack [12].

Singh and Sharma [13] proposed an access control authentication scheme— SWAS (Secure WLAN Authentication Scheme). The scheme introduces the concept of delegation in WLANs and provides access to clients only upon authentication. SWAS provides authentication of all parties (STA, AP and AS) and evolves a fresh key for securing the data sessions. In addition, it provides security to all messages by utilizing cryptographic primitives, such as encryption and Message Integrity Code (MIC). The proposed scheme reduces the length and complexity compared to IEEE 802.11i authentication and key deriving process. The use of cryptographic techniques does not increase the authentication time of the proposed method. The scheme reduces the communication cost, network overhead and is also resilient against DoS attacks. Therefore, the main contribution of SWAS is to provide a secure and efficient authentication mechanism that evolves fresh communication keys.

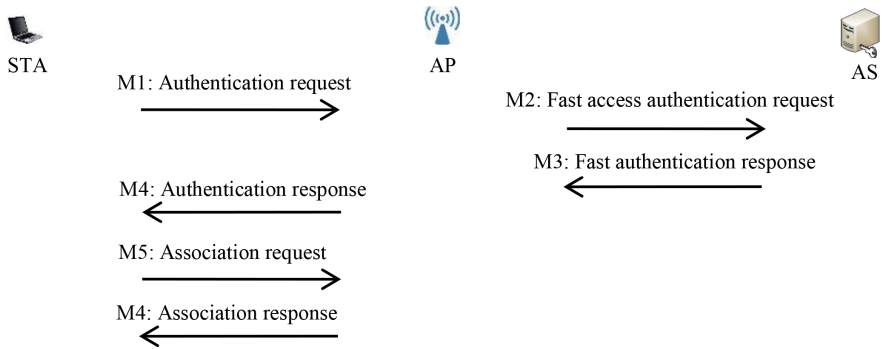

**Figure 1.**
*A simplified overview of initial access authentication protocol (FLAP).*

The SWAS scheme involves three parties: STA, AP and AS. It has three phases: registration phase, request phase and authentication phase. Initially, STA registration is performed at AS and is required only once in a given network. In registration, AS utilizes delegation concept, and generates shared secret key (σ) for AS and STA [14]. The registration phase is followed by the request phase, where the existing 802.11 probe requests, and the probe response messages are utilized by the STA to request the network connection and access. After the request phase, SWAS authentication is performed for authentication and to derive a new communication key that is used to protect the data packets in subsequent sessions.

Both online and offline authentications are used in the SWAS scheme. Online authentication provides authentication and security to all messages among STA, AP and AS. The online authentication utilizes three random numbers (r1, r2, r3) and a sequence number (s1) to ensure proper encryption, authentication and key freshness. In addition, it maintains a key hierarchy similar in purpose to 802.11i with a Master Session Key (MSK), Pairwise Master Key (PMK) and Pairwise Transient Key (PTK). The PTK evolved on the STA and AP during the authentication process is used to encrypt the data packets between them. A simplified view of the SWAS online authentication message exchanges (M1, M2, M3 and M4) is shown in **Figure 2**. In this figure it is clearly visible that each one among STA, AP and AS authenticates each other through various passcode/digital signature verification. The passcode is nothing but protected information (secured through cryptographic means) for the other party. Offline authentication is required whenever a new session key between the same STA and AP is required. This does not involve AS for authentication rather it uses prior stored information at STA and AP. The offline authentication is done via a re-association request and utilizes loosely synchronized sequence number scheme [15].

The salient features of SWAS include: (1) Resistance to DoS attacks in almost all the phases, (2) Less communication and computation time as compared with IEEE 802.11i standard, (3) authentication of all the associated parties i.e., STA, AP and AS by each other and, (4) authentication of all the messages used during all the protocol communication phases. The shortcomings include: (1) lack of practical demonstration of the protocol and (2) no extension of the scheme under the handoff situations is provided till date.

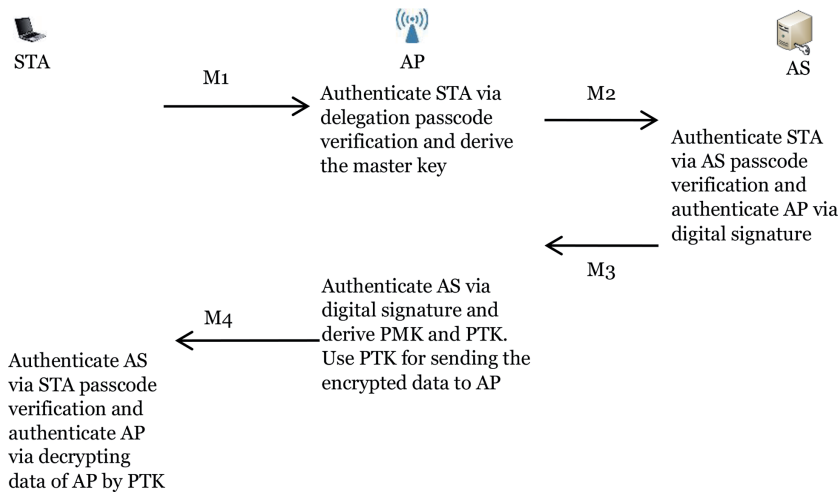

**Figure 2.**
*A simplified overview of online authentication phase of SWAS scheme.*

Authentication per frame and symmetric key based encryption is an implicit necessity for security in Wireless Local Area Networks (WLANs). Singh and Sharma [16] proposed a novel symmetric key based Access Control and per frame authentication scheme for WLANs termed as Key Hiding Communication (KHC) scheme. KHC scheme has two phases: initial phase and communication phase. Former is utilized for sharing and evolving the master key (MK) between STA and AP whereas latter is utilized for onwards data frame communication using the (refreshed) keys. The major establishment of this scheme is the introduction of novel concepts of refreshing the key, protecting the key and initial vector (IV) using different counters and then mixing the bytes of protected key and IV together for each communicating frame. The mixing is based upon the shared secret key and hence only the two communicating parties i.e., STA and AP can mix and separate the bytes of key and IV. The protected mixed bytes are termed as codeword while the concept of mixing the protected key and IV bytes is termed as key hiding. The codeword is added in the WLAN frame. This addition of codeword to the existing WLAN frame occupies extra space and hence the scheme has extra space overheads. Integrity to the frame is provided via MIC. A new key and new IV for the new frame to be transmitted is evaluated based upon existing secret key and existing IV. Evaluation of new key and new IV is termed as key and IV refreshing. The refreshed new key and new IV are first protected using incremented values of counters and then mixed together to form new codeword. The verification and separation of the key and IV from the transmitted codeword provides frame authentication. Once the frame is authenticated, its integrity is verified through MIC verification involving key. The frame authentication is lightweight in KHC as it involves trivial increment, XOR and modulus operations. Thus, KHC follows the notion of frame authentication first and then checking the frame integrity for protection against computation DoS attacks. The separated key and IV are used to decrypt the frame contents and are also used to confirm the frame integrity via MIC. The simplified overview of KHC communication process is shown stepwise in **Figure 3**.

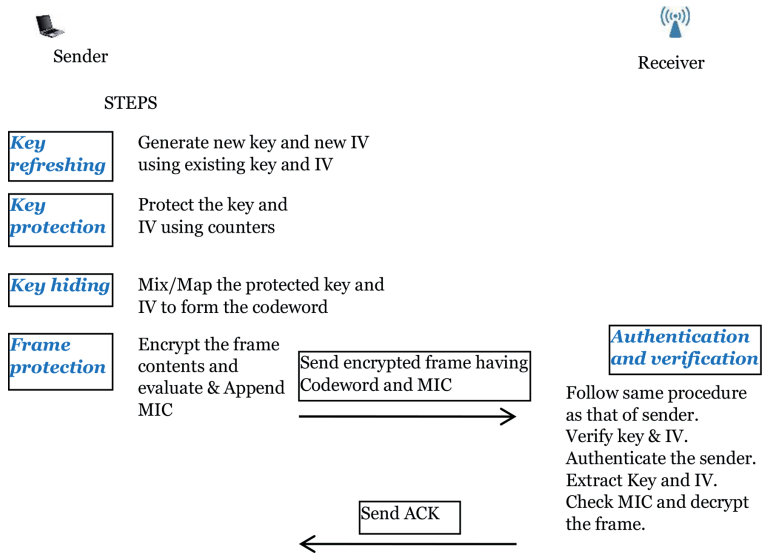

**Figure 3.**
*A simplified overview of communication phase of KHC scheme.*

In nutshell, KHC introduces the concept of key hiding which involves protecting the key using counters followed by mixing of refreshed key & IV i.e., mapping of refreshed key & IV. Through this process of formation of the codeword, the secret symmetric key remains concealed from the attacker. The recipient extracts the key from the codeword, compares it with its own evaluated key, thereby authenticating the sender. Key along with IV, is then used to decrypt the data frame of the sender. Thus, KHC is a useful WLAN communication scheme that is not only secure but is also efficient. The major contributions made by KHC are: (1) lightweight WLAN communication methodology, (2) utilization of symmetric key based encryption/decryption, (3) Per frame Key refreshment, (4) protection against computation DoS attacks and, (5) comparable security as that of 802.11i.

## 2.1. Comparisons of various WLAN access control mechanisms

A property wise comparison between prominent WLAN access control security mechanism is presented in **Table 1**. WEP is though deprecated but mentioned here for the sake of completeness. It can be noted that WEP provides weak authentication, integrity and encryption support. Further, WEP does not consider key and IV refreshing. IEEE 802.11i is a strong protocol as it maintains strong authentication, integrity and encryption. It involves large number of messages and hence consumes times during initial authentication. For key refreshing, it involves 4-way handshake having 4 message exchanges between STA and AP. This 4-way handshake is the major concern in 802.11i. It is prone to DoS attacks and KRACK attacks. FLAP and SWAS both enjoys features similar to that of 802.11i with a difference that the messages exchanged for symmetric key evaluation are less in FLAP and SWAS. In FLAP, very few i.e., approx. 6 messages are exchanged for the key evaluation (including those between STA and AP). In SWAS, only four (4) initial messages are required during online authentication (including those between STA and AP) for sharing the PTK. During offline authentication for refreshing the shared symmetric key only two messages are required. The KHC scheme adopts an interesting methodology which is different from the other access control protocol.

**WLAN access control—Security mechanisms**

| Property | WEP | 802.11i [1] | FLAP [11] | SWAS [13] | KHC [16] |
|---|---|---|---|---|---|
| Authentication | Yes, weak | Yes, strong, initial entity authentication followed by MIC based per frame auth. | Yes, strong, initial entity authentication followed by MIC based per frame auth. | Yes, strong, initial authentication followed by MIC based per frame auth. | Yes, strong, initial entity authentication followed by continuous, lightweight per frame auth. |
| Integrity support | Yes, weak, CRC based | Yes, strong, MIC based | Yes, strong, MIC based | Yes, strong, MIC based | Yes, strong, MIC based |
| Encryption support for confidentiality, strength of encryption | Yes, low, RC4 algorithm | Yes, high, TKIP and AES based | Yes, high, TKIP and AES based | Yes, high (Once shared key is evolved, rest process is same as that of 802.11i) | Yes, high, any one of RC4/ TKIP/ AES can be used |
| Synchronization Algorithm | No | No | No | No | Yes |
| Initial message Exchange for symmetric key exchange | No, done manually | Yes, large | Yes, few – 06 messages (two round trip times) | Yes, few -only four (4) initial messages during online authentication | Yes, few |
| Key freshness | No | Yes | Yes | Yes | Yes |
| IV freshness | No | N.A.* | N.A.* | N.A.* | Yes |
| Messages exchange for key renewal | N.A.* | Yes, four, explicitly | Yes, four (between STA and AP), explicitly | Yes, two using offline authentication | No, done implicitly |

*Not Applicable in this mechanism.*

**Table 1.**
*Property wise comparison of WLAN access control security mechanisms [16].*

It does not use any third party like AS in the authentication process and hence involves less number of messages. It provides an implicit key hiding per frame authentication procedure that is capable of communicating the key to the other entity and is able to refresh not only the shared key but also the IV for encrypting each frame. Thus, least messages are required for key refreshing among all the access control WLAN security mechanisms. Also, the adopted methodology of key refreshing, protection and mapping makes the cracking of key difficult for the attacker. In contrast to WEP, IV is hidden and not visible to the attacker. Other access protocols do not have the notion of IV.

As shown in **Table 2**, memory requirements of WEP is least. 802.11i has more memory requirements than WEP but less than others. Among others, SWAS has highest while FLAP has lowest memory requirements. Communication overhead analysis shows that (1) KHC and WEP involves per frame overheads whereas in others it is done implicitly and, (2) KHC is efficient in key refreshing as compared to others. For key refreshing each of 802.11i and FLAP requires 4 frames,

| **WLAN access control—Security mechanisms** | | | | | |
|---|---|---|---|---|---|
| **Overheads** | **WEP** | **802.11i [1]** | **FLAP [11]** | **SWAS[13]** | **KHC [16]** |
| *Memory requirements***\*\*** | Storing key and IV | Storing Master Key, Refreshed key | Storing Master Key, Refreshed key and counter | Storing delegation key, public key pairs, Symmetric keys: MK, PMK, MSK, PTK, two counters, one sequence number. (Also pool of random numbers at AP) | Storing Master Key, Refreshed key, IV and two counters |
| *Communication overheads* | | | | | |
| For per frame authentication | IV (128 bits) per frame | Implicitly by MIC | Implicitly by MIC | Implicitly by MIC/ authentication information | 256 bits per frame |
| For key refreshing | N.A.**\*** | 4 data frames | 4 data frames | 2 data frames | implicit |

*\*Not Applicable in the scheme.*
*\*\*Considered per participating node.*

**Table 2.**
*Performance comparison of WLAN access control security mechanisms [16].*

SWAS requires 2 frames whereas it is handled implicitly in KHC. In [11], the average authentication delays of the EAP-TLS and FLAP are evaluated as 260.253 and 13.884 ms, respectively. In [13], the total time for SWAS authentication is found to be of the order of 26.46 ms (including time for DoS protection). In [16] Key refreshing timings of 802.11i and KHC are shown as 13.5 ms and 7.5 ms, respectively.

The security comparison shown in **Table 3** clearly indicates that SWAS and KHC scheme provides almost equivalent and better security. 802.11i is prone to DoS attacks whereas FLAP is prone to replay and man-in-middle attacks. Obviously, security of FLAP is least and hence it is not much used presently.

In most of the WLAN access control mechanisms (except KHC), authenticity to the data frame is usually provided by MIC. The MIC based per frame authentication may lead to computation DoS. Hence, lightweight per frame authentication solution is required. It is discussed next.

| **Attacks** | **WEP** | **802.11i [1]** | **FLAP [11]** | **SWAS[13]** | **KHC [16]** |
|---|---|---|---|---|---|
| Possibility of frame contents overwritten by attacker | Yes | No | No | No | No |
| Possibility of modification of authentication bits | N.A. as authentication is implicit | No | No | N.A.**\*** | No |
| Man-in-middle attack | Yes | No | Yes | No | No |
| Replay attack | Yes | No | Yes | No | No |
| Reduce DoS attacks | No | No | No | Yes | Yes |

*\*Not applicable in this mechanism.*

**Table 3.**
*Comparison of WLAN access control security mechanisms under attacks [16].*

## 3. Frame authentication

In WLANs, a two layer redundant security exists. One at the Medium Access Control (MAC) layer while other at the higher layer dealing with End to End security. In former, 802.11i provides security while in latter, higher layer protocols like IPSec, SSL-TLS etc. provides security. Hence, it is suggestive that lightweight authentication and symmetric key based cryptographic measures per frame should be used.

For providing individual frame level protection, two kinds of per frame authentication exist in WLANs: MIC based authentication and lightweight authentication. MIC based frame authentication for data frames is utilized by standard WLAN protocols like IEEE 802.11i, FLAP etc. In these protocols, each frame is accompanied by a unique MIC calculated using sender's shared secret key. The receiver verifies it by recalculating and matching using its share secret key. The MIC calculations and verification consume computation time of the order of 1.5 ms and as shown in Section 2 for FLAP protocol, computation DoS attacks are a possibility [12, 17, 18]. Main reason for computation DoS attack is attributed to the fact that MIC is serving two purposes: authentication and message integrity. Instead, first lightweight authentication should be used. If it succeeds, frame integrity (MIC) should be checked only for those frames whose authentication has succeeded. This will reduce the DoS attacker chances. Thus, lightweight authentication techniques which uses less computation time may prove useful.

The lightweight authentication schemes [19–25] generate the random authentication bits at sender and receiver using random bit generator with commonly shared secret seed as input. These authentication bits are inserted into the WLAN frames. Upon verification of the authentication bits, the frame is accepted at the receiver. Though such schemes provides authentication but they usually lack other security measures like key freshness, secrecy and integrity. A brief tabulation of these schemes is presentation in **Table 4**, showing advantage and disadvantage of each.

### 3.1 Comparisons of various lightweight authentication mechanisms

All the schemes considered in **Table 4** provide per frame continuous authentication. Schemes of Pepyne et al. [25] and Singh and Sharma [26] supports integrity. Former supports CRC based weak integrity while latter supports MIC based strong integrity. Schemes of Pepyne et al. [25] and Singh and Sharma [26] supports encryption. Former supports RC4 based weak encryption while latter supports TKIP/AES based strong encryption. All the schemes considered use their own synchronization algorithm, in fact scheme by Wang et al. [22] uses three different synchronization algorithms. Schemes by Ren et al. [23], Lee et al. [24], Pepyne et al. [25] and Singh and Sharma [26] involves initial message exchanges. Key freshness is incorporated by Pepyne et al. [25] and Singh and Sharma [26]. None of these involves extra messages for evolving new symmetric key (key renewal).

Considering the memory requirements of these schemes Singh and Sharma [26] has the greatest (912 bits) while Lee et al. [24] has the lowest (24 bits). Others except Pepyne et al. [25] have 256 bits memory requirements. Pepyne et al. [25] has 384 bits memory requirements. As far as communication overheads are concern, Johnson et al. [19, 20] and Ren et al. [23] have requirements of 3 bits per frame and 7 bits per ACK frame for counter. Wang et al. [21, 22] has no extra bit requirements as these keep the authentication bits in the unused type and subtype fields of 802.11 frame. Lee et al. [24] requires four extra frames, each having 3 authentication bits. Pepyne et al. [25] has requirements of keeping 128 bits per frame for keeping counter. ASN based scheme by Singh and Sharma [26] has no explicit requirements but requires 48 bits per ACK for synchronization.

| Light weight authentication schemes | Features | Advantage(s) | Disadvantage(s) |
| --- | --- | --- | --- |
| Johnson et al. [19] Wu et al. [20] | Only one bit from the authentication stream generator is placed in the link layer data frame | • scheme provides originator sender identity authentication<br><br>• has low communication overhead<br><br>• as one bit can easily be damaged, synchronization algorithm is also proposed | • attack leading to non-synchronization can easily be launched via successive frame authentication failures<br><br>• The number of bits used for authentication purpose is too less due to which attacker has 50% chances<br><br>• the data packets are not encrypted in SOLA nor MIC per frame is provided, hence payload may be changed (overwrite attack) |
| Wang et al. [21] | • the sender and the receiver generates an authentication stream using same seed value<br><br>• The bit from the authentication stream is put in the frames by the sender and are verified by the receiver using its authentication stream | lightweight protocol with synchronization algorithm and low communication overhead | • The authentication bits are not bound to the frame contents<br><br>• synchronization process is affected by flooding DoS attack where the attacker confuses the sender via unauthenticated ACK frames<br><br>• long authentication bits of continuous 0's or 1's by attackers in the frames can cause confusion |
| Wang, et al. [22] | • single bit lightweight authentication solution<br><br>• Concept of discrimination among legitimate STAs and attacker nodes is used | efficient in terms of computation cost, communication cost and synchronization efficiency | Possibility of authentication bit manipulation by attacker exists |
| Ren et al. [23] | 3 bit authentication solution | Has synchronization algorithm that uses 7 bit counter value put in the ACK frame by the receiver for attaining synchronization | still utilizes less number of bits and therefore high probability of attacks |
| Lee et al. [24] | Scheme selects 3 bits for authentication of management frames | Protection from DoS attack performed by unauthenticated management frames | • scheme protects only the management frame whereas the data frame are not protected<br><br>• DoS attack is still possible by using frames other than the management frames |
| Pepyne et al. [25] | • based upon improvising the WEP protocol<br><br>• uses random stream generator for generating the authenticator variables and fresh encryption keys | Frame counter 'k' is used for synchronization purpose | attacker can easily modify 'k' and launch the attack leading to non-synchronization and Denial of Service |

| Light weight authentication schemes | Features | Advantage(s) | Disadvantage(s) |
|---|---|---|---|
| Singh and Sharma [26] | • utilizes sequence number of the frame along with the authentication stream generators for authentication<br><br>• provides authentication by modifying sequence number of the frame by trivial math operations by sender such that the modification is verified at the receiver | • it requires no extra bits or messages for authentication purpose and also no change in the existing frame format is required<br><br>• lightweight authentication<br><br>• helps in protecting against computation DoS attacks<br><br>• prohibits replays and maintains the synchronization | AP maintains sequence numbers per STA |

**Table 4.**
*Comparison of per frame WLAN authentication solutions.*

On comparing the computational performance of the lightweight authentication schemes mentioned in **Table 4**, it is found that Pepyne et al. [25] and Singh and Sharma [26] take more computational time as compared with others. Singh and Sharma [26] takes more computational time due to the fact that it involves MIC evaluation and encryption of frame for enhancing the security. It is shown in [26] that considering only the authentication the time taken for computational cost for is 0.5 micro seconds which implies that it is same as that of other lightweight solutions.

Except, Pepyne et al. [25], the chances of Brute Force attacks on authentication bits embedded in the frames are quite high in these schemes. Except Pepyne et al. [25] and Singh and Sharma [26] the possibilities of frame contents modification, man-in-the middle attack, replay attacks and DoS attacks are quite high. Pepyne et al. [25] and Singh and Sharma [26] do not allow frame contents modifications and DoS attacks. Pepyne et al. [25] suffers under man-in-the middle attack and replay attacks.

Though KHC is considered in this chapter initially under the Access control mechanisms, it involves lightweight per frame authentication also and needs a special mention in this sub-section. In comparison with the schemes mentioned in **Table 4**, KHC has longer initial entity authentication process. KHC also has raised memory requirements but meets important security features like forward secrecy, key refreshing, lightweight per frame authentication, per frame encryption etc. required by any WLAN security protocol.

Apart from the two main authentication types i.e., MIC based authentication and lightweight authentication, the others are password key exchange mechanisms and layered authentication. The password key exchange mechanisms [27, 28] provide mutual authentication between client and authentication server (AS), identity privacy, half forward secrecy and low computation cost for a client. These mechanisms lack some of the mandatory and recommended requirements for the key exchange methods [29]. Also, these schemes provide authentication at the AS level only while ignoring the authentication at the AP level. The layered authentication achieved by EAP which acts as basis for higher layer authentication protocols, contains certain vulnerabilities e.g. no identity protection, no protected cipher suite negotiation, and no fast reconnection capability [29].

## 4. Secure handoff

WLANs handoffs are essential for providing continuous mobility to a wireless Station in an Enterprise LAN. Two important requirements of the handoff are: (1) establishment of a secure connection of the roaming STA with new access point (AP) and (2) completion of handoff within time limits such that the undergoing communication remains unaffected. The time limit on handoff for multimedia and real time WLAN applications is approximately 50 ms [30]. During this period no data packets transfer occurs. As per the 802.11i WLAN security standard, the complete secure STA authentication (default Full EAP/TLS) via AS evolving shared secret key between STA and AP takes time of the order of 300 ms to 4 s [12] and hence is unfit for the handoffs. For reducing this time, notion of pre-authentication is introduced wherein full 802.1X authentication involving AS is done utilizing old AP and candidate AP (new AP). Hence, at the time of handoff only 4-way hand-shake is required between STA and candidate AP. In this pre-authentication process, an inaccurate candidate AP prediction has associated resource wastage issues as full 802.1X will again be required [31]. Researchers have considered predictive authentication and proactive key distribution for reducing the handoff times. Former involves predicting the candidate AP whereas latter involves locating a group of candidate APs. Thus, in former the problem of inaccurate candidate AP prediction exists whereas in latter the problem of extra communication overhead for authentication with group of APs exists.

Researchers have also worked towards reactive solutions wherein the candidate AP is selected by STA and then the security context is transferred to this AP. In such solutions, STA requests to AS via old AP, then AS transfer security context and material to the candidate AP. Singh and Sharma [32] proposed one such novel secure handoff scheme that maintains security properties while evolving and transferring the security context (key and initial vector) to the candidate AP. The scheme is light-weight and uses reactive method for handoff. Two kinds of APs are defined in the scheme: normal AP and Domain Controller AP (DCAP). STA request DCAP through AP by putting ID of the candidate AP. DCAP in turn distributes the STA context (key and initial vector) to the candidate AP. Thus, when STA roams into the area of candidate AP, less time is involved in the STA authentication at the candidate AP.

For providing fast and secure handoff for the mobile STA in WLANs, standard bodies IEEE and IETF have defined protocols like Control and Provisioning of Wireless Access Points (CAPWAP), HandOver Keying (HOKEY) and IEEE 802.11r (Task group r) [5]. CAPWAP supports centralized management of APs. HOKEY extends the Authentication, Authorization and Accounting (AAA) architecture to support key deriving and distribution with involving full EAP authentication. 802.11r depends upon passing credentials directly between APs for handover. Though CAPWAP takes very less time, it is more or less re-authentication with centralized Access Controller (AC), followed by key transfer to new Wireless Termination Points (WTP). HOKEY is successful in multidomains but it takes more communication time. Among these three (CAPWAP, HOKEY and 802.11r), 802.11r is more efficient in terms of communication overheads. It still has issues concerning the safe transfer of key between APs.

### 4.1 Comparisons of various handoff mechanisms

CAPWAP and HOKEY does not change the existing 802.11 frame structure. 802.11r is a separate protocol and hence has different frame structure. All except CAPWAP scheme generates fresh session keys. Fresh traffic keys are generated by all the schemes. Communication overhead of KHC based handoff scheme is less as compared to any other scheme. This handoff scheme shortens the handoff latency by initiating a key transfer process prior to moving to the new AP and performing handoff. It strengthens the security by (1) protecting STAs from re-associating to Malicious APs, (2) evolving fresh keys even during handshake, (3) authenticating all the frames during the handoff and, (4) safeguarding against DoS attacks and, (5) providing continuous authentication during communication.

## 5. Conclusions

This chapter discusses about the present WLAN security environment. It is clear that the WLAN security environment till date is dominated by WPA2 (IEEE 802.11i) standard. Researchers have pointed out regarding length and complexity of the WPA2. The major point of concern in WPA2 is key refreshing mechanism i.e., 4-way handshake due to which the WLAN security is considered vulnerable. Researchers, hence target to reduce the length of this handshake while maintaining the security properties intact.

The chapter also studies other WLAN security mechanisms proposed by researchers and categories them into: (i) access control, (ii) per frame authentication and (iii) secure handoff mechanisms. It provides category wise comparative analysis of these mechanisms. Three mechanisms are considered in the access control category. Among them Key Hiding Communication (KHC) is the most attractive but it requires changes in the existing WLAN frame structure. Per frame category is further sub-categorized into: (a) per frame authentication mechanisms utilizing MIC and (b) lightweight per frame authentication mechanisms. For enhancing the security, most of the per frame authentication solutions rely on MIC for both authentication and integrity of frame. It is shown that this MIC verification involves computation time and large number of such verifications may result in computation DoS attack on the receiver. The researchers hence advocate separating the authentication and integrity parts in per frame authentication. The lightweight per frame authentication mechanism are though lightweight in nature but lacks security properties like key refreshing, secrecy and integrity. In this chapter, several handoff mechanisms for WLAN environment are also discussed and it is accomplished that none guarantees to maintain required level of security during the specified handoff time limits.

WLAN security is having a transformation from WPA2 to WPA3. WLAN security is strengthened in the upcoming standard i.e., WPA3. It is very early to comment on the effectiveness of WPA3 and it is evident that the existing WLAN devices will continue to use WPA2. The new upcoming WLAN devices will obviously follow the backward compatibility towards WPA2. Thus, researchers can still target to test the implementation of 802.11i with the novel ideas like MIC reduction, 4-way handshake reduction and blockchain application in WLANs [33]. In wireless medium, per frame lightweight authentication mechanisms will prove an edge and in future, researchers may consider developing such solutions. For maintaining uninterrupting communication quick, secure, accurate and secure handoff is the need of the hour. Hence, researchers in future may consider implementation of efficient and secure handoff mechanisms using WPA3.

## Acknowledgements

The authors acknowledge and express the gratitude towards their parent Institutes for the support.

## Conflict of interest

The authors declare no conflict of interest.

### Author details

Rajeev Singh[1]* and Teek Parval Sharma[2]

1 G.B. Pant University of Agriculture and Technology, Pantnagar, Uttarakhand, India

2 National Institute of Technology, Hamirpur, Himachal Pradesh, India

*Address all correspondence to: rajeevpec@gmail.com

IntechOpen

# Analysis of Network Protocols: The Ability of Concealing the Information

*Anton Noskov*

## Abstract

In this chapter, we consider the possibility of hidden data. Since today all network services rely on the basic protocols, the use of untestable and redundant fields may become a big problem. All of the modern data protocols have vulnerabilities. An attacker can use the reserved fields or field use undocumented way. Depending on the data transmission method and detection mechanisms, the technology for assessing the possibility of transmitting hidden information is changing. The work is of great practical interest for the implementation of systems to detect and prevent intrusions and data leaks in it. The authors determine the possibility of transmission and detection sends using a comparative evaluation of the fields in the packet with the values recommended in the standard protocol.

**Keywords:** network protocols, transport protocols, network analyze, network security

## 1. Introduction

Network steganography—type of steganography, in which secret data carriers use the network protocols of the OSI reference model—the open systems interconnection network model. In general, network steganography is a family of methods for modifying data in the headers of network protocols and in the payload fields of packets, changing the structure of packet transmission and hybrid methods in a particular network protocol (and sometimes several at once).

The transfer of hidden data in network steganography is carried out through hidden channels. The term "covert channel" introduced by Simmons in 1983 determined that the problem of information leakage is not limited to the use of software. A covert channel can exist in any open channel in which there is some redundancy. The hidden data is called steganogram. They are located in a specific carrier (carrier).

In network steganography, the role of the carrier is carried out by the packet transmitted over the network. The main parameters of network steganography are the bandwidth, covert channel, probability of detection, and steganographic cost. Bandwidth is the amount of secret data that can be sent per unit of time. The probability of detection is determined by the possibility of detecting a steganogram in a particular carrier. The most popular way to detect a steganogram is to analyze the statistical properties of the data obtained and compare them with typical values for this carrier. Steganographic cost characterizes the degree of change in the carrier after exposure to the steganographic method.

## 1.1 Network steganography methods

Baseline data for consideration classifications of methods and means of network steganography come from the materials of Polish scientists Mazurczyk and Szczypiorski and reports on the experiments of Canadian scientists Ahsan and Kundur, scientists Cauich and Gomez of the University of California at Irvine, and researchers Handel and Sandford at the National laboratory at Los Amos. All materials are freely available. Network steganography methods can be divided into three groups [1]:

- Steganography methods, whose essence is in changing data in the fields of the network protocol headers and in the packets payload fields.

- Steganography methods, in which the structure of packet transmission changes, for example, the sequence of packet transmission or the intentional introduction of packet loss during transmission.

- Mixed (hybrid) methods of steganography—when they are used, the contents of the packages, the delivery times of the packages, and the order of their transfer change.

Each of these methods is divided into several groups; for example, package modification methods include three different methods:

- Methods for changing data in protocol header fields: they are based on modifying the IP, Transmission Control Protocol (TCP), SCTP header fields, and so on.

- Packet payload modification methods; in this case, various watermark algorithms, speech codecs, and other steganographic techniques for hiding data are used.

- Methods of mixed techniques.

Methods for modifying the structure of gears and packages include three guidelines:

- Methods in which the order of the sequence of packets is changed.

- Methods that change the delay between packets.

- Methods, the essence of which is to introduce intentional packet loss by skipping sequence numbers at the sender.

Mixed (hybrid) methods of steganography use two approaches: methods of audio packet loss (LACK) [2] and packet retransmission (RSTEG) [1].

The main idea of methods for modifying header fields is to use some header fields to add steganogram to them [3, 4]. This is possible due to some redundancy in these fields, that is, there are certain conditions in which the values in these fields will not be used in the transmission of packets. The most commonly used header fields are IP and TCP protocols.

Consider an example of a similar method based on modifying unused IP protocol fields to create a hidden channel [4].

The value of the "Identification" field of the IP packet is generated to the sender side. This number contains a random number that is generated when a package is created. The "Identification" field is used only when fragmentation is used. Therefore, to use this method, you need to know the MTU value in the transmitted network and not exceed it, so that the packet is not fragmented during transmission. In the absence of the need for packet fragmentation, a certain redundancy occurs in the "Flags" field, in the second bit, which is responsible for setting the Don't Fragment (DF) flag. It is possible to specify a flag notifying the sender's unwillingness to fragment a packet. If the steganogram package is not fragmented due to its size, you can hide the information in the "DoNotFragmentBit" flag field. Using this method provides bandwidth of 1 bit.

The advantage of this method is the transmission of unchanged information from the sender to the recipient, but it also limits the amount of information sent. Steganography based on this method is easily implemented; has a good bandwidth, since you can send a lot of IP packets with the changes; and is low cost due to the use of fields that do not violate the functionality of the packet. Among the shortcomings it should be noted that the transmitted data is contained in the open form and can be easily read by the observer (although it is possible to strengthen the protection using additional cryptography).

Another method of modifying network packets that alters the payload of a VoIP packet can be widely used in practice with the popularity of programs that provide voice and video communications over the Internet. The network steganography method designed to hide VoIP messages is called Transcoding Steganography (TranSteg), a network steganography method that compresses the payload of a network packet by transcoding. TranSteg can be used in other applications or services (e.g., streaming video), where there is a possibility of compression (with or without losses) of open data. In TranSteg, data compression is used to make room for the steganogram: transcoding (lossy compression) of voice data from a high bitrate to a lower bitrate occurs with minimal loss of voice quality, and after compression, data is added to the free space in the payload package [5]. In general, the method allows to obtain more or less good steganographic bandwidth of 32 kb/s with the smallest difference in packet delay. Experiments of Polish scientists have shown that the delay in transmitting a VoIP packet using TranSteg increases by 1 ms, in contrast to a packet without a steganogram. The complexity of detection directly depends on the choice of the scenario and the conditions of the outside observer (e.g., its location). Among the shortcomings worth mentioning is the fact that this method is difficult to implement. It is necessary to find out which codecs the program uses for voice communication, to choose codecs with the smallest difference in speech quality, while giving more space for embedding steganograms. During compression, the quality of the transmitted speech information is lost.

Also interesting is the direction using the mechanisms of the SCTP protocol. Stream control transport protocol (SCTP) [6] is a packet-based transport protocol, a new-level transport protocol that will replace TCP and User Datagram Protocol (UDP) in future networks. Today, this protocol is implemented in operating systems such as BSD, Linux, HP-UX, and SunSolaris, supports network devices of the Cisco IOS operating system, and can be used in Windows. SCTP steganography uses new features of this protocol, such as multi-threading and the use of multiple interfaces (multi-homing).

The methods of SCTP steganography can be divided into three groups [7]:

- Methods in which the contents of SCTP packets change.

- Methods in which the sequence of transmission of SCTP packets is changed.

- Methods that affect both the content of packages and their order when transfer (hybrid method).

Methods for changing the contents of SCTP packets are based on the fact that each STCP packet is made up of parts and each of these parts can contain variable parameters. Regardless of the implementation, a statistical analysis of the addresses of the network cards used for the forwarded blocks can help in detecting hidden connections. Eliminating the possibility of applying this method, steganography can be achieved by changing the source and destination addresses in randomly selected packet, which is contained in the re-expel e PTO unit.

The essence of the hybrid method based on the SCTP protocol is to use certain protocol mechanisms that allow you to organize the intentional passing of packets in a stream without resending it. Later a steganogram is added to this packet, and it is resubmitted [7]. Modification of packages using a hybrid method can be presented on the Hidden Communication System for Corrupted Networks (HICCUPS), which uses the imperfections of data transmission in a network environment, such as interference and noise in a communication environment, as well as the usual susceptibility of data to distortion. HICCUPS is a steganographic system with bandwidth allocation in a public network environment. Wireless networks are more susceptible to data corruption than wired ones, so the use of noise and noise in the communication environment during system operation looks very tempting. "Listening" of all the frames with the transmitted data in the environment and the ability to send damaged frames with incorrectly corrected code values are two important network features necessary for the implementation of HICCUPS. In particular, wireless networks use an air connection with a variable bit error rate (BER), which makes it possible to introduce artificially damaged frames. This method has low bandwidth (network dependent), cumbersome implementation, low steganographic cost, and high detection complexity. However, the frame analysis does not involve checksum may lead to the discovery of the use of Nogo given method.

The RSTEG method is based on the packet resending mechanism, the essence of which is as follows: when the sender sends a packet, the recipient does not respond with a confirmation flag; thus the packet resending mechanism should work, and the packet with the steganogram inside will be sent again, but confirmation does not come. The next time this mechanism is triggered, the original packet is sent without hidden attachments, to which the packet arrives with confirmation of successful receipt.

The performance of an RSTEG depends on many factors, such as the details of the communication procedures (in particular, the size of the packet payload, the frequency with which segments are generated, and so on).

The investigated method of steganography using packet retransmission RSTEG is a hybrid. Therefore, its steganographic bandwidth is approximately equal to the bandwidth of the methods with packet modification and at the same time higher than the methods of changing the order of packet transmission. The complexity of detection and throughput is directly related to the use of the implementation mechanism of the method. RSTEG based on RTO is characterized by high detection complexity and low bandwidth, while SACK has the maximum bandwidth for RSTEG, but is also more easily detected. The use RSTEG utilizing TCP protocol is a good choice for IP networks. Among the shortcomings, it should be noted that this method is difficult to implement, especially its scenarios, which are based on interception and correction of packets transmitted by ordinary users. Due to the dramatically increased frequency of retransmitted packets or the unusual occurrence of delays in the transmission of steganograms, a casual observer may be suspicious.

Lost audio packets steganography (LACK)—steganography of deliberate delay of audio packets [2]. This is another method implemented via VoIP. Communication over IP telephony consists of two parts: signaling (dialing) and conversational. Both parts of the traffic are transmitted in both directions. The signaling protocols used are SIP and RTP (with RTCP acting as the control protocol). This means that during the signaling phase of the call, the SIP endpoints (called user SIP agents) exchange some SIP messages. Usually SIP messages pass through SIP servers: proxy or redirected, which allows users to search and find each other. After this stage, the conversation phase begins, where the audio (RTP) stream goes to both directions between the caller and the callee. This method has certain advantages. The bandwidth is not less and sometimes higher than the other algorithms that use audio packets. But if you intentionally cause losses, the quality of the connection deteriorates, which can become suspicious for both ordinary users and listeners. Based on the presented steganalysis LACK methods, it can be concluded that the method has an average detection complexity. The implementation of the method is too complex, but may not be possible within certain operating systems.

**Table 1** shows a comparison of methods and their main characteristics and implementation. The position of each method in this table shows how much its characteristics are superior or inferior to the others. The higher the method displayed at the table, the more indicators of its characteristics. In the "Implementation" field, the simplicity of the organization of this method is considered. The less time and effort required by the implementation of this method, the higher its position in this title. Based on the data from **Table 1**, it can be concluded that the main characteristics are directly dependent on each other.

| No | Throughput ability steganography | Complexity discoveries | Steganography cost | Implementation |
|----|----------------------------------|------------------------|--------------------|----------------|
| 1 | TranSteg | HICCUPS | HICCUPS | Modification header fields TCP and IP packets |
| 2 | LACK | TranSteg | LACK | Modification data blocks in SCTP protocols |
| 3 | HICCUPS | LACK | RSTEG | TranSteg |
| 4 | RSTEG | RSTEG | TranSteg | Using SCTP multi-homing |
| 5 | Modification fields in TCP headers and IP packets | Using SCTP protocol (hybrid) | Protocol use SCTP (hybrid) | Using SCTP protocol (hybrid) |
| 6 | Modification data blocks in SCTP protocols | SCTP multi-homing | Modification of blocks data in SCTP protocols | LACK |
| 7 | Using SCTP protocol (hybrid) | Modification fields in TCP headers and IP packets | SCTP multi-homing | RSTEG |
| 8 | Using SCTP multi-homing | Modification data blocks in SCTP protocols | Modifying fields in TCP and IP headers packages | HICCUPS |

**Table 1.**
*Comparison of network steganography methods.*

## 2. The combined method using modification of the fields IP and TCP

As mentioned earlier, the methods for modifying the IP and TCP header fields have certain features that make them stand out from the rest of the methods:

• The most common and standard protocols are used as carriers of the steganogram.

• Total gives bandwidth of 49 bits per 1 packet.

• Implemented on any operating system, the implementation does not require long adjustments and preparations.

• Changes in the package will not affect its behavior on the network, in case it will not be fragmented.

Despite the many advantages of both methods, there are some flaws, and the main one, to which attention is immediately drawn, is the obviousness of data transfer, i.e., any statistical analysis allows us to calculate both the hidden communication channel itself and the information transmitted in it.

The method proposed by Rowland [3] is as follows: to generate a value in the "Sequence Number" field, the plaintext character is encoded in accordance with the ASCII table, and the resulting value is multiplied by a certain number multiple of two. The resulting value is entered in the "Sequence Number" field and sent to the recipient. The recipient, knowing the key (divider), should check all incoming TCP packets for the subject of the steganogram, dividing the value of the "Sequence number" field by the key.

On the one hand, this method allows you to create a data channel through which you can transmit secret data in front of a passive observer. But the existence of a single key is a disadvantage, since, based on a dozen of such packages, it can be concluded that the sequence numbers of all packages have a common factor, which is the key. Thus, the proposed method is easy to detect.

Based on the source data and analysis of the disadvantages of network steganography methods with modification of the IP and TCP packet header fields, we can propose a modified method that will be based on the simultaneous use of the IP and TCP protocol header fields. The key needed to decrypt the transmitted message will also be transmitted as a steganogram, only in encrypted form in the "Identifier" field of the IP header, while the encrypted steganogram will be transmitted in the "Sequence number" field of the TCP header.

The implementation of this method is divided into two parts:

• Preparing data for the transfer, which includes generating the key k, converting the transmitted secret symbol or number into its corresponding code in the ASCII table, and calculating the value of the carrier C, which is an encrypted steganogram.

• Entering data into the corresponding TCP and IP header fields.

The first block consists of the following steps:

• Generation of the key k, which will be used in the future. The key can be any number that is a multiple of two. To generate a key, take two numbers x and y and raise the first to the power of the second.

- The conversion of secret data—a character or number that must be transferred to the corresponding code in the ASCII table. The coded number is denoted by S, since it is our steganogram.

- Getting the media C as the product of the key value by the value of a secret character.

$$C = S{*}k_{10}$$

- Checking the number C—it must meet the requirement $2^{28} < C < 2^{33}$. This condition is necessary so that the value of the "Sequence number" field does not look suspicious. If the value of C does not meet the requirements, the numbers x and y need to be changed to others, and repeat steps 1–2. Further studies will be conducted on the automatic formation of x, y.

- The value of the numbers x and y is written together into the number z and is flipped so that the previous values can only be read from right to left.

Then the data is converted from decimal to hexadecimal. Thus, we get a three-digit hexadecimal number inv. (z) 16.

Then, at the second stage, you need to put the obtained values of the encrypted key and steganogram into the TCP and IP header fields.

We briefly describe the network steganography method with a modification of the fields in the TCP header, since in it we will transmit the secret message itself. For the purpose of steganography, the header of this protocol usually uses some fields that can be changed without losing the functionality of the package. For the purpose of our research, we will focus on the "Sequence Number" field (SN, SequenceNumber). This field performs two tasks. The first is the following: if the SYN flag is set, then this initial value of the sequence number is ISN (InitialSequenceNumber), and the first byte of data that will be transmitted in the next packet will have a sequence number equal to ISN + 1. Otherwise, if SYN is not set, the first byte of data transmitted in this packet has this sequence number. For our case it is important to know that this value will not change during the path of the packet from the sender to the recipient.

The "Sequence Number" field allows you to create a 32-bit length sequence. According to the Rowland method, the transmitted message is encoded in accordance with the ASCII table and multiplied by a certain number (the key), a multiple of two to reduce the detection probability, then entered into the generated TCP packet in the "Sequence number" field, and the packet is sent. When the packet reaches the destination address, the recipient must save all incoming TCP packets, from which he must remove the value in the "Sequence number" field and then divide by the key he knows in advance. But, as it was said before, this method is extremely easy to detect based on the analysis of a number of TCP packets due to a permanent key. In the proposed modification of the method, this key will be transmitted simultaneously with the TCP packet, in the IP header. This will increase the difficulty of detecting the steganogram.

The next step is to add the value of the C media in the "Sequence Number" field of the TCP header.

Next, you must enter the value of the encrypted key (inv (z)) 16 in the IP header field. To organize such an operation, you should return to the network steganography method with modification of the IP header fields. During the packet path, only the "Identifier" field remains unchanged; its length is 16 bits and 1 bit in the "Flags" field, which is responsible for the DF flag. Changing these fields does not carry changes in the package, in case the package is not fragmented, but it should not be, since by condition we need to know the minimum MTU value and not exceed it when creating and sending the package.

At the "Identifier" field, 16 bits is available to us for adding a steganogram; the information in it is displayed in the form of four numbers in hexadecimal number system. Thus, we have 65,535 possible values that can be used both for transmitting the steganogram and for the key, which in turn is also a steganogram. In order not to transmit the key in such an explicit form, it is proposed to use only three numbers out of four, while reading them from right to left. In this case, the number can be odd with its standard reading from left to right. The fourth unused number can take any value. Thus, we can use only 16 of the 17 bits available in a packet. It is proposed to use the second bit in the "Flags" field—DF—as a specific label, the presence of which allows you to expand the key extraction algorithm: whether you need to read the value from the first or from the second number in the "Identifier" field to extract the key.

Thus, the next step is to enter (inv (z)) 16 in the "Identifier" field of the IP header. At the same time, we must set the value of "1" to the second bit in the "Flags" field if we enter the key in the first 12 bytes of the "Identifier" field or 0 if we fill the first 4 bytes of the field with random values and in the remaining 12 bytes our key.

Next, we send a packet with modified fields to the recipient, where he must carry out the procedure inversely described in the framework of this algorithm [8].

We calculate the bandwidth of the proposed method.

Since the "Identifier" field in the IP header can contain 16 bits of information, 1 bit is available in the "Flags" field, and in the "Sequence number" field, a 32-bit information is available in the TCP header; we can conclude that the total throughput of steganography is 49 bits. But it should be noted that in this method we use the "Identifier" field to transmit the encrypted key in the steganogram, which is used to extract secret information from the "Sequence number" field, and the bit in the "Flags" field is used as a label. Thus, to transfer the encrypted key, we allocate 12 bits of information available in the "Identifier" field, and in the remaining 4 bits, we enter a random number from 0 to 16 in the hexadecimal number system (from 1 to F) and use 1 bit as a label, necessary for more organization more flexible operation of the algorithm. Based on this, we can conclude that for transmitting specific information, we have 32 bits left in the "Sequence number" field, and 3 bits of secret information can be transmitted, which is encrypted in 32 bits of information hiding the secret.

## 2.1 Intercomputer exchange

The exchange of computer networks is based on the Open System Interconnection (OSI) reference model.

Studying hidden information flows with computer interaction on networks of interest will include information about the services that are added to the network traffic data. As part of the protocol, headings are assessed at two levels: network and transport. We will address network protocols (IPv4 and IPv6) and transport protocols (TCP and UDP).

Further, we are considering the reports and the possibility of more detailed manipulation.

## 2.2 IPv4

IPv4 is the most popular protocol of network level; see more information in RFC791.

The header size of IPv4 is 20 bytes; using specialized field in header—"Options" field—can increase it. When the amount of the header is less than 20 bytes, it is likely damaged and has to be discarded.

### 2.3 Header of IPv4

The format of IPv4 header is presented in **Figure 1**.
IPv4 header field analysis shows the following results:

1. "Internet Header Length" field. Ability to increase the size of the Internet Header Length field to extend the original header. This change allows you to add data to the next two "Options" and "Padding" fields.

2. "Type of Service" field

   Bits from 0 to 2 are set for priority and 6 to 7 set to reserved.

   -0-2:

   The value "111" should not appear on the networks of provider; it could be appearing only for local networks, which leads to the point that the capture of this value in the network provider is a mark of malicious information injection.

   -6-7:

   By default, these bits are reserved and must be set to 0; the result is that the other value is possible injection information.

3. "Identification" field

   You can change the value of the identification field. The point is that the field is used to build correctly after fragmentation, but there is a DF flag that rejects fragment packets, so if the flag is set to "1" this ID is not required, and this field could be used to pass hidden information.

4. "Flags" field

   As the standard requires, the first bit is reserved and should be set to "0"; if the result is different, it is mark of injection information.

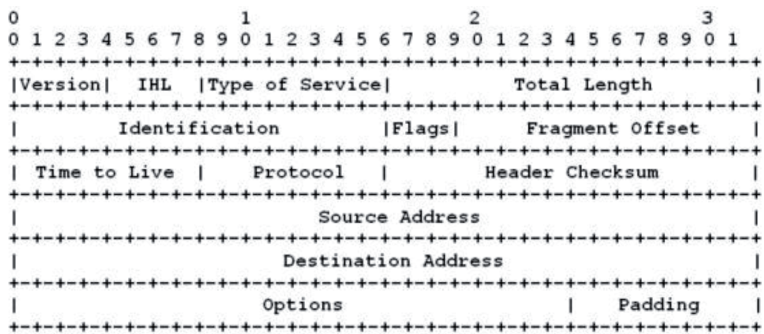

**Figure 1.**
*Header of IPv4.*

5. "Fragment Offset" field

   You can change the value of the "Fragment Offset" field. The best option is when the DF flag is set to "0," since the fragmentation strategy is designed so that an unfragmented datagram in all fields related to fragmentation has zero values. This means, despite the fact that the flag prevents fragmentation, we can still implement it in the offset of the fragment, but the fact of identification of the manipulation becomes more detectable.

6. "Source Address" field

   You can change the "Source Address" field value.

7. It should be noted that manipulation is possible only on the condition that the package consists of hidden source data. Since the manipulation will not be caused by the source of the information, the receiving site could not properly build the packets.

8. "Destination Address" field

   IPv4-in IPv6 headers can be encapsulated using the IPv4 Destination Address field to insert information into it. In this case, the IPv6 header will be responsible for delivering the package.

9. "Options" field

   The value of the options field is limited in the IPv4 header, and as a result of the analysis, we are trying to determine any field value that may appear in this type of field. So we may try to determine the incorrect significant of this field, the appearance of which indicates the possible malicious activity on the injection of information.

10. "Padding" field

    This field goes after value 0x00 of the "Options" field; the value is the EOL and takes up to 32-bit header boundaries. The interest in this manipulation is that after the optional EOL, the equipment does not examine headers on 32-bit boundaries; this means that these bytes are invisible to network devices and sniffer. Although the analysis of this field is simple enough, the EOL up to 32-bit header boundaries must be set to "0" at the standard behind the "Options" field, causing any other value of this field to indicate that the data is being injected.

**2.4 Injection's result**

The standard IPv4 header size with options and fields with padding is 320 bits. Two different options need to be considered:

1. IPv4 is a carrier and is responsible for packet addressing. Due to manipulation, 182 bits can be used, which is 56.88% of the total number of bits. This volume allows you to insert 22 symbols from 8 bits in ASCII encoding into the header.

So after calculations we have got a value up to 4 bits. This remainder is part of the other 8 bits of the transmitted information.

2. IPv4 is a passenger, it's an IPv4 encapsulated header in other headers, such as IPv6 or GRE. In this case, the method for implementing the target address can be used. As a result, handling bits 214, 66% of the total number of bits can be used. This volume allows you to implement a 26-character header with 8 bits in ASCII encoding. Thus, after calculations, a value of 6 bits is obtained. The treated residue was included in an additional 8 bits of the transmitted symbol.

## 2.5 IPv6

### 2.5.1 Header of IPv6

The header's format of IPv6 is presented in **Figure 2**.

1. "Traffic Class" field

You can change the "Traffic Class" value arbitrarily. This manipulation cannot be detected by analysis.

2. "Flow description" field

You can change the value of the "Flow Label" field.

This manipulation cannot be detected by the packet sniffer.

3. "Load Length" field

It is possible to increase the size of this field when adding data to the end of the original IP packet, like IPv4. This modification cannot be detected by the packet sniffer.

```
 0                   1                   2                   3
 0 1 2 3 4 5 6 7 8 9 0 1 2 3 4 5 6 7 8 9 0 1 2 3 4 5 6 7 8 9 0 1
+-+-+-+-+-+-+-+-+-+-+-+-+-+-+-+-+-+-+-+-+-+-+-+-+-+-+-+-+-+-+-+-+
|Version| Traffic Class |                Flow Label             |
+-+-+-+-+-+-+-+-+-+-+-+-+-+-+-+-+-+-+-+-+-+-+-+-+-+-+-+-+-+-+-+-+
|         Payload Length        |  Next Header  |   Hop Limit   |
+-+-+-+-+-+-+-+-+-+-+-+-+-+-+-+-+-+-+-+-+-+-+-+-+-+-+-+-+-+-+-+-+
|                                                               |
+                                                               +
|                                                               |
+                        Source Address                         +
|                                                               |
+                                                               +
|                                                               |
+-+-+-+-+-+-+-+-+-+-+-+-+-+-+-+-+-+-+-+-+-+-+-+-+-+-+-+-+-+-+-+-+
|                                                               |
+                                                               +
|                                                               |
+                     Destination Address                       +
|                                                               |
+                                                               +
|                                                               |
+-+-+-+-+-+-+-+-+-+-+-+-+-+-+-+-+-+-+-+-+-+-+-+-+-+-+-+-+-+-+-+-+
```

**Figure 2.**
*Header format of IPv6.*

4. "Source Address" field

You have the possibility to change the data of this field at IPv4 format, but international standards from the IPv6 community do not recommend using it as a source address.

5. "Destination Address" field

In this protocol, you can use the IPv6 "Destination Address" field in the IPv4 encapsulation header to load information into it. In this case, the IPv4 header will be responsible for the packet delivery.

This manipulation cannot be detected by the packet sniffer.

## 3. Result of injection

The standard IPv6 header size with options and fields with padding is 320 bits. Two different options need to be considered:

1. IPv6 is a carrier, that is, it is responsible for addressing the package. As a result of the manipulations described above, 156 bits can be used, which is 48.75% of the total number of bits. This volume allows you to insert a caption with 19 characters from 8 bits into the ASCII character set. Thus, after calculations get a value of 4 bits. The treated residue was included in an additional 8 bits of the transmitted symbol.

2. IPv6 is a passenger and is transmitted by IPv6 encapsulation header to other headers, such as IPv4 or GRE. In this case, the method for implementing the target address can be used. As a result of the manipulations described above, it is possible to use 284 bits, which is 88.75% of the total number of bits. This volume allows you to implement a 35-character header with 8 bits in ASCII. Thus, after calculations, we get a possible value of 4 bits. The processed remainder will be added as an additional 8 bits of transmitted characters.

### 3.1 TCP

Transmission Control Protocol is a reliable protocol of transport layer. TCP is oriented to establish a logical connection, that is, the hosts negotiate and create a session and then begin to transfer data. Every time a package is sent, the sender is awaiting acknowledgement of delivery receipt. This protocol is standardized by RFC 793.

*3.1.1 Header of TCP*

Header's format of TCP is presented in **Figure 3**.
"Source Port" field

1. You can change the "Source Port" field value. Processing is only possible when the package was a hidden data source. Due to the manipulations that occur on the source host, the receiving party will not be able to properly assemble the original packet. This manipulation cannot be detected by the packet sniffer.

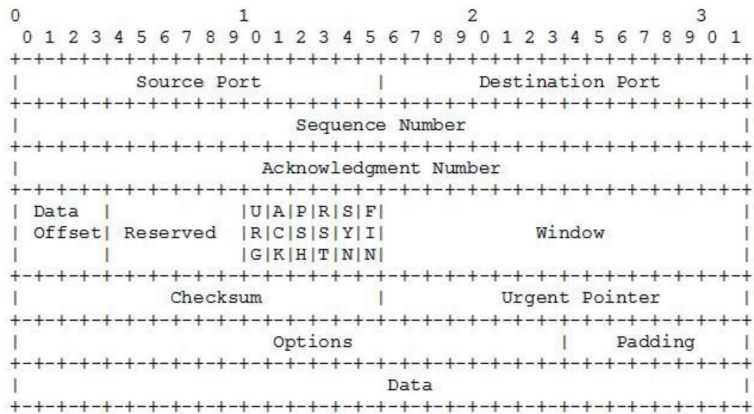

**Figure 3.**
*Header format TCP.*

2. "Destination Port" field

You can change the value of "Destination Port" field. Handling is only possible when the packet was a hidden data source. Due to the manipulations that occur on the source host, the receiving party will not be able to properly assemble the original packet. This modification cannot be detected by sniffer.

3. "Sequence Number" field

We could modify the information in this field. Processing is only possible when the package was a source of hidden data. Because of the modification that had occurred at the source device, the receiving PC cannot correctly build the original packets. This modification cannot be observed by the network sniffer.

4. "Acknowledgment Number" field

We could change the contents of this field. Modification is allowed provided that the package was made up source of hidden data. Due to the modification that occurred on the source host, the receiving party will not be able to properly assemble the original packet.

5. "Data Offset" field

The manipulation is as follows: this increases the size of the "Data Offset" field, expands the TCP header, and adds a parameter field. In the options you can add data after byte 0x00 EOL.

At standard byte 0x00 EOL, bytes with a value of "0" should be due to some other value that indicates that a data injection has occurred.

6. "Reserved" field

You can modify the value of this field.

By default, the values of all standard bits must be set to "0" as a result of some other values that indicate that a data injection has been occurred.

7. "Window" field

   You can modify the value of the "Window" field. Handling is only possible when the packet was built at a hidden data source. Due to the manipulations that occur on the source host, the receiving party will not be able to properly assemble the original packet

8. "Pointer Urgent" field

   You can modify the value of this field. This injection is only possible if all URG options are present.

   So, if the Urgent Pointer is filled in and the flag of URG is not setting, it means that the Urgent Pointer is not used correctly.

9. "Options" field

   We could modify the data of this field. In the options, you can realize the data after value 0x00, but it is not considered after this byte header data.

   TCP header option values are limited, and network analysis results in attempting to identify a possible option that attempts to identify incorrectly filled options or unknown options whose appearance indicates a possible injection of information.

10. "Padding" field

    It is possible to fill the field of any padding.

    It should be noted that manipulation is only possible if the package is made up of hidden source data. Because of the manipulation that occurs at the source, the receiving party cannot properly collect packets.

Handling "Padding" is one of the most interesting. The "Padding" field starts after the 0x00 in the "Options" field; the value is the EOL option and takes up to 32-bit header boundaries. Interest in this manipulation is contained in the following text after the EOL does not produce a 32-bit header, which means that these bytes are invisible to network devices and sniffer. Although the analysis of this field is simple enough, the EOL up to 32-bit header boundaries must be set to "0" at the standard behind the "Options" field, causing any other value of this field to indicate that the data is being injected.

## 4. Result of injection

The standard TCP header field with options and fall is 192 bits. As a result of the above actions, you can use up to 150 bits, which is 78.13% of the total number of bits in the original, unmodified header. This amount of data allows the use of 18 characters in an 8-bit header in the standard ASCII character set. Therefore, after all the calculations, we get the maximum possible amount equal to 6 bits. The processed piece of information was included in the next 8 bits of the transmitted symbol.

### 4.1 UDP

User Datagram Protocol is a connectionless transport layer protocol. No connection setup is created before transferring between hosts. This protocol is less reliable than TCP, but gives a higher transfer rate with less overhead. This protocol is standardized by RFC 768.

#### 4.1.1 Header field of UDP

Header's format of UDP is presented in **Figure 4**.
"Source Port" field

1. You can change the "Source Port" field value. Processing is only possible when the package was a hidden data source. Because of the manipulation that had occurred at the source device, the receiving party cannot properly assemble the original packets. This manipulation cannot be detected by the packet sniffer.

2. "Destination Port" field

   You can change the "Target Port" field value. Processing is only possible when the package was created by a hidden data source. Due to the manipulation of the device generating the packages, the receiving party cannot correctly assemble the source packages. This modification cannot be detected by the network sniffer.

3. "Length" field

   We could change the significance of the "Length" field. Increasing the value of this field has also increased the size of the package, so we can change the fields of data octets by appending to the end of datagram.

So processing is possible when the package was a source of hidden data. So if the modification had occurred at the source device, the receiving host cannot properly assemble the original packets. This manipulation cannot be detected by the packet sniffer.

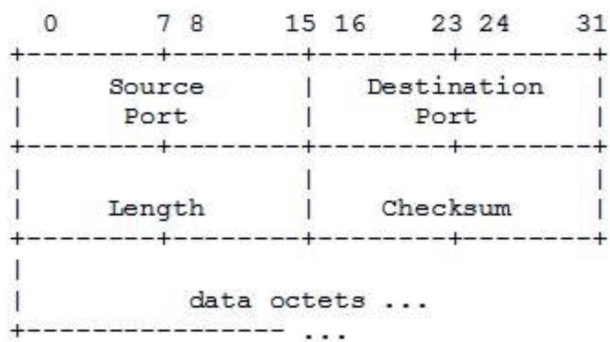

**Figure 4.**
*Header of UDP.*

## 5. Result of injection

The size of the UDP header of the datagram is 64 bits; as a result of the described changes, you can use 32 bits, which is 50% of the total number of bits in the header, which allows you to implement a 4–8-bit header in the ASCII character set.

## 6. Conclusion

In this work, we began to develop methods and special software for generating bitstreams in order to organize a secure connection.

This software method was implemented in software, ensuring secure network communication. The main part of the program model is a detector program for analyzing network traffic to search for possible hidden transmissions. The analysis is implemented by checking the header in compliance with the standards, which is needed to identify unauthorized values for specific areas of the PDU.

The surveys revealed possible vulnerabilities that could embed relevant information in the puncture headers we reviewed. **Table 2** presents the quantification of the study results, showing the remainder is the number of bits that are part of the next 8 bits of the transmitted symbol.

It should be noted that TCP was created as a reliable protocol for delivery, but after entering the hidden data by the TCP header proposed above, changes made to the header fields of the TCP lead to the loss of the functionality of a reliable protocol, making it similar to the UDP.

In the created model, the transmission of one packet is realized, that is, the full message is embedded in all possible of headers at only one datagram. In order to see the maximum feasible messaging, we chose the following protocols: IPv4, IPv6, and TCP. Note that for simplicity, TCP header data is not included in the fragment offset. Thus, thanks to the proposed manipulation, the programming model uses 603 bits, which is 74.04% of the total number of bits in the order of three headers. This volume allows you to enter 75 characters out of 8 bits in ASCII encoding.

| Protocol | Size of injection information (bits) | Percentage of the total header size (%) | The number of symbols | Rest bits |
|---|---|---|---|---|
| IPv4 (carrier) | 182 | 56.88 | 22 | 6 |
| IPv4 (passenger) | 214 | 66 | 26 | 6 |
| IPv6 (carrier) | 156 | 48.75 | 19 | 4 |
| IPv6 (passenger) | 284 | 88.75 | 35 | 4 |
| TCP | 150 | 78.13 | 18 | 6 |
| UDP | 32 | 50 | 4 | 0 |

**Table 2.**
*The quantification of the study results.*

## Author details

Anton Noskov
Yaroslavl State Technical University, Yaroslavl, Russia

*Address all correspondence to: anton.noskov@gmail.com

IntechOpen

# Multifactor Authentication Methods: A Framework for Their Comparison and Selection

*Ignacio Velásquez, Angélica Caro and Alfonso Rodríguez*

## Abstract

There are multiple techniques for users to authenticate themselves in software applications, such as text passwords, smart cards, and biometrics. Two or more of these techniques can be combined to increase security, which is known as multi-factor authentication. Systems commonly utilize authentication as part of their access control with the objective of protecting the information stored within them. However, the decision of what authentication technique to implement in a system is often taken by the software development team in charge of it. A poor decision during this step could lead to a fatal mistake in relation to security, creating the necessity for a method that systematizes this task. Thus, this book chapter presents a theoretical decision framework that tackles this issue by providing guidelines based on the evaluated application's characteristics and target context. These guidelines were defined through the application of an extensive action-research methodology in collaboration with experts from a multinational software development company.

**Keywords:** security, authentication scheme, multifactor authentication method, action-research, decision framework

## 1. Introduction

Generally, to protect the personal information of users in software applications, distinct authentication techniques are utilized to prevent intruders from accessing to it. Authentication is, thus, the process of verifying the identity of a user as part of a system's access control to protect the information stored within them [1]. Various authentication techniques have been proposed in literature, such as text passwords [2, 3], smart cards [4, 5], and biometrics [6–8]. All of the mentioned techniques belong to distinct authentication factors. An authentication factor is a piece of information that can be used to verify the identity of a user [9]. There are three main groups or factors of authentication techniques [10, 11]: (i) knowledge-based, that is, based on something that the user knows, such as text passwords; (ii) possession-based, that is, based on something that the user possesses, such as smart cards; and (iii) inherence-based, that is, something that the user is, such as biometrics. Two or more of these techniques can be combined to increase security, which is known as multifactor authentication [1].

In this book chapter, to differentiate between single-factor and multifactor authentication techniques, the former will be referred to as **authentication schemes**, whereas the latter will be referred to as **multifactor authentication methods**.

Nowadays, the decision of what authentication scheme or method to implement in a software application resides within the software development team. However, the experience of the involved developers can vary from team to team, which could affect in the decision of what authentication technique to implement. Due to the importance of security [12], selecting the wrong authentication technique could potentially be a fatal mistake [13].

The above statement creates the necessity of a method that systematizes the task of comparing and selecting the authentication schemes and methods. A few frameworks in literature partially help to achieve this [14, 15]; however, they do not present the adequate characteristics for their application in distinct application contexts or do not consider all authentication techniques or multifactor authentication. Thus, this book chapter presents a decision framework that covers the observed gap. This framework has been generated through the application of an action-research methodology [16]. This action-research has been performed in collaboration with a multinational software development company and contemplates the utilization of other research methodologies that support it.

The remainder of this book chapter is organized as follows. The methodology utilized for the research is presented in Section 2. Section 3 is focused on obtaining of the knowledge base utilized for the research. In Section 4, the generated decision framework is presented. Section 5 consists on the validation of the framework. Finally, the conclusions and future work of the research are given in Section 6.

## 2. Methodology

The realization of this research is within the scope of an action-research methodology that was carried for over a year in collaboration with a software development company. The objective of action-research is to provide a benefit for the research's "client" while also generating relevant "research knowledge" [16, 17]. This kind of collaboration allows to study complex social processes, such as the use of information technologies in organizations, by introducing changes in them and observing their effects [18].

There are four roles involved in action-research [19]. These roles are as follows:

- The **researcher(s)** who undertake(s) the action-research. In this case, the researchers are the book chapter's authors.

- The **studied object**, that is, the problem to solve. In this case, the studied object is the comparison and selection of authentication schemes and methods.

- The **critical group of reference** that has a problem that needs to be solved and also participates in the research process. In this case, the critical group of reference is composed by the employees of the partnered software development company (PSDC).

- The **beneficiary** who can receive benefits from the research results, without directly participating in its process. In this case, the main beneficiary is the PSDC, but other software developers can also benefit from this research.

During the realization of this action-research, multiple activities were performed in conjunction with the PSDC. These activities helped to generate and validate the proposed decision framework for solving the need of automatizing the comparison and selection of authentication techniques. These activities were performed utilizing the iterative process of action-research, which considers, for every cycle, the following four phases [20]: (i) the planning phase, which considers the elaboration of a research question to be answered through the iteration; (ii) the action phase, where distinct research methodologies are applied to address the posed research question; (iii) the observation phase, where the results of the interventions from the previous phase are processed; and (iv) the reflection phase, where the researchers shares their finding with the group of reference to generate feedback; it is also possible to transversely perform this phase instead of cyclically [19], as it was done in this action-research through the realization of weekly progress meetings.

In this work, the action-research methodology was applied through three cycles. The objective of the first cycle was to obtain the required knowledge base for creating the framework. To achieve this, two strategies were applied: first, a systematic literature review (SLR) [21] was performed to obtain the existing knowledge in literature, and secondly, a number of surveys and interviews [16, 22] were conducted to learn the perceptions of the industry through the PSDC's employees. The second cycle was centered on the creation of the decision framework. During this cycle, an expert panel [23] was held to validate the initial draft of the framework. Finally, the third cycle focused on validating the final framework through the application of case studies [24].

## 3. Identification of the knowledge base

To construct the decision framework, it was necessary to obtain an adequate knowledge base regarding the topic at hand. To achieve this, two methodologies were applied. The first was the realization of a systematic literature review to identify the existing knowledge in related academic publications. The second corresponds to the application of a survey and interviews (S&I) to employees of the PSDC to learn the perceptions of the industry. The combined usage of these methods allowed the procurement of a knowledge base useful both for the academic and industrial sectors.

### 3.1 Systematic literature review

A systematic literature review has been carried out with the objective of "identifying authentication schemes proposed in literature and their possible combinations for their use as multifactor authentication methods, while also detecting criteria used for their comparison and selection and the existence of frameworks that handle such a task." Based on this objective, the following four research questions were formulated:

1. Which are the main authentication schemes that exist in the literature?

2. What combinations of these schemes can be found that can be used as multifactor authentication methods?

3. What criteria can be used to compare and/or to select between authentication schemes and/or multifactor authentication methods?

4. Are there frameworks that help to compare and/or to select authentication schemes or multifactor authentication methods? What are their characteristics?

The planning and results of the SLR have already been published in literature [25]. Additionally, a list containing the publications accepted during the SLR can be found in http://colvin.chillan.ubiobio.cl/mcaro/. Next, a brief summary of the main results of the SLR for every research question is presented.

### 3.1.1 Authentication schemes

A total of 515 publications regarding the proposal of authentication schemes were found. Their distribution among the authentication factors is as shown in **Figure 1**. Additionally, the context for which these schemes were proposed was recorded as well; this is presented in **Table 1**, including the publication's origin (journal article, conference article, or book chapter). It is important to mention that only 233 of the publications indicated a context.

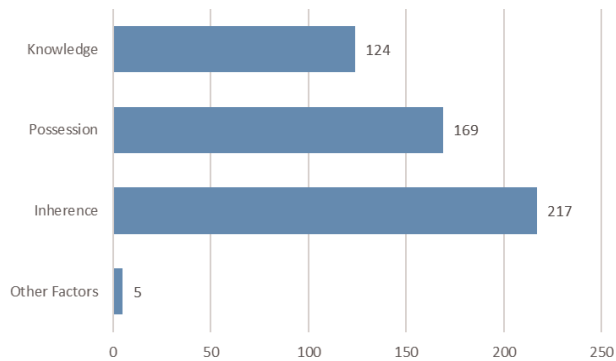

**Figure 1.**
*Number of publications proposing authentication schemes for every authentication factor.*

| Context | Journal | Conference | Book | Total |
|---|---|---|---|---|
| Mobile environment | 38 | 43 | 0 | 81 |
| Remote authentication | 31 | 11 | 0 | 42 |
| Healthcare/telecare | 23 | 1 | 0 | 24 |
| Multi-server environment | 15 | 2 | 0 | 17 |
| Continuous authentication | 9 | 2 | 0 | 11 |
| Wireless sensor networks | 8 | 2 | 0 | 10 |
| Cloud computing | 3 | 4 | 2 | 9 |
| Banking and commerce | 2 | 6 | 0 | 8 |
| Smart environment | 2 | 5 | 0 | 7 |
| Login protocols | 5 | 0 | 0 | 5 |
| Web applications | 4 | 1 | 0 | 5 |
| Other contexts | 7 | 7 | 0 | 14 |
| Total | 147 | 84 | 2 | 233 |

**Table 1.**
*Number of publications proposing authentication schemes for every context.*

### 3.1.2 Multifactor authentication methods

Four hundred forty-two publications proposing the combination of two or more authentication schemes in a multifactor manner were identified. Their distribution among the distinct authentication factor combinations is as shown in **Figure 2**. Similarly to the previous research question, the context for which these methods were proposed was recorded as well; this is presented in **Table 2**, including the publication's origin (journal article, conference article, or book chapter). In this case, 272 of the publications did indicate a context.

### 3.1.3 Comparison and selection criteria

Only 17 publications presented criteria for the comparison and selection of authentication schemes and methods. The presented criteria in the distinct publications can be categorized based on the kind of criteria proposed. Every publication considered one or more criteria categories; however, only three of them could be identified in more than one publication. The most identified categories of criteria are usability, security, and costs. The first two were identified in nine publications each, whereas the latter was found in five publications.

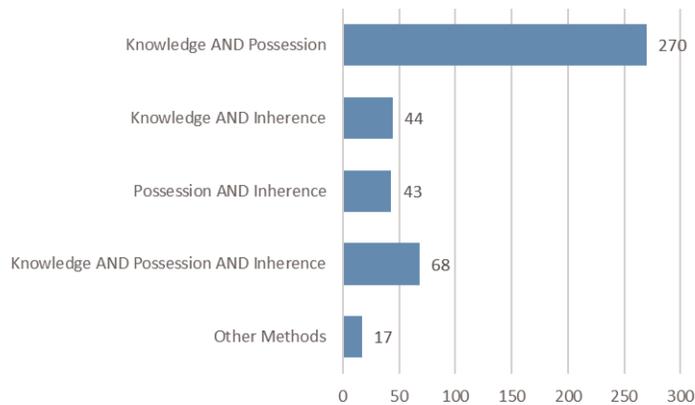

**Figure 2.**
*Publications proposing authentication methods for every factor combination.*

| Context | Journal | Conference | Book | Total |
|---|---|---|---|---|
| Remote authentication | 52 | 12 | 0 | 64 |
| Healthcare/telecare | 45 | 3 | 0 | 48 |
| Wireless sensor networks | 29 | 4 | 0 | 33 |
| Multi-server environment | 22 | 7 | 0 | 29 |
| Mobile environment | 10 | 11 | 0 | 21 |
| Cloud computing | 12 | 5 | 0 | 17 |
| Banking and commerce | 6 | 5 | 0 | 11 |
| Web applications | 5 | 6 | 0 | 11 |
| Wireless networks | 6 | 2 | 0 | 8 |
| USB devices | 1 | 5 | 0 | 6 |
| Insecure environment | 3 | 2 | 0 | 5 |
| Other contexts | 15 | 3 | 1 | 19 |
| Total | 206 | 65 | 1 | 272 |

**Table 2.**
*Number of publications proposing authentication methods for every context.*

Moreover, it could be observed that most of these articles highly considered the importance of the use context for comparing and selecting schemes and methods. This was mainly done by the publication addressing specific contexts or considering the context itself as another criterion.

### 3.1.4 Decision frameworks

Eight decision frameworks that help in the comparison and selection of authentication schemes and methods were identified. Through the analysis of these frameworks, it could be observed that multifactor authentication is not often considered, whereas proposals that do consider it utilize a limited number of criteria. Thus, no decision framework that considered multifactor authentication and enough criteria for a detailed comparison and selection of authentication schemes and methods could be found.

### 3.2 Survey and interviews

A survey and interviews have been applied to the PSDC's employees with the objective of learning the perceptions of people from the industry regarding authentication and the comparison and selection of distinct schemes and methods. The interviews were realized as a pilot application of the survey. A total of 12 employees were interviewed. In addition, 45 valid responses, out of a sample of 83 people ranging from developers to project leads, were received through the survey. Out of the 57 respondents, over two thirds of them held a senior position in the PSDC, as well as having over 6 years of working experience.

Four main questions were posed to the respondents, whose contents can be summarized as follows:

Q1. What authentication schemes do you know?

Q2. What multifactor authentication methods do you know?

Q3. What authentication schemes or multifactor authentication methods have you implemented in applications that you have developed?

Q4. What is the importance that you give to distinct factors when deciding what authentication scheme or method should be implemented in an application?

In http://colvin.chillan.ubiobio.cl/mcaro/ it is possible to find the questionnaire used for the survey. A summary of the responses obtained for every question is provided next.

### 3.2.1 Authentication schemes known by the respondents

For this question, respondents were asked to mark from a list the authentication schemes that they knew. The most known schemes were text passwords, one-time passwords (OTP, tokens), and mobile-based authentication. All respondents answered this question. The complete results of this question can be observed in **Table 3**, which shows the number of survey respondents and interviewed people that know each authentication scheme.

| Authentication scheme | Interviewees | Survey respondents |
|---|---|---|
| Text passwords (TP) | 10 | 40 |
| Graphical passwords (GP) | 1 | 20 |
| Cognitive authentication (CA) | 0 | 10 |
| OTP (tokens) | 7 | 38 |
| Smart cards (SC) | 3 | 24 |
| Mobile-based (MB) | 8 | 31 |
| Biometrics (B) | 5 | 30 |
| Federated single sign-on (FSSO) | 4 | 22 |
| Proxy-based (PB) | 1 | 8 |
| Others | 0 | 2 |

**Table 3.**
*Number of respondents that know each authentication scheme.*

### 3.2.2 Multifactor authentication methods known by the respondents

For the second question, respondents were given a brief explanation about multifactor authentication. Afterward, they were asked what multifactor authentication methods they knew. The combination of text passwords and OTP was the most known among them. A total of 27 out of the 45 survey respondents answered this question. The complete results of this question can be observed in **Table 4**, which shows the number of survey respondents and interviewed people that know each multifactor authentication method.

| Combination | Method | Interviewees | Survey respondents |
|---|---|---|---|
| Knowledge + possession | TP + OTP | 7 | 15 |
| | TP + SC | 2 | 8 |
| | TP + MB | 6 | 6 |
| | Others | 0 | 1 |
| | **Total** | 15 | 30 |
| Knowledge + inherence | TP + B | 0 | 15 |
| | Others | 0 | 3 |
| | **Total** | 0 | 18 |
| Possession + inherence | OTP + B | 0 | 6 |
| | MB + B | 0 | 3 |
| | SC + B | 0 | 3 |
| | **Total** | 0 | 12 |
| Knowledge + possession + inherence | TP + SC + B | 0 | 7 |
| | TP + OTP + B | 1 | 2 |
| | Others | 0 | 2 |
| | **Total** | 1 | 11 |
| **Grand total** | | 16 | 71 |

**Table 4.**
*Number of respondents that know each authentication method.*

### 3.2.3 Authentication schemes and methods implemented by the respondents

Next, the respondents were asked what authentication techniques they had implemented in applications developed by them and the kind of application. Most applications were either web-based or for banking and commerce. A total of 23 out of the 45 survey respondents answered this question. The complete results of this question can be observed in the graphs of **Figures 3** and **4**, which show the implemented authentication schemes and methods and the contexts of the applications that were being developed, respectively.

### 3.2.4 Comparison and selection criteria used by the respondents

For the last question of the S&I, distinct strategies were applied between the interviewees and the survey respondents. In the case of the former, they were directly asked what criteria they utilized for the comparison and selection of authentication schemes and methods. In the case of the latter, the responses from the interviewees, coupled with the results of the previously performed SLR, were used to generate a list of comparison and selection criteria that respondents were asked to value from 1 to 5. A higher value meant that the respondent gave a higher importance to the criterion. A total of 29 out of the 45 survey respondents answered this question. The complete results of this question can be observed in **Table 5** and in **Table 6**, which show the responses given by the interviewees and the survey respondents, respectively.

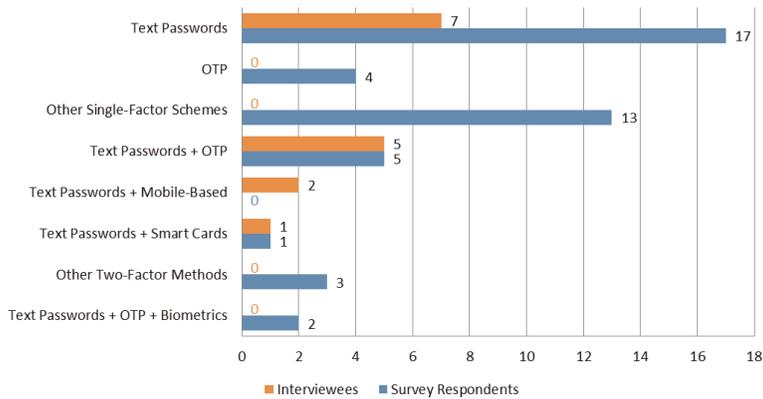

**Figure 3.**
*Authentication schemes and methods implemented by the respondents.*

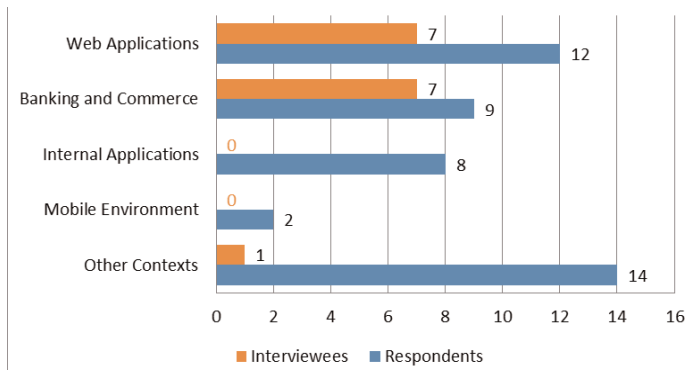

**Figure 4.**
*Contexts of the applications developed by the respondents.*

| Criterion | Interviewees that consider the criterion |
|---|---|
| Client's requirements | 11 |
| Application context | 11 |
| Usability-related criteria | 9 |
| Security-related criteria | 11 |
| Cost-related criteria | 8 |
| Other criteria | 2 |

**Table 5.**
*Comparison and selection criteria considered by the interviewees.*

Finally, survey respondents were asked what other comparison and selection criteria they would consider. The received answers include the ease of authentication information recovery, the registration method, and the sensitivity of the information.

### 3.3 Short survey

A second survey was later applied to nine employees of the PSDC. These employees were selected among the most experienced developers of the company, based on their years of experience and positions. The single aim of this survey was to ascertain the importance that the respondents would assign to an application's security and usability based on the target context. The importance was valued in percentages, with the sum of usability and security being 100% for every context. **Table 7** presents the results of this survey.

The obtained values were used afterward as part of the input for the decision framework.

| Category | Criterion | Value |
|---|---|---|
| Usability | Ease of use | 3.31 |
| | Ease of learning | 3.28 |
| | Need of using a device | 3.10 |
| | Method's reliability | 4.10 |
| Security | Importance of security | 4.41 |
| | Resistance to well-known attacks | 4.21 |
| Costs | Implementation costs | 4.07 |
| | Costs per user | 4.00 |
| | Server compatibility | 3.69 |
| | Need of acquiring licenses | 3.86 |
| | Available technologies | 3.93 |
| Others | Client's requirements | 4.17 |
| | Application context | 4.41 |
| | Norms and legislation | 3.90 |

**Table 6.**
*Comparison and selection criteria valued by the survey respondents.*

| Context | Importance of security (%) | Importance of usability (%) |
|---|---|---|
| Mobile environment | 45.56 | 54.44 |
| Remote authentication, multi-server environment, cloud computing | 64.44 | 35.56 |
| Healthcare/telecare | 57.78 | 42.22 |
| Wireless sensor networks | 63.33 | 36.67 |
| Banking and commerce | 73.33 | 26.67 |
| Web applications | 28.89 | 71.11 |

**Table 7.**
*Importance given to security and usability in distinct contexts by the respondents.*

## 4. The framework

This section describes the decision framework constructed through the knowledge base acquired by using the methodologies presented above. It has been given the name of Kontun framework, which means "to enter foreign property" in Mapudungún, an indigenous language from Chile, which is what it aims to prevent. **Table 8** shows a summary of the main findings during the knowledge base gathering and their origin (either the SLR or the S&I).

A summary of the constructed framework's characteristics is provided next. A complete description can be found in [26].

First, the framework considers a number of criteria obtained from the knowledge base, divided among the three most observed categories: security, usability, and costs. Each criterion is then given distinct possible importance values and a weight based on the findings from the knowledge base. To illustrate the above criterion, **Table 9** shows the usability-related criteria, their importance values, and their weights.

| Most reported knowledge-based schemes | • Text passwords (SLR, S&I)<br>• Graphical passwords (SLR) |
|---|---|
| Most reported possession-based schemes | • Smart cards (SLR)<br>• OTP (S&I)<br>• Mobile-based (S&I) |
| Most reported inherence-based schemes | • Face biometrics (SLR, S&I)<br>• Behavioral biometrics (SLR)<br>• Palm print (SLR)<br>• Fingerprints (SLR, S&I)<br>• Vein biometrics (SLR)<br>• Iris biometrics (SLR, S&I) |
| Multifactor authentication | • Prevalence of the combination of knowledge- and possession-based authentication schemes (SLR, S&I) |
| Most observed application contexts | • Mobile environment (SLR)<br>• Remote authentication (SLR)<br>• Multi-server environment (SLR)<br>• Cloud computing (SLR)<br>• Healthcare/telecare (SLR)<br>• Wireless sensor networks (SLR)<br>• Banking and commerce (S&I)<br>• Web applications (S&I) |
| Comparison and selection criteria | • Criteria are mainly related to usability, security, and costs (SLR)<br>• Identified criteria are valued positively by the industry (S&I)<br>• High importance observed regarding application context (SLR, S&I) |

**Table 8.**
*Summary of the acquired knowledge base.*

Every criterion has two or more importance values between 20 and 100, and the sum of all the weights of the criteria belonging to the same category is 100%. In this manner, when using the framework, a person must select the importance values that best describe their application and then calculate the average values of security (S), usability (U), and costs (C) using the following equations:

$$S = \sum_{\text{for each criterion of } S} AssessmentValue * CriterionWeight \tag{1}$$

$$U = \sum_{\text{for each criterion of } U} AssessmentValue * CriterionWeight \tag{2}$$

$$C = \sum_{\text{for each criterion of } C} AssessmentValue * CriterionWeight \tag{3}$$

The framework also considers a number of common contexts identified through the knowledge base. These contexts were given distinct weights based on the importance of security and usability in the context itself. Here, a term known as the security/usability value (SUV) is presented. The knowledge base allowed to ascertain the fact that, generally, the more secure an authentication scheme or method is, it has a lower usability and vice-versa. The SUV is used to denote this. Based on the calculated average values of S, U, and C, coupled with the selected application context (Ct), the SUV is calculated as follows:

$$SUV = A * S + B * (100 - U) \tag{4}$$

A and B are constants defined based on the importance given to S and U, respectively, in the selected context. A high SUV value thus indicates that more secure authentication methods should be implemented in the application, whereas a low SUV indicates that more usable authentication schemes or methods should be implemented in the application.

| Criterion | Importance | Value | Weight |
|---|---|---|---|
| Ease of use | The method necessarily needs to be easy to use | 100 | 25% |
| | The method preferably needs to be easy to use | 60 | |
| | It is not necessary for the method to be easy to use | 20 | |
| Ease of learning | A user should not take longer than a day to get used | 100 | 25% |
| | A user should not take longer than a week to get used | 60 | |
| | The time it takes to get used is not relevant | 20 | |
| Authentication information recovery | The recovery process should be simple | 100 | 10% |
| | The recovery process should be complex | 20 | |
| Need of using a device | It does not need to use a device | 100 | 10% |
| | It can use a possession or biometric device | 60 | |
| | It can use both a possession and a biometric device | 20 | |
| Authentication method's reliability | It should never or hardly fail during authentication | 100 | 30% |
| | It should not fail occasionally during authentication | 75 | |
| | It can fail occasionally during authentication | 45 | |
| | It does not matter how often it fails | 20 | |

**Table 9.**
*Criteria considered by the framework.*

Having calculated the SUV and also considering the average value given to C, the framework is able to provide a suggestion on what authentication schemes or methods to implement in the evaluated application. The recommendation is as follows: for a SUV of 65 or higher, the framework will suggest the implementation of highly secure authentication methods; for a SUV of 35 or lower, the framework will suggest the implementation of highly usable authentication schemes; and for a SUV between 35 and 65, the framework will suggest the implementation of averagely secure and usable authentication methods. Moreover, for a value of C of 60 and above, the framework will suggest the implementation of more affordable authentication schemes or methods; for a value of C below 60, the framework will suggest the implementation of more expensive authentication schemes or methods. The recommendations are also different based on the target Ct. Thus, for every Ct, the framework will give six possible recommendations based on the calculated SUV and C. **Table 10** illustrates the above framework for the context of mobile environment.

Finally, the person utilizing the framework must decide the authentication scheme or method to implement in their application, taking into consideration the recommendations given by the framework.

## 4.1 Tool prototype

To facilitate the use of the framework in software development environments, a tool prototype has been constructed that allows its utilization in a semiautomatic manner. This tool has also supported the validation process of the framework. With the tool prototype, the person in charge only needs to indicate the evaluated application's features and target context through a radio form. Afterward, the tool prototype automatically calculates the values of average S, U, and C and the SUV. The tool prototype is available for download in http://colvin.chillan.ubiobio.cl/mcaro/.

| | |
|---|---|
| *SUV* 65<br>*C* < 60 | Graphical passwords + smart cards + behavioral biometrics<br>Text passwords + OTP + behavioral biometrics<br>Graphical passwords + OTP + behavioral biometrics<br>Graphical passwords + OTP + face biometrics |
| *SUV* 65<br>*C* 60 | Text passwords + smart cards + behavioral biometrics<br>Text passwords + smart cards + face biometrics |
| 35 < *SUV* < 65<br>*C* < 60 | Graphical passwords + behavioral biometrics<br>OTP + behavioral biometrics<br>Text passwords + palm print/fingerprints<br>Graphical passwords + OTP |
| 35 < *SUV* < 65<br>*C* 60 | Text passwords + behavioral biometrics<br>Text passwords + smart cards |
| *SUV* 35<br>*C* < 60 | Behavioral biometrics<br>Graphical passwords<br>Face biometrics<br>Palm print/fingerprints |
| *SUV* 35<br>*C* 60 | Behavioral biometrics<br>Text passwords<br>Graphical passwords |

**Table 10.**
*Recommendation given by the framework for the context of mobile environment.*

The tool prototype has been developed using the model view controller (MVC) design pattern, with the Java programming language and supported by the Spring Framework. PostgreSQL has been used as the database management system.

The main screens of the tool prototype can be observed in **Figures 5**–7. They show the procedures for the criteria selection, the context selection, and the framework's recommendation, respectively.

**Figure 5.**
*Criteria selection in the tool prototype.*

**Figure 6.**
*Context selection in the tool prototype.*

**Figure 7.**
*Framework's recommendation in the tool prototype.*

The tool prototype also has additional features that facilitate its use in software development companies. Specifically, it has a user registration feature which allows maintaining a registry of its usage and a functionality for adapting its preferences based on the software development company's needs.

## 5. Validation through the industry

Through the creation of the framework, its adequacy was repeatedly validated using strategies associated to the application of the action-research methodology. Specifically, the validation was ascertained through the realization of an expert panel and the application of case studies. These are detailed in remainder of this section.

### 5.1 Expert panel

An expert panel was held in collaboration with five experts from the PSDC that consisted of four sessions with the aim of ascertaining their perceptions regarding an initial draft of the framework, so that it was more adequate to the real requirements observed in a software development environment. The activities during every session of the expert panel are described next.

#### 5.1.1 Presentation of the initial draft of the framework

The first session consisted on the presentation of the initial draft of the framework, with the purpose of helping the experts to have a general notion of the aim of this research.

#### 5.1.2 Validation of comparison and selection criteria

The preliminary list of criteria, their categorization, their values, and their weights were presented to the experts for their validation. This allowed to discard the least adequate ones and to generalize those that were too specific for the needs of a software development team.

#### 5.1.3 Validation of the considered contexts

The contexts considered by the framework were presented to the experts. Similarly to the previous session, this allowed to make the appropriate modifications to the currently selected contexts. Additionally, the SUV was presented to the experts, who generally agreed to the adequacy of its use.

#### 5.1.4 Validation of the framework's recommendations

The authentication schemes and methods recommended for every situation were presented to the experts. This allowed to ascertain the adequacy of every recommendation. The experts were generally in agreement with the recommendations.

### 5.2 Case studies

After its construction, the validation of the framework's recommendations was realized through the application of a case study methodology in collaboration with the PSDC. Specifically, the framework's recommendations were compared with the authentication schemes or methods implemented in existing applications developed by the PSDC or with the recommendations that their experts would give for hypothetical situations. The case studies are described in detail in [26]. Next, a brief summary of their application is provided.

The case studies are split in three categories: (i) those that were realized by comparing the framework's recommendation against the implemented scheme or method on an existing application, (ii) those that were realized by comparing the framework's recommendation against the recommendations given by experts for hypothetical applications, and (iii) those that were realized by comparing the framework's recommendation against the implemented scheme or method on an existing application and also against the recommendation given by experts for hypothetical applications with nearly the same features as the existing ones. These case studies are presented in **Tables 11–13**, respectively, presenting the implemented scheme or method in the existing application, the framework's recommendation, the most recommended scheme or method by the experts, and the acceptance rate of the framework's recommendation, as appropriate.

In general, the results of the case studies are favorable for the framework. It is important to mention that, where discrepancies are observed, there was often a reasoning behind them. For example, for case study 3 (existing application), the implemented scheme was demanded by the client and not selected by the software development team.

| ID | Implemented scheme or method | Framework's recommendation |
|----|------------------------------|----------------------------|
| 1 | Two-factor authentication (text passwords + smart cards) | Three-factor authentication (text passwords + OTP + behavioral biometrics) |
| 2 | Two-factor authentication (text passwords + mobile-based) | Two-factor authentication (text passwords + mobile-based) |
| 3 | OTP (demanded by client) | Behavioral biometrics |

**Table 11.**
*Case studies based on existing applications.*

| ID | Experts' recommendation | Framework's recommendation | Acceptance rate of framework's recommendation |
|----|-------------------------|----------------------------|-----------------------------------------------|
| 4 | Two- or three-factor authentication | Three-factor authentication | 100% |
| 5 | Text passwords | Two-factor authentication | 80% |

**Table 12.**
*Case studies based on hypothetical applications.*

| ID | Implemented scheme or method | Experts' recommendation | Framework's recommendation | Acceptance rate of framework's recommendation |
|----|------------------------------|-------------------------|----------------------------|-----------------------------------------------|
| 6 | Two-factor authentication | Text passwords | Text passwords | 100% |
| 7 | Text passwords | Two-factor authentication | Two-factor authentication | 90% |

**Table 13.**
*Case studies based on existing applications with a hypothetical counterpart.*

## 6. Conclusions

The research presented in this book chapter summarizes the definition of a theoretical framework. This framework will help in the comparison and selection of the most appropriate authentication schemes or multifactor authentication methods for applications created by software developers. It has been created through the application of an action-research methodology that considered the utilization of various other research methodologies that helped to contribute in distinct ways to the research objective.

On the one hand, a systematic literature review, coupled with surveys and interviews, was performed to obtain the required knowledge base for generating the framework. The utilization of these two methodologies allowed to ascertain the perceptions on authentication from both the academy and the industry.

On the other hand, an expert panel and several case studies were realized to validate the adequacy of the framework. This permitted to obtain feedback from the end users of the framework so that it would provide adequate authentication scheme or method recommendations and have an appropriate usability.

Thus, this experience allowed to observe the usefulness of performing a research in collaboration with the industry, as it permits obtaining results that align more adequately with their needs while also providing more refined academic results.

Several future work lines can be followed based on this research. Namely, the framework could be adapted to work as a recommendation system so that its recommendations get refined through its usage. For the industry, it would be of interest that the framework not only recommends an authentication technique but that it also provides the required code for its implementation. Finally, the last cycle of the action-research, that is, the realization of case studies, could be replicated in other software development companies to further validate the adequacy of the framework.

## Acknowledgements

This research is part of the following projects: DIUBB 144319 2/R and BuPERG (DIUBB 152419 G/EF).

## Author details

Ignacio Velásquez, Angélica Caro* and Alfonso Rodríguez
Computer Science and Information Technologies Department, University of Bío-Bío, Chillán, Chile

*Address all correspondence to: mcaro@ubiobio.cl

IntechOpen

Section 2

# Cryptography

# Secure Communication Using Cryptography and Covert Channel

*Tamer S.A. Fatayer*

## Abstract

The keys which are generated by cryptography algorithms have still been compromised by attackers. So, they extra efforts to enhance security, time consumption and communication overheads. Encryption can achieve confidentiality but cannot achieve integrity. Authentication is needed beside encryption technique to achieve integrity. The client can send data indirectly to the server through a covert channel. The covert channel needs pre-shared information between parties before using the channel. The main challenges of covert channel are security of pre-agreement information and detectability. In this chapter, merging between encryption, authentication, and covert channel leads to a new covert channel satisfying integrity and confidentiality of sending data. This channel is used for secure communication that enables parties to agree on keys that are used for future communication.

**Keywords:** encryption, authentication, dynamically, covert channel, confidentiality, algorithm, undetectability, fake key

## 1. Introduction

Encryption is considered the main key factor of security to achieve confidentiality and to protect data from disclosure [1]. Encryption is not efficient to achieve integrity. It needs another factor called authentication [2]. Covert channel is created to transfer data indirectly between client and server; it was created by Lampson [3].

Before the client and server use the covert channel, they must agree on a pre-agreement knowledge. Also, they agree on how to send that knowledge. A good example is they agree on even word meaning "00" and odd word meaning "11." If the client sends "communication channels," the server will know that the client's message is "1100" [4, 5].

Covert channel cannot be detected if the following two factors exist: plausibility and bit distribution. Plausibility deceives the attacker who thinks that the channel is normal channel and it is not used to send secret information. On the other hand, the bit distribution of normal channel must same distribution of covert channel [5].

The technique is worked as shown in **Figure 1**, where it shows the general idea of the proposed technique. Covert channel needs shared information. In my protocol, it is considered as a table shared between the client and server. This table contains characteristics of the client such as the name which represents the original key. Each original key has fake keys. I used encryption algorithm to guarantee the confidentiality. HMAC is used to check integrity. Finally, the time that is needed for the client and server to agree on secret information (e.g., secret keys) is measured.

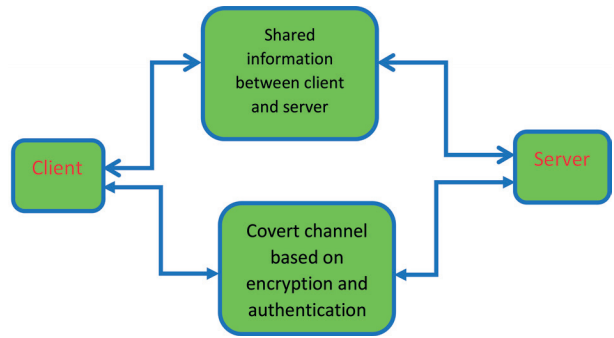

**Figure 1.**
*Secure communication using covert channel, encryption, and authentication.*

In this chapter, secure communication channel for transferring data is implemented. The channel between the client and server is considered a covert channel that depends on authentication and encryption.

## 2. Background

Lampson was the first to introduce the idea of a covert channel [3]. Transferring data between two entities indirectly through a channel is called a covert channel. Before the client and server use the channel to transfer data, they must agree on a pre-agreement knowledge (e.g., shared memory, table). For example, a word containing "mm" means bit "0" other than this means bit "1." So, if the client wants to send "10" to the server indirectly, the client will send "secure communication" to the server. The attacker hardly breaks the covert channel and it is considered to be more secure if it is undetectable [3, 4].

### 2.1 Covert channel characteristics and properties

Although a covert channel transfers information in a hidden way, it has the same characteristics as other communication channels. These characteristics are:

- Capacity: the amount of data that can be transmitted through the channel. From security viewpoint, increasing channel capacity leads to more information leakage. The covert channel capacity is measured in bits/second. To obtain maximum bandwidth through a covert channel, encoding schemes must be chosen between the sender and receiver.

- Noise: transmitted data through a covert channel are exposed to an amount of perturbations that makes the transmitted and received information between two entities not the same.

- Transmission mode: the transmission of information in covert channels (as in normal channels) can be synchronous or asynchronous. The sender and receiver in synchronous mode should manage their transmission based on a condition or a specific event. On the other hand, in asynchronous mode, the transmission occurs without a prior condition.

## 2.2 The covert channel is more private and undetectable if it satisfies the following

1. Plausibility: the TCP is usually used for Internet traffic, and it always employs using time stamp option. As a result, TCP using time stamp is a plausible covert channel because the majority of users using TCP will not use it for sending covert data. So, the adversary will believe that TCP time stamps will not be used for sending data covertly.

2. Undetectability: in order for a channel to be more undetectable, the channel must satisfy that the distribution of bits with covert data must be similar to the distribution of the normal channel. If an adversary notices that there are differences (using statistical tests) in bit distribution, then he will detect that the channel is a covert channel. Also, to achieve undetectability, the channel's bits must be random; otherwise, it will be noticed by the adversary.

3. Indispensability: Lampson [3] reports that a communication channel is a covert channel if it is neither designed nor intended to transfer information at all. The channel should introduce several benefits to the users besides sending data covertly; thus, the adversary cannot or will not close off that channel.

## 2.3 Covert channel classification

Covert channels can be classified as storage or timing channels, noisy or noiseless channels, and program-flow channels.

### 2.3.1 Storage channels and timing channels

The covert storage channel depends on a shared variable or a storage location, whereby one process (sender) can be allowed to write directly or indirectly to the storage location and the other process (receiver) reads from that storage location. On the other hand, the covert timing channel enables senders to send information to the receiver through signals, whereby the sender manages the time that is needed to perform some operation in such a way that when the receiver observes the time, it will understand a special event or a special piece of information. The main disadvantage of the timing channel is that it is considered very noisy because of the several external factors that affect the execution time of a process. Covert storage channels and timing channels need a synchronization process, which enables the sender and receiver to synchronize with each other to send and receive information. The storage covert channel uses a data variable to enable the sender and receiver to communicate. Therefore, a synchronization variable, called sender-receiver, is needed by the sender to notify the receiver that he has completed reading or writing a data variable. The covert channel uses another synchronization variable, called receiver-sender. To distinguish between storage and timing channels, if a channel uses a storage variable to transfer data between the sender and receiver, it is considered a storage channel. On the other hand, a covert timing channel uses time reference (e.g., a clock) to transfer data between the sender and receiver, whereby the sender and receiver use a common time reference.

### 2.3.2 Noisy and noiseless channels

I discussed previously that the characteristics of the covert channel are similar to any communication channel. One of these characteristics is that the channel may be noisy. The covert channel can be noiseless if the transmitted data by the sender and received data by the receiver are the same with probability 1; otherwise, the channel is noisy. Usually, data transmitted through a covert channel is represented by bit "0" or "1." Nevertheless, if the receiver decodes every bit transmitted by the sender correctly, then the covert channel is considered noiseless. Thus, to reduce error rate, which is produced by noise, correction codes are used [6].

### 2.3.3 Program-flow channels

I present a new type of covert channel, which is program-flow. The program-flow covert channel depends on the flow of program execution to convey information. In our proposed covert channel, the sender tries to guess the correct delta_mmap (encoded information) of the vulnerable server program. The server code which executes in case of successful guess differs from which executes in a failed guess. The receiver distinguishes between server code executed in successful and failed guesses.

## 2.4 Authentication and key exchange

Authentication process identifies entities that are attempting to access some resources. Diffie-Hellman (DH) algorithm is used as method of public key exchange.

### 2.4.1 Authentication process

Authentication is a process of checking whether someone or something is authorized or not to access some resources. Authentication can be computer to computer or process to process and mutual in both directions [7, 8]. Bob can authenticate Alice's identity depending on four factors [7, 9], which are:

#### 2.4.1.1 Something you know

Alice sends a request to the server to access some resources; Bob authenticates Alice by asking her about a secret thing that she knows, such as password. If Alice issues a correct password, then Bob will accept her request for accessing some resources. Fortunately, a password is needed to login into the system and access its resources. Yet, unfortunately, the user is always asked to reuse the password when he wants to log into the system, which gives attackers opportunities to hack the password and reuse it. The solution for this problem is to use a onetime password (OTP) so that the user each time she logs into a system needs a new password.

#### 2.4.1.2 Something you have

One of the disadvantages of the first authentication factor (something you know) is that the user may forget his password. Thus, the second authentication factor (something you have) overcomes this problem, whereby the user has an object (e.g., automatic teller machine (ATM) cards, OTP cards [7], and smart cards [9]) to access the system. Unfortunately, the objects may get stolen by attackers.

#### 2.4.1.3 Something you are

The third authentication factor is based on the measurements of the user's physical characteristics such as the fingerprints, iris, and voice. The techniques that measure the behavioral characteristics of the user are called biometrics [7, 8]. This factor overcomes the problems of the previous factors because it does not depend on a password or a token.

### 2.4.1.4 Somebody you know

Brainard et al. [9] proposes a fourth factor of authentication that is dependent on emergency authenticator, and it is used when the primary authenticator is unavailable to a user. A good example for emergency system is email; thus, when a user forgets his password, he often has the option of having password reset instructions. A system called "vouching" is introduced. A voucher system permits swapping of the roles of the token and PIN to deal with the case when the user has forgotten his PIN but still has his token.

### 2.4.2 Message authentication

When Alice and Bob want to exchange messages, they do not want an attacker to modify the contents of their messages. This can be achieved by using message authentication odes (MACs), where the MAC is a tag, attached to the message by Alice to Bob or vice versa. If Bob validates this tag, the request of Alice will be accepted by Bob; otherwise, it is rejected [7]. MAC that is based on cryptographic hash functions is called HMAC [10]. There are many hash functions, such as message digest 5 (MD5) and Secure Hash Algorithm 1 (SHA1). When HMAC is used with MD5, it is called Hashed Message Authentication Code-Message Digest 5 (HMAC-MD5), and when it is used with SHA1, it is called Hashed Message Authentication Code-Secure Hash Algorithm 1 (HMAC-SHA1) [7, 10, 11]. In our dissertation, we use the secure hash algorithm SHA256 with pre-shared key to form HMAC-SHA256, where the secure hash algorithm SHA256 takes a message of 512-bit blocks as input and returns a digest message with 256 bits as output [7].

### 2.4.3 Diffie-Hellman (DH) key exchange algorithm

Key exchange algorithms are cryptographic methods that generate cryptographic shared keys that are shared among users. After Alice and Bob agree on a shared key, they can use it in HMAC, and they can also use it in symmetric encryption algorithms to encrypt or to decrypt files. Alice encrypts file using one of the symmetric algorithms and sends it to Bob. Bob in turn uses the same symmetric algorithm to decrypt the file. Note that Alice and Bob must agree on a shared key before using symmetric algorithm. Many key agreement protocols have been proposed. The Diffie-Hellman (DH) algorithm [12] is a very popular example that introduces a key exchange protocol using the discrete logarithm problem [13]. DH algorithm enables Alice and Bob to exchange secure keys over an insecure channel.

**Figure 2** shows the mechanism of DH. The values g and p are public parameters known to Alice and Bob, whereby p is a prime number and g (generator of p) is an integer less than p. This means that for all every number n between 1 and p-1, there is a power k of g such that $n = g^k$ mod p. Both Alice and Bob choose a secret random integer number, a and b, respectively. After that, Alice sends to Bob ($g^b$ mod p), and Bob sends to Alice ($g^b$ mod p). Finally, they agree on a secret key by using this formula (($g^a)^b$ mod p).

**Figure 3** shows that a "man-in-the-middle" (Mallory) can listen and modify the conversation messages between Alice and Bob. In so doing, she can convince Alice and Bob that they are communicating with each other while in fact both are communicating with Mallory [7]. Moreover, **Figure 3** shows that the main vulnerability in the DH protocol is that it does not have an authentication process. Several versions of DH protocol exist to overcome this problem, for example, by using DH with digital signature [11]. Diffie et al. [14] enhance a Diffie-Hellman protocol with an authentication process, whereby Alice and Bob must authenticate themselves using a digital signature. Alice and Bob must have a pair of keys (public and private key) and a certificate for the public key. So, during execution of DH protocol, Alice and Bob transmit massages with signature; Mallory cannot forge the signature because she needs to share Alice's private key and Bob's private key.

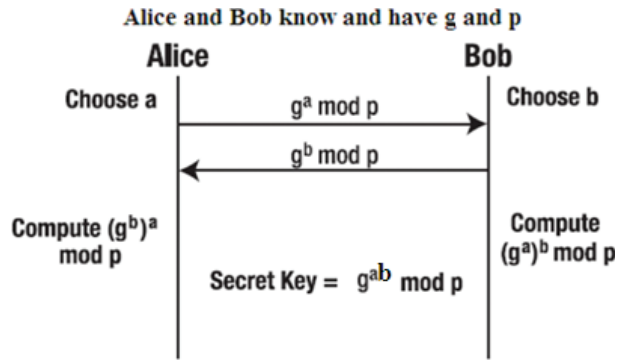

**Figure 2.**
*Diffie-Hellman key exchange algorithm. Alice and Bob agree on a secret key over an insecure channel. The secret key that they agree on is computed as (ga)b mod p.*

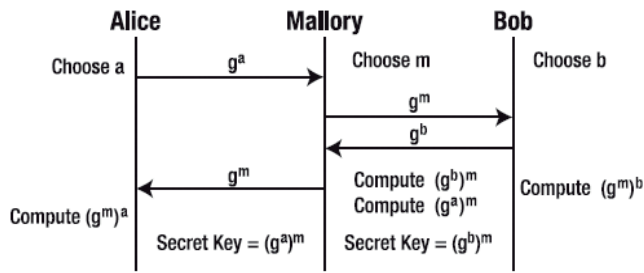

**Figure 3.**
*A Diffie-Hellman weakness. A man-in-the-middle (e.g., Mallory) impersonates Alice to agree on a shared key with Bob. Also, she impersonates Bob to agree on a shared key with Alice.*

The DH algorithm relies on heavy computation, which may not be suitable to resource-constrained platforms.

## 3. Related work

There are many researches that target covert channel undetectability [15–17], but most of the works have drawbacks and lack in channel detectability (Girling [18] in 1987). He creates three covert channels through a local area network (LAN): two of them are storage channels, and the third is a timing channel. The two storage channels depend on "what-is-sent" strategy, whereby one of them depends on the frame size, which is sent by the sender. If the frame size equals to 256, the amount of covert information decoded by receiver, who monitors the sender activity on the LAN, would be 8 bits. On the other hand, the timing channel depends on the time that represents the time interval between successive sends. The time difference between successive sends may be odd or even, and the prior agreement between the sender and receiver (who monitors the time between successive sends) is such that the odd time means bit "0" and even time means bit "1." So, the timing channel obviously depends on "when-is-sent" strategy.

TCP/IP protocol is used to create covert channel that is targeted by many researchers, where they used TCP to hide information [19–22]. Zhang et al. [6] propose covert channel to transfer messages to control (increasing or decreasing) the period of silence in traffic of VoLTE traffic. Create covert channel through hiding information in IP fields [20, 23]. Mead et al. [24] propose timing covert channel for wireless communication; they developed android application to communicate through local area network and mobile network. The results show that the channel is very undetectable in spite of the existence of malware and intrusion detection system.

Some researches, Fatayer et al. [15–17], try to use covert channel as benign channel, and it can be used to send legal information between the client and server. They used gaps in memory to create covert channel. Also, they used the channel to send text and audio files in acceptable time. The proposed technique depends on pre-agreement database which consists of original keys and its corresponding fake keys. Each original key has multiple fake keys. The database consist of the characteristics of clients; each feature represents an original key, and it has multiple fake keys. **Figure 4** figures out the pre-agreement between the client and server before using covert channel.

Customer asks cloud provider to access his resources. The summarization of approach is as follows. First, pre-agreement between the server and client is shown in **Figure 4**. Second, the customer sends a packet which contains "Fakei" attribute belongs to a specific customer (e.g., name) to the cloud provider. Third, the cloud provider will analyze the packet and make sure that the "Fakei" belongs to which customer. If yes, the provider goes to next step. Fourth, the cloud provider will ask for extra information to verify the customer and then he sends a packet that contains another fake key to the customer. Fifth, the customer receives the packet and he verifies the packet. The customer will send the required information to cloud provider such as one of the fake keys of email. Seventh, the cloud provider will analyze the packet to make sure that the "Fakei" (email) belongs to which customer. If yes, the cloud provider will accept the request. Eight, steps from 4 to 7 are considered the first level of security, so if these steps are repeated more than one time, it can achieve multilevel of security.

A new detection approach of covert timing channel is proposed by Fahimeh et al. [25], where this approach enables to detect covet time channel through traffic distribution. They used statistical test to measure the network traffic online.

| Characteristics/ fake keys | Original | Fake $_{i=0}$ | ... | Fake $_{i=n}$ |
|---|---|---|---|---|
| Name | tamer | 00001111 | ... | 000000001 |
| id | 93/98 | ........ | ... | ... |
| Email | ........ | ........ | ... | .... |
| ...... | ..... | .... | | |

**Figure 4.**
*Database as pre-agreement between the client and server.*

## 4. Covert channel with authentication leads to secure communication

The proposed technique is responsible for creating secure communication channel using covert channel, encryption, and authentication. In the first step before using covert channel, the client and server must agree on pre-agreement table as shown in **Figure 4**, where it depicts out the pre-agreement table consisting of original key (OK) and its corresponding fake keys. The key point here is that the fake key is used in communication channel and the original key is kept secret in both sides (client and server). **Figure 5** illustrates the proposed technique, where the network consists of the client and server. The client agrees on shared information (e.g., table as database) with the server.

The client wants to send to the server a message. Then, he encrypts the message using the original key. After that the encrypted message is attached with the fake key to be a parameter to HMAC function. HMAC depends on shared key between the client and server. Then the client sends (HMAC + encrypted message + fake key) to the server. Where, HMAC is used for integrity and fake key for server to know the original key to decrypt the encrypted message. On the other hand, the server receives the client's message. After that he separates the message to the HMAC, encrypted message, and fake key. Then, he checks the integrity of the message. After that he gets the original key from its corresponding fake key to decrypt the encrypted message.

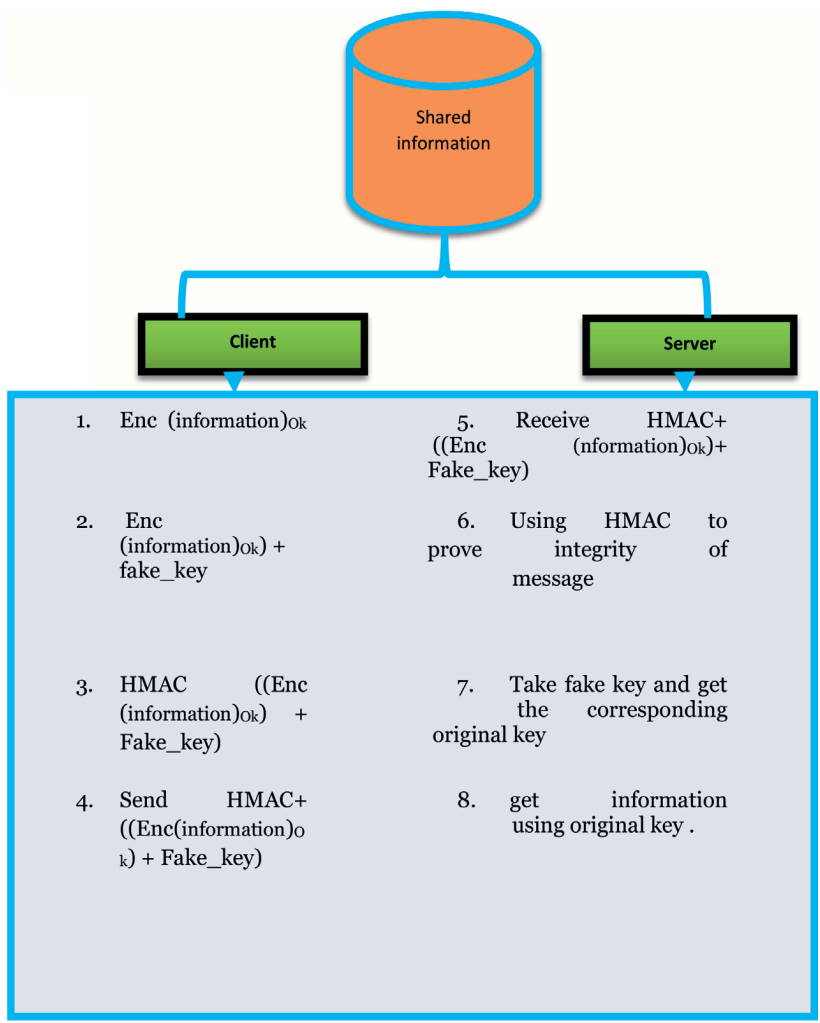

**Figure 5.**
*Technique for transferring secure data through covert channel using covert channel, encryption, and authentication.*

## 5. Performance discussion

The implementation of technique is done by: first create a network as client and server with implemented java application. Client and server machines are 32 bit ×86, CPU Core (TM) i5 2.40 GHz, and Ram 4GB. Advanced Encryption Standard (AES) [26] algorithm is used in the implementation. Hashed message authentication code (HMAC) is used to guarantee the integrity [7, 11, 12]. The following issues are satisfying in this technique:

1. Confidentiality: the technique guarantees that the messages are protected from disclosure, which is done by encrypting the messages with original keys that do not send through communication channel. Instead, the original keys are sent encrypted by fake keys.

2. Integrity: the information is protected from being changed by unauthorized parties through using HMAC function which checks if the content of the message is altered or not.

3. Undetectability: undetectability is achieved depending on two conditions: first, plausibility, the messages that are sent through the covert channel are protected from adversary by making the covert channel appear like a normal channel through sending normal encrypted messages, and, second, hiding the fake key inside the message which does not affect bit distribution, especially when the size of the fake key is small.

4. Comparative analysis: My technique is used in two ways, malicious and benign usages. Also, encryption and authentication are used that differ from other techniques.

5. Dynamically: the generating keys between the client and server have a flexible length. Because when you repeat the scenario in **Figure 5** several times, you get new keys with different sizes. If you repeat the technique four times and each time generates key with 16-bit size, then you will get a key with 64-bit size.

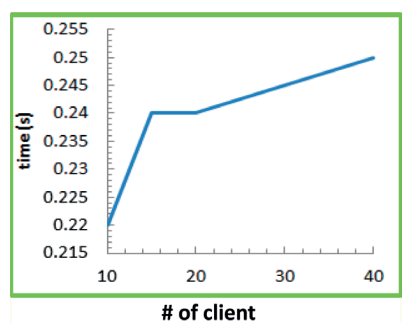

**Figure 6.**
*The technique can handle multiple users in acceptable time.*

The time that is needed by the server to serve the client is measured. The technique can handle several clients with acceptable time as shown in **Figure 6**. The server needs less than 0.25 s to deal and handle 30 clients.

## 6. Conclusion and future work

Encryption is used to achieve confidentiality to protect data from stealing from the third party (e.g., attacker). If users use encryption, they cannot achieve integrity. They need authentication and covert channel besides encryption technique to achieve integrity. In this chapter, we used encryption, authentication, and covert channel to produce a secure communication between eligible parties. This technique satisfies confidentiality through encryption and integrity using authentication algorithm. Finally, this technique generates undetectability covert channel through normal distribution of bits. Secure communication channel between the client and server that enables them to transfer data securely and to agree on keys that are used for future communication. The technique needs pre-shared table that consists of original and fake keys.

## Acknowledgements

This is to happily express my sincere thanks and appreciation to the following for their support and guidance throughout the chapter writing. I would like to thank my friends who stood beside me and helped me pursue my work. This chapter is dedicated to my parents, without whom, after the blessings of Allah, all this work would not be possible. They have been a source of endless love, encouragement, and support. They believed in me and in whatever decision I took and are proud of me on whatever achievement I may have.

## Thanks

Your support means a lot of thanks.

## Author details

Tamer S.A. Fatayer
Computer Science and Information Technology, Al-Aqsa University, Gaza, Palestine

*Address all correspondence to: ts.fatayer@alaqsa.edu.ps

**Chapter 8**

# High-Speed Area-Efficient Implementation of AES Algorithm on Reconfigurable Platform

*Altaf O. Mulani and Pradeep B. Mane*

**Abstract**

Nowadays, digital information is very easy to process, but it allows unauthorized users to access to this information. To protect this information from unauthorized access, cryptography is one of the most powerful and commonly used techniques. There are various cryptographic algorithms out of which advanced encryption standard (AES) is one of the most frequently used symmetric key cryptographic algorithms. The main objective of this chapter is to implement fast, secure, and area-efficient AES algorithm on a reconfigurable platform. In this chapter, AES algorithm is designed using Xilinx system generator, implemented on Nexys-4 DDR FPGA development board and simulated using MATLAB Simulink. Synthesis results show that the implementation consumes 121 slice registers, and its maximum operating frequency is 1102.536 MHz. Throughput achieved by this implementation is 14.1125 Gbps.

**Keywords:** cryptography, AES, FPGA, VLSI, system generator

## 1. Introduction

NIST has started a development process of FIPS for AES algorithm stating that this is the replacement for data encryption standard (DES) algorithm. Alternatively, this algorithm is also known as Rijndael algorithm. Rijndael algorithm has the advantages like resistance against all recognized attacks, code and speed compactness, and simple design. Cryptography is a process in which the information to be sent is added with secret key so as to transmit the data securely at the destination. There are two types of cryptography based on the type of key applied: symmetric key cryptography and asymmetric key cryptography. In symmetric key cryptography, equal key is utilized for encryption as well as decryption, whereas in asymmetric key cryptography, different keys are required in encryption and decryption. AES algorithm is selected for implementation because it is secure and its components and design principles are completely specified. AES is a symmetric key block cipher. The design of AES algorithm is based on linear transformation. Due to the use of Rijndael algorithm, different block and key sizes can be selected which was not possible in DES algorithm. Block and key size can be selected from 128/160/192/224/256 bits and need not be the same. According to AES standard, this algorithm can only accept 128 bits of block, and key size can be selected from 128/192/256 bits. Based on the key size, the number of rounds will vary. For example, if key size is 128, 192, or 256, then the number of rounds will be 10, 12, and 14, respectively. The structure of AES algorithm is shown in **Figure 1**. In this chapter, this algorithm is designed with 128 bits of block size and key size, respectively, that is, AES generates cipher text of 128 bits for 128 bits of plaintext. After the initial round, plaintext processes through ten rounds. Each round contains processes like byte substitution, shift rows, mix columns, and add round key.

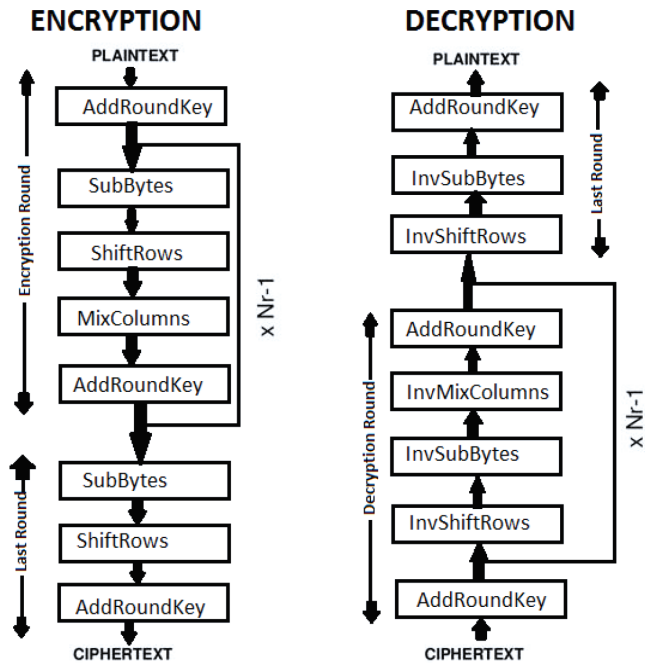

**Figure 1.**
*Structure of AES algorithm.*

**1.1 Byte substitution**

The 16 input bytes are substituted by using fixed lookup table known as s-box. **Figure 2** shows s-box of AES algorithm. This s-box consists of all possible combinations of 8-bit sequence. The resulting new 16 bytes are organized in a matrix having four rows and four columns.

**Figure 3** shows byte substitution stage in AES algorithm.

|   | 0 | 1 | 2 | 3 | 4 | 5 | 6 | 7 | 8 | 9 | A | B | C | D | E | F |
|---|---|---|---|---|---|---|---|---|---|---|---|---|---|---|---|---|
| 0 | 63 | 7C | 77 | 7B | F2 | 6B | 6F | C5 | 30 | 01 | 67 | 2B | FE | D7 | AB | 76 |
| 1 | CA | 82 | C9 | 7D | FA | 59 | 47 | F0 | AD | D4 | A2 | AF | 9C | A4 | 72 | C0 |
| 2 | B7 | FD | 93 | 26 | 36 | 3F | F7 | CC | 34 | A5 | E5 | F1 | 71 | D8 | 31 | 15 |
| 3 | 04 | C7 | 23 | C3 | 18 | 96 | 05 | 9A | 07 | 12 | 80 | E2 | EB | 27 | B2 | 75 |
| 4 | 09 | 83 | 2C | 1A | 1B | 6E | 5A | A0 | 52 | 3B | D6 | B3 | 29 | E3 | 2F | 84 |
| 5 | 53 | D1 | 00 | ED | 20 | FC | B1 | 5B | 6A | CB | BE | 39 | 4A | 4C | 58 | CF |
| 6 | D0 | EF | AA | FB | 43 | 4D | 33 | 85 | 45 | F9 | 02 | 7F | 50 | 3C | 9F | A8 |
| 7 | 51 | A3 | 40 | 8F | 92 | 9D | 38 | F5 | BC | B6 | DA | 21 | 10 | FF | F3 | D2 |
| 8 | CD | 0C | 13 | EC | 5F | 97 | 44 | 17 | C4 | A7 | 7E | 3D | 64 | 5D | 19 | 73 |
| 9 | 60 | 81 | 4F | DC | 22 | 2A | 90 | 88 | 46 | EE | B8 | 14 | DE | 5E | 0B | DB |
| A | E0 | 32 | 3A | 0A | 49 | 06 | 24 | 5C | C2 | D3 | AC | 62 | 91 | 95 | E4 | 79 |
| B | E7 | C8 | 37 | 6D | 8D | D5 | 4E | A9 | 6C | 56 | F4 | EA | 65 | 7A | AE | 08 |
| C | BA | 78 | 25 | 2E | 1C | A6 | B4 | C6 | E8 | DD | 74 | 1F | 4B | BD | 8B | 8A |
| D | 70 | 3E | B5 | 66 | 48 | 03 | F6 | 0E | 61 | 35 | 57 | B9 | 86 | C1 | 1D | 9E |
| E | E1 | F8 | 98 | 11 | 69 | D9 | 8E | 94 | 9B | 1E | 87 | E9 | CE | 55 | 28 | DF |
| F | 8C | A1 | 89 | 0D | BF | E6 | 42 | 68 | 41 | 99 | 2D | 0F | B0 | 54 | BB | 16 |

**Figure 2.**
*S-box of AES algorithm.*

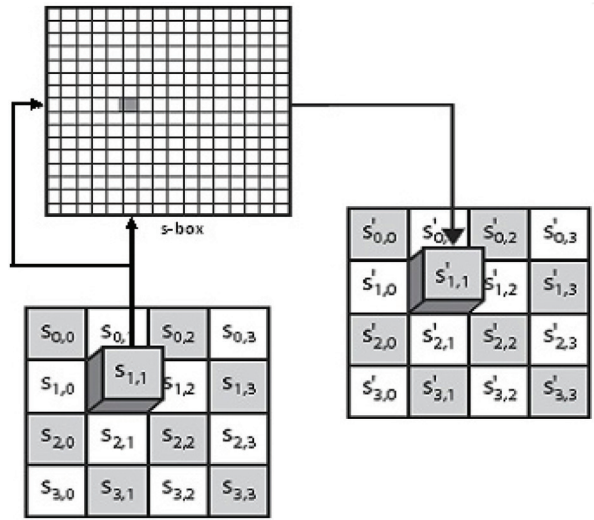

**Figure 3.**
*Byte substitution stage.*

### 1.2 Shift row

Each row from the matrix generated from the byte substitution is cyclically shifted to the left. Any entry that is dropped off is reinserted to the right side. The first row is kept as it is, the second row is shifted by one-byte position to the left, the third row is shifted by two-byte position to the left, and the fourth row is shifted by three-byte position to the left. The resultant matrix consists of same 16 bytes but at different position. **Figure 4** shows shift row stage in AES algorithm.

### 1.3 Mix column

Each column of four bytes is now transformed using special arithmetical function of Galois field (GF) 28. This function takes four bytes of the column as input and outputs completely new four bytes that replaces the original four bytes. **Figure 5** shows mix column stage in AES algorithm.

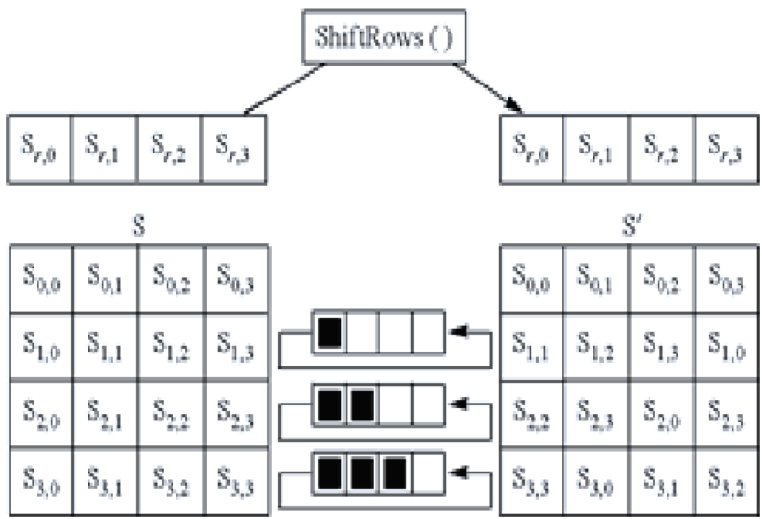

**Figure 4.**
*Shift row stage.*

## 1.4 Add round key

The 16 bytes of the resultant matrix generated from mix column stage are then considered as 128 bits. In add round key stage, 128 bits of state are bitwise EX-ORed with 128 bits of round key. If this result belongs to the last round, then the output is cipher text else the resulting 128 bits is considered as 16 bytes, and another round is started with new byte substitution process. This is a column-wise operation between four bytes of state column and one word of round key. In the last round, there is no mix column step. **Figure 6** shows add round key stage in AES algorithm.

Decryption of cipher text, generated from AES encryption, contains all the stages in encryption but in reverse order. AES decryption starts with inverse initial round. The remaining nine rounds in decryption consist of processes like add round key, inverse shift rows, inverse byte substitution, and inverse mix columns.

Add round key: Add round key has its own inverse function since XOR functions its own inverse and the round keys should be selected in reverse order.

Inverse shift rows: Inverse shift rows functions exactly in the same way as shift row stage but in opposite direction. The first row is kept as it is, the second row is shifted by one-byte position to the right, the third row is shifted by two-byte position to the right, and the fourth row is shifted by three-byte position to the right. The resultant matrix consists of same 16 bytes but at different position. **Figure 7** shows inverse shift row stage in AES algorithm.

Inverse byte substitution: Inverse byte substitution is done using predefined substitution table known as inverse s-box. **Figure 8** shows inverse s-box in AES algorithm.

Inverse mix column: Transformation in inverse mix column is done using polynomials of degree less than 4 over Galois field (GF) 28 in which coefficients are the elements from the column of the state.

The rest of the chapter is organized as follows:

Section 2 presents the survey based on the various kinds of implementation of AES algorithm on reconfigurable platform. In Section 3, implementation of AES algorithm using the proposed approach is discussed. In Section 4, experimental results achieved using the proposed method along with the comparative analysis with existing methods are discussed.

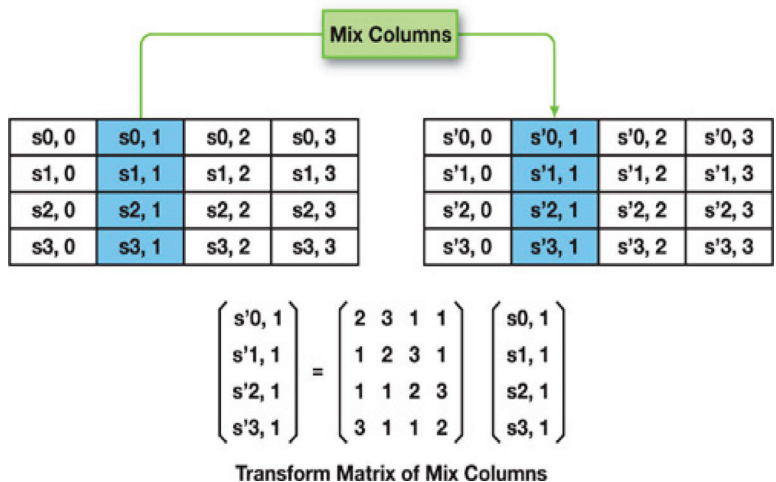

**Figure 5.**
*Mix column stage.*

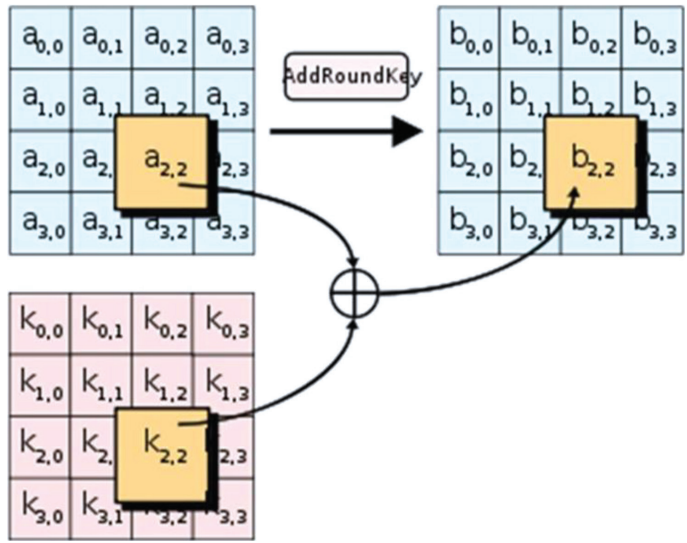

**Figure 6.**
*Add round key stage.*

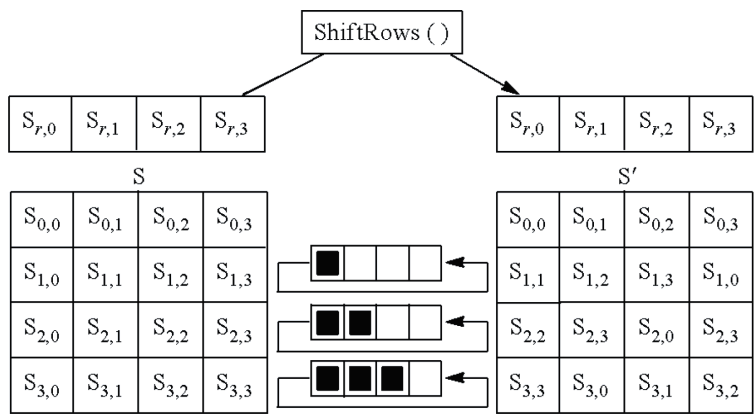

**Figure 7.**
*Inverse shift row.*

|   | 0 | 1 | 2 | 3 | 4 | 5 | 6 | 7 | 8 | 9 | A | B | C | D | E | F |
|---|---|---|---|---|---|---|---|---|---|---|---|---|---|---|---|---|
| 0 | 52 | 09 | 6A | D5 | 30 | 36 | A5 | 38 | BF | 40 | A3 | 9E | 81 | F3 | D7 | FB |
| 1 | 7C | E3 | 39 | 82 | 9B | 2F | FF | 87 | 34 | 8E | 43 | 44 | C4 | DE | E9 | CB |
| 2 | 54 | 7B | 94 | 32 | A6 | C2 | 23 | 3D | EE | 4C | 95 | 0B | 42 | FA | C3 | 4E |
| 3 | 08 | 2E | A1 | 66 | 28 | D9 | 24 | 82 | 76 | 5B | A2 | 49 | 6D | 8B | D1 | 25 |
| 4 | 72 | F8 | F6 | 64 | 86 | 68 | 98 | 16 | D4 | A4 | 5C | CC | 5D | 65 | B6 | 92 |
| 5 | 6C | 70 | 48 | 50 | FD | ED | 89 | DA | 5E | 15 | 46 | 57 | A7 | 8D | 9D | 84 |
| 6 | 90 | D8 | AB | 00 | 8C | BC | D3 | 0A | F7 | E4 | 58 | 05 | B8 | B3 | 45 | 06 |
| 7 | D0 | 2C | 1E | 8F | CA | 3F | 0F | 02 | C1 | AF | BD | 03 | 01 | 13 | 8A | 6B |
| 8 | 3A | 91 | 11 | 41 | 4F | 67 | DC | EA | 97 | F2 | CF | CE | F0 | 84 | E6 | 73 |
| 9 | 96 | AC | 74 | 22 | E7 | AD | 35 | 85 | E2 | F9 | 37 | E8 | 1C | 75 | DF | 6E |
| A | 47 | F1 | 1A | 71 | 1D | 29 | C5 | 89 | 6F | 87 | 62 | 0E | AA | 18 | BE | 1B |
| B | FC | 56 | 3E | 4B | C6 | D2 | 79 | 20 | 9A | DB | C0 | FE | 78 | CD | 5A | F4 |
| C | 1F | DD | A8 | 33 | 88 | 07 | C7 | 31 | B1 | 12 | 10 | 59 | 27 | 80 | EC | 5F |
| D | 60 | 51 | 7F | A9 | 19 | B5 | 4A | 0D | 2D | E5 | 7A | 9F | 93 | C9 | 9C | EF |
| E | A0 | E0 | 38 | 4D | AE | 2A | F5 | B0 | C8 | EB | BB | 3C | 83 | 53 | 99 | 61 |
| F | 17 | 2B | 04 | 7E | BA | 77 | D6 | 26 | E1 | 69 | 14 | 63 | 55 | 21 | 0C | 7D |

**Figure 8.**
*Inverse S-box of AES algorithm.*

## 2. Literature survey

In this section, focus is given on the work done by various researchers on FPGA-based implementation of AES algorithm. There are various researchers which have either concentrated on area optimization or speed optimization. Mulani and Mane [1] discussed integrating of DWT and AES algorithm for implementation of watermarking on FPGA. The design was implemented on xc6vcx75t-2ff484, and it utilizes 2117 slices at maximum operating frequency of 228.064 MHz. Ratheesh and Narayanan [2] proposed implementation of AES algorithm with low-power MUX LUT-based s-box on FPGA. This design achieved total power distribution of 0.55 W. Agarwal et al. [3] suggested implementation of AES algorithm using Verilog on Spartan-3E FPGA. This design utilizes 1464 slices. Farooq and Faisal Aslam [4] discussed implementation of AES algorithm on FPGA device using five different techniques which are suitable for area critical applications and speed critical applications. This design was implemented on Spartan-6 FPGA device, and it utilizes 161 slices at maximum operating frequency which is 886.64 MHz. The throughput of this system is 113.5 Gbps. Sai Srinivas and Akramuddin [5] proposed less complex hardware implementation of AES Rijndael algorithm on Xilinx Virtex-7 XC7VX90T FPGA. In the proposed design, synthesis tool was set to optimize speed, area, and power. Mathur and Bansode [6] proposed a cryptosystem, which is a combination of AES algorithm and ECC. This is a hybrid encryption scheme and the key size is 192 bits and there are 12 numbers of iterations in this system. Kalaiselvi and Mangalam [7] proposed a low-power and high-throughput FPGA implementation of AES algorithm using key expansion technique. This design accepts key size of 256 bits for both encryption and decryption. This design utilizes 5493 slices, and its maximum operating frequency is 277.4 MHz. The throughput of this system is 0.06 Gbps. Deshpande et al. [8] suggested BRAM-based and FPGA-based implementation of AES algorithm. Due to the use of BRAMs for implementing s-box, this design utilizes less number of slices. The design was implemented on XC3S1400AN and it utilizes 3376 slices. Ibrahim [9] presented FPGA implementation of AES encryption core that is suitable for limited resource-limited applications. This design was implemented on Spartan-3, and it utilizes 150 slices at maximum operating frequency of 90 MHz. Khose and Raut [10] proposed implementation of AES algorithm on FPGA in order to achieve high speed of data processing and also to reduce time for generating key. This design utilizes 201 slices and 2 BRAMs at maximum operating frequency of 70 MHz. Mulani and Mane [11] proposed FPGA implementation of DES algorithm. The design was implemented on XC2S200, and it utilizes 2118 slices and 97 IOBs. Yewale Minal and Sayyad [12] proposed implementation of AES encryption using VHSIC hardware description language VHDL) and decryption using Visual Basic. With this approach, 1403 slices are utilized at maximum operating frequency of 160.875 MHz, and it has a throughput of 2.059 Gbps. Deshpande et al. [13] discussed FPGA-based optimized architecture that utilizes less area. This design was intended for plaintext of 128 bits and key of 128 bits. Tonde and Dhande [14] discussed FPGA-based implementation of AES algorithm using iterative looping approach for 128 bits of block and key size. Varhade and Kasat [15] proposed a FPGA-based AES algorithm, which utilizes 1746 logic elements and 32,768 memory bits. This design was synthesized on Cyclone-II using Altera. Wadi and Zainal [16] proposed some modifications like decreasing number of rounds and replacing S-box with new s-box to reduce hardware requirements in order to enhance the performance of AES algorithm in terms of time ciphering and pattern appearance. Wang et al. [17] suggested high-speed implementation of AES algorithm on FPGA to transmit the data securely using pipelining and parallel processing methods. Shylashree et al. [18] focused on various novel FPGA architectures of AES algorithm. Borkar et al. [19] proposed iterative design approach for FPGA implementation of AES algorithm using VHDL. This design utilizes 1853 slices, and its operating frequency is 140.390 MHz. Deshpande et al. [20] presented very low complexity FPGA-based architecture for integrated AES encryptor and decryptor. This design is synthesized on Spartan-3 XC3S400 FPGA. Kaur and Vig [21] suggested an efficient implementation of AES algorithm on FPGA in which multiple rounds are processed simultaneously. Due to this implementation, speed is increased but it increases area. This design utilizes 6279 slices and 5 BRAMs, and its operating frequency is 119.954 MHz. Samanta [22] proposed fast and efficient reconfigurable platform-based implementation of AES algorithm using pipelining. This design utilizes 1051 slices and 11 BRAMs, and its operating frequency is 76.699 MHz. Good and Benaissa [23] discussed hardware implementation of fastest and slowest AES algorithm which utilizes 16,693 slices at maximum operating frequency of 184.8 MHz.

From the literature survey, it is clear that many researchers have either worked on optimizing the area or speed. Few researchers have concentrated on optimizing the speed as well as area. Implementation of AES algorithm, which is optimized in speed as well as area, is discussed in this chapter.

## 3. Implementation of AES algorithm

The proposed design is implemented with the aim to achieve both area and speed optimization. In the proposed design, keys for each round are initially generated by using MATLAB code, and then those keys are used in the design. Due to this approach, the design occupies less number of slices, and also the speed is faster than the normal approach. The design is implemented using Xilinx system generator. **Figure 9** shows Xilinx system generator-based model for AES algorithm.

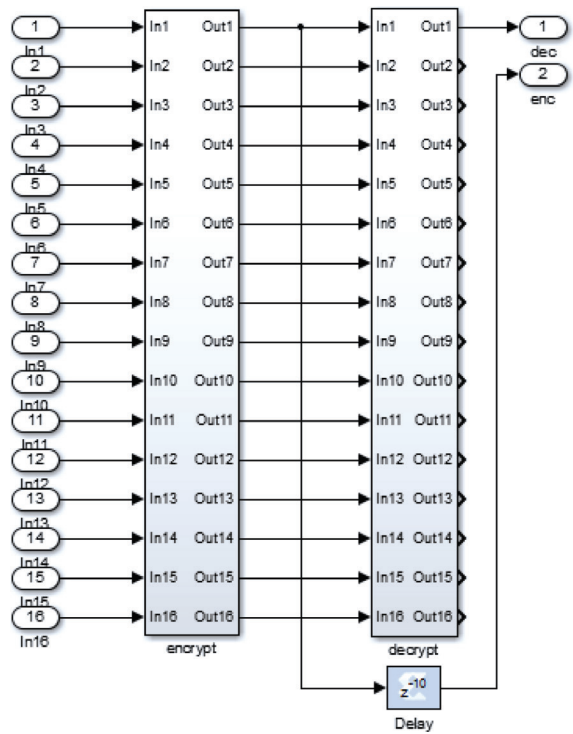

**Figure 9.**
*System generator model for AES algorithm.*

### 3.1 AES encryption

A plaintext of 128-bit is processed through 10 rounds. Each round contains processes like byte substitution, shift rows, mix columns, and add round key. As keys are generated using MATLAB code, only remaining system generator-based models like byte substitution, shift rows, and mix columns are discussed in this section.

Round function is one of the important processes in AES algorithm. **Figure 10** shows system generator-based model for implementing round0 function.

Round function consists of s-box, shift row, and mix column as shown in **Figure 11**.

**Figure 12** shows implementation of s-box.

**Figure 13** shows implementation of shift row.

**Figure 14** shows implementation of mix column.

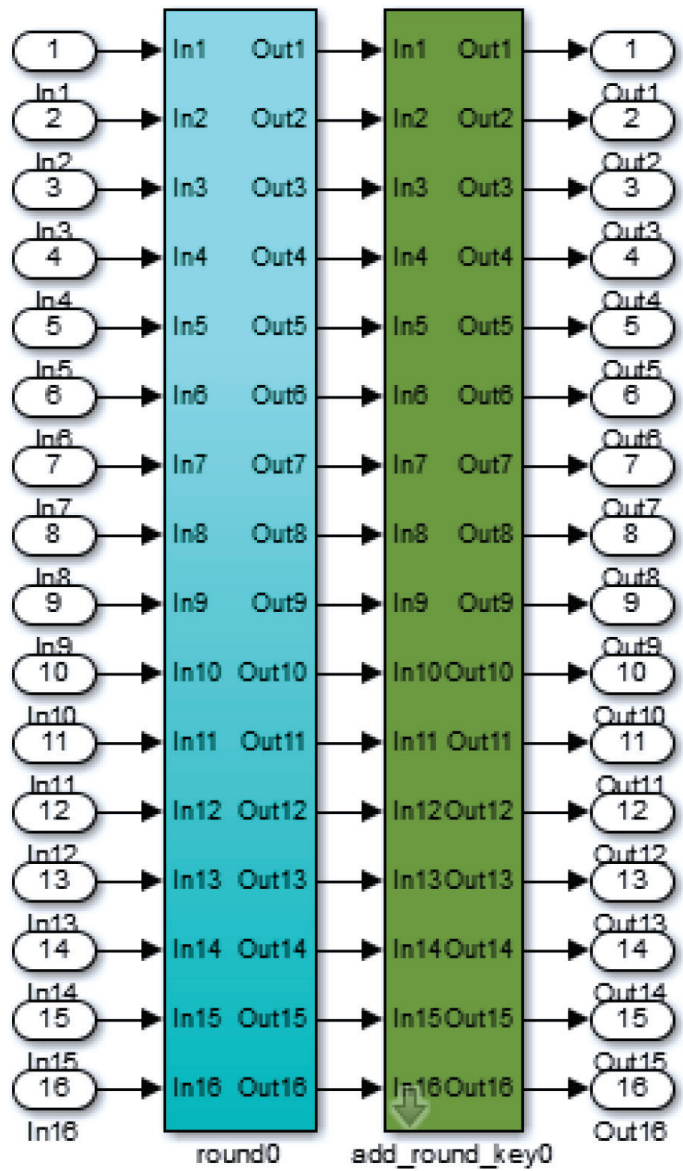

**Figure 10.**
*System generator-based model of round function.*

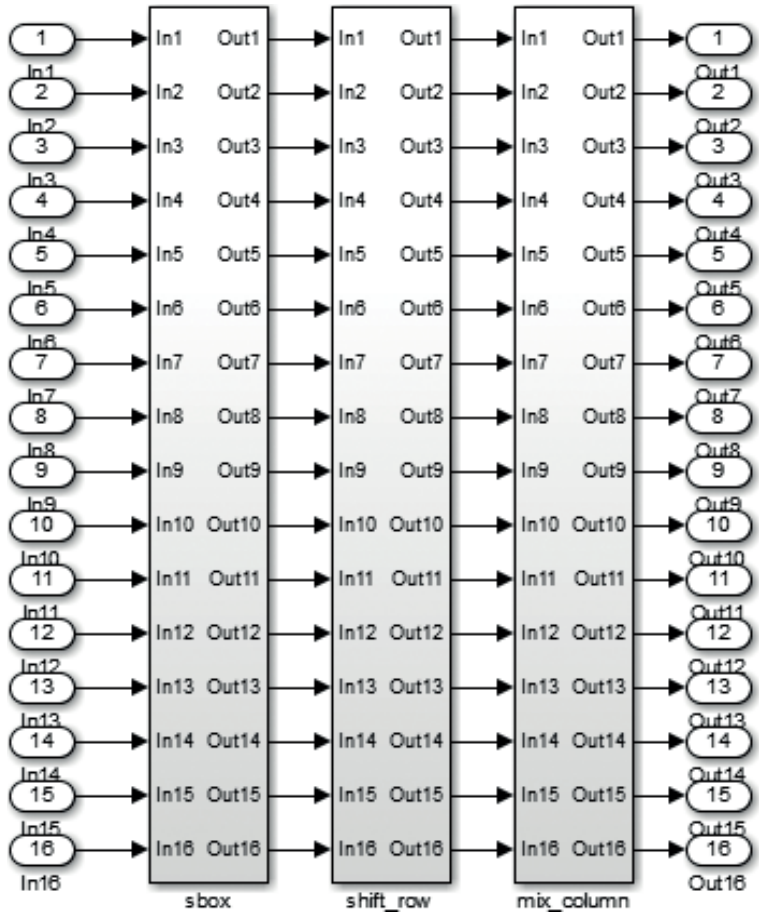

**Figure 11.**
*Round0.*

Mix column consists of group_1, group_2, group_3, and group_4. **Figure 15** shows implementation of group. Further each group consists of four multiplication blocks such as mul_blk, mul_blk1, mul_blk2, and mul_blk3. **Figure 16** shows implementation of multiplication block.

### 3.2 AES decryption

A cipher text of 128-bits is processed through 10 inverse rounds. Each round contains processes like inverse byte substitution, inverse shift rows, inverse mix columns, and add round key.

**Figure 17** shows implementation of inverse round function.

Inverse round function consists of inverse s-box, inverse shift row, and inverse mix column as shown in **Figure 18**.

**Figure 19** shows implementation of inverse mix column.

Inverse mix column consists of four groups, i.e., group_1, group_2, group_3, and group_4. **Figure 20** shows implementation of group. Each group consists of multiplication blocks like mul_blk, mul_blk1, mul_blk2, and mul_blk3. **Figure 21** shows implementation of multiplication block.

Each multiplication block consists of three multipliers mul_2, mul_4, and mul_8 and EX-OR operations. **Figure 22** shows implementation of multipliers.

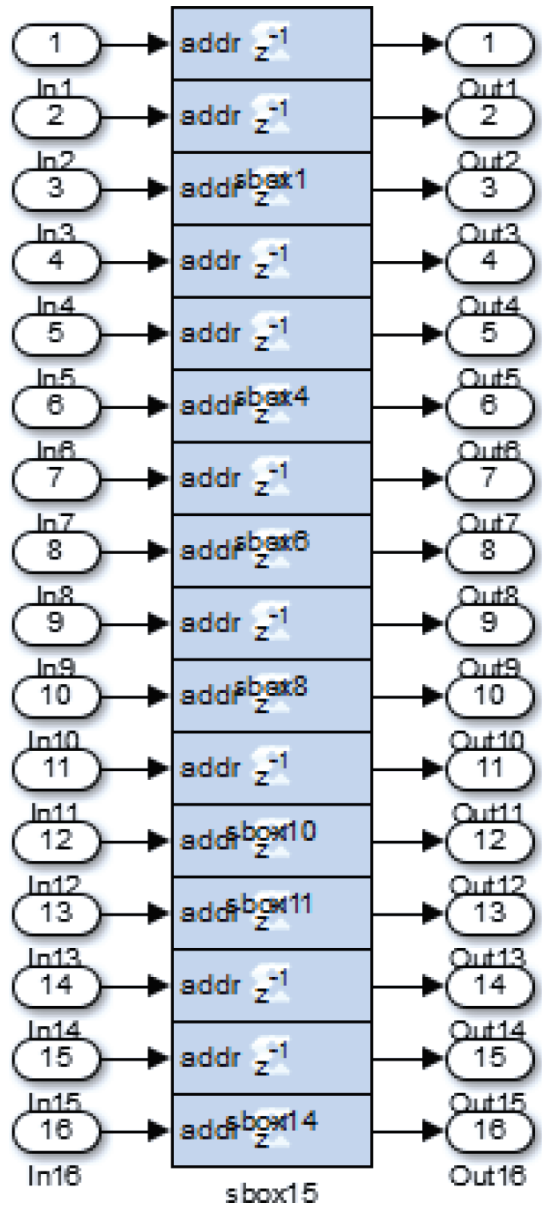

**Figure 12.**
*Implementation of s-box.*

**Figure 23** shows implementation of inverse shift row.
**Figure 24** shows implementation of inverse s-box.

### 3.3 Tools utilized

*3.3.1 Software utilized*

For implementing the proposed design, MATLAB 2013a and Xilinx ISE Design Suite are used. MATLAB is used for generating the keys and also to get the results in terms of images, whereas Xilinx ISE Design Suite is used to get the synthesis result, RTL schematic, and throughput of this implementation.

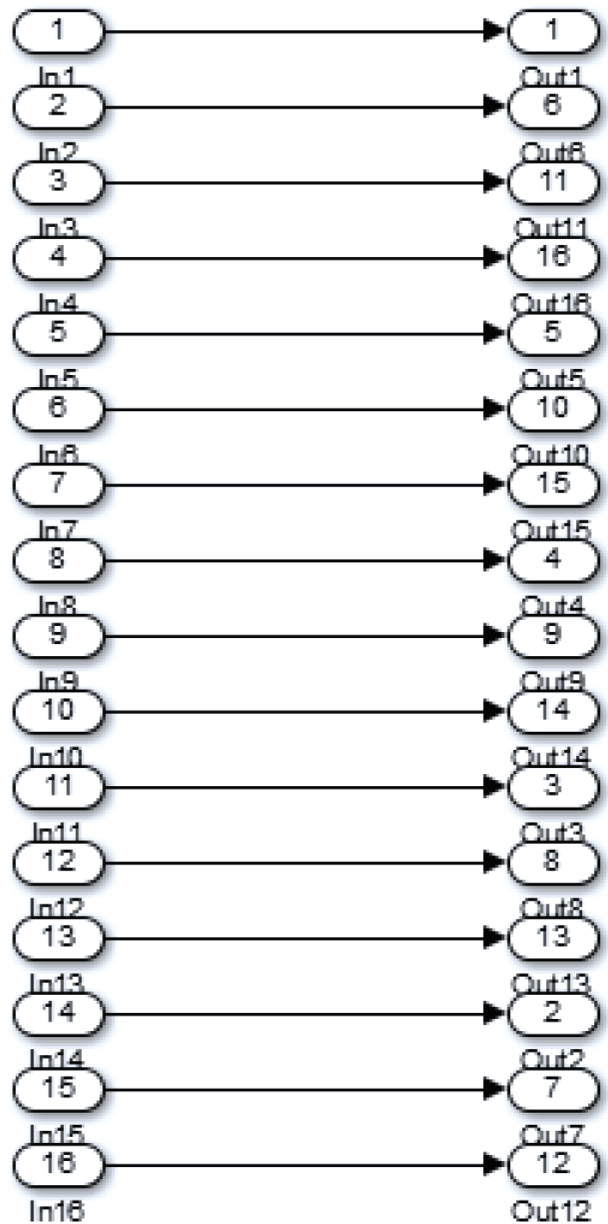

**Figure 13.**
*Implementation of shift row.*

### 3.3.2 Hardware utilized

Nexys-4 DDR development board is used for implementation. This board has the following features:

a. Xilinx Artix-7 FPGA XC7A100T-1CSG324C

b. 15,850 logic slices, each with four 6-input LUTs and 8 flip-flops

c. 4860 Kbits of fast block RAM

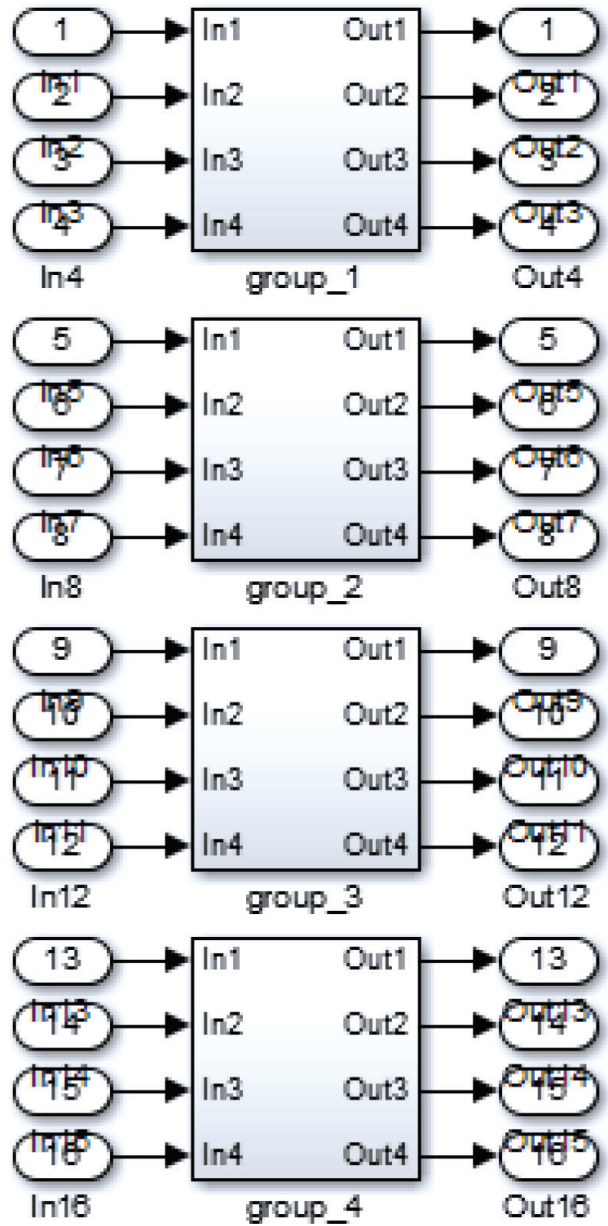

**Figure 14.**
*Implementation of mix column.*

d. Six clock management tiles, each with phase-locked loop (PLL)

e. 240 DSP slices

f. Internal clock speeds exceeding 450 MHz

g. On-chip analog-to-digital converter (XADC)

h. 128 MiB DDR2

i. Serial Flash

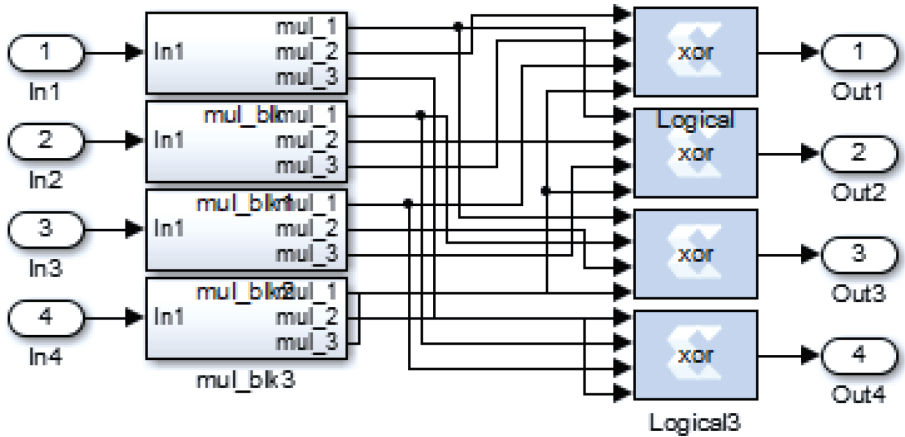

**Figure 15.**
*Implementation of group.*

j. Digilent USB-JTAG port for FPGA programming and communication

k. MicroSD card connector

l. Ships with rugged plastic case and USB cable

m. USB-UART Bridge

n. 10/100 Ethernet PHY

o. PWM audio output

p. 3-axis accelerometer

q. 16 user switches

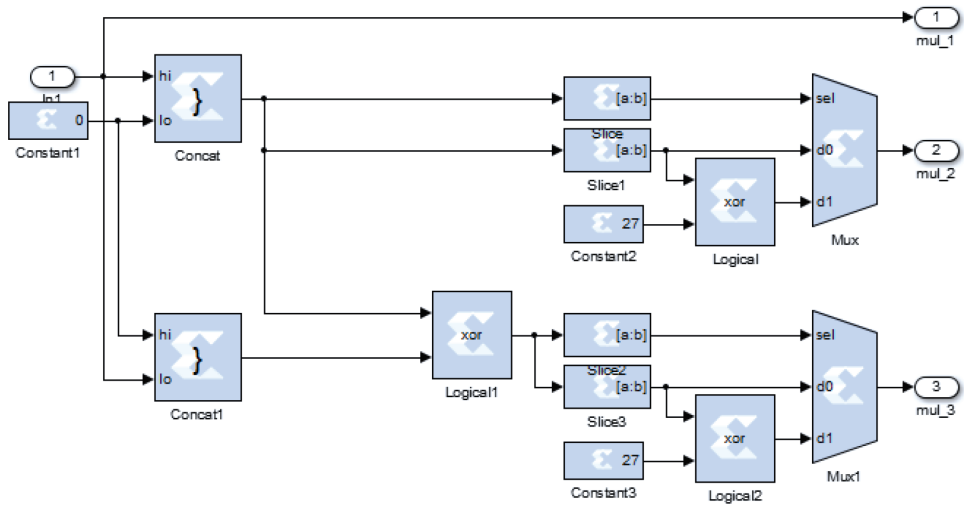

**Figure 16.**
*Implementation of multiplication block.*

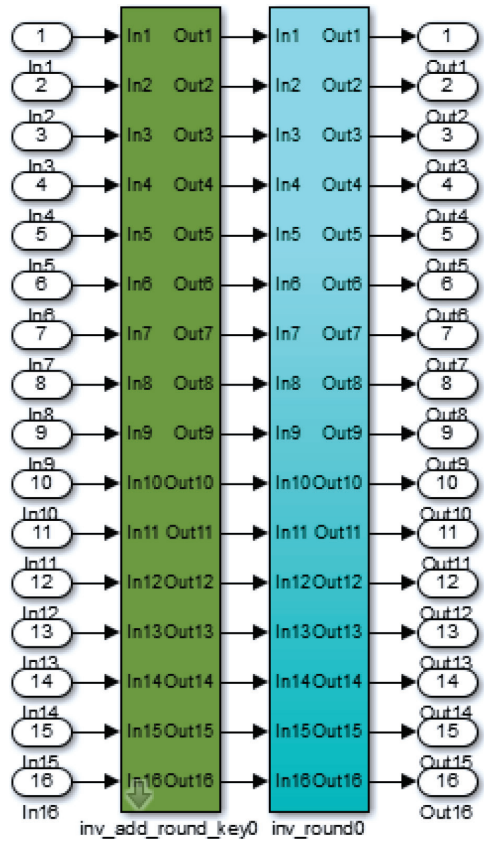

**Figure 17.**
*System generator-based model of inverse round function.*

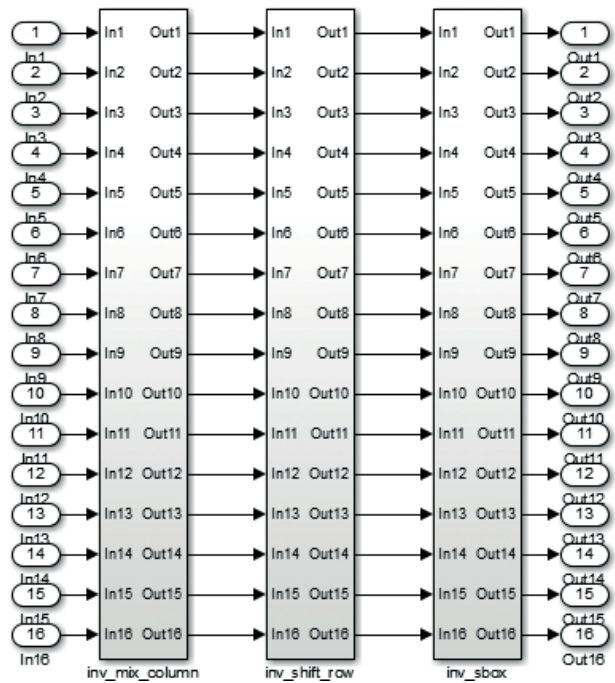

**Figure 18.**
*Inverse round0.*

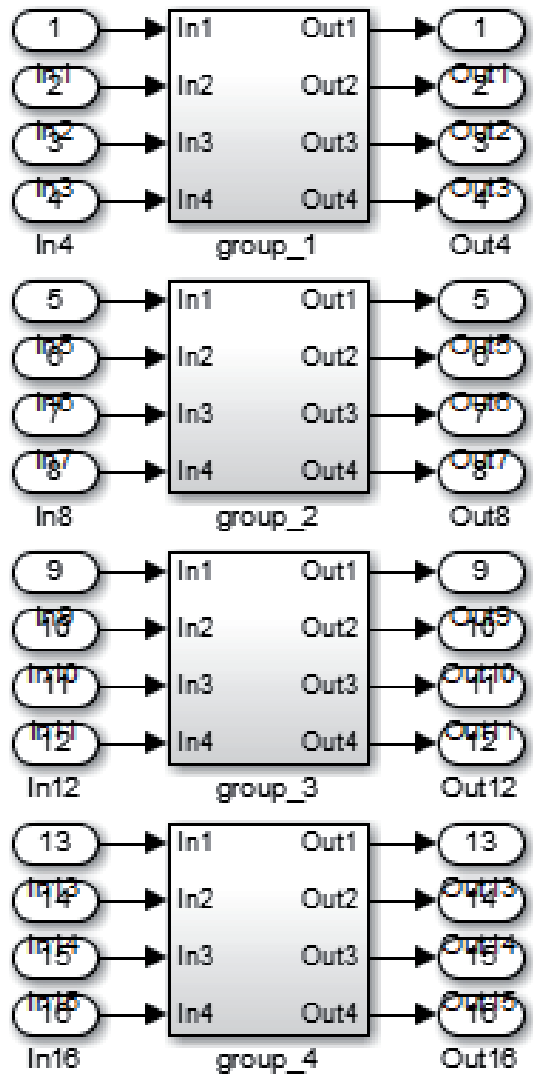

**Figure 19.**
*Inverse mix column.*

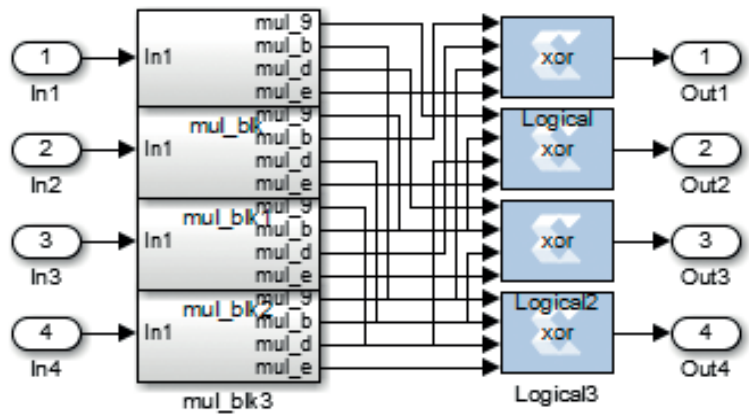

**Figure 20.**
*Implementation of group.*

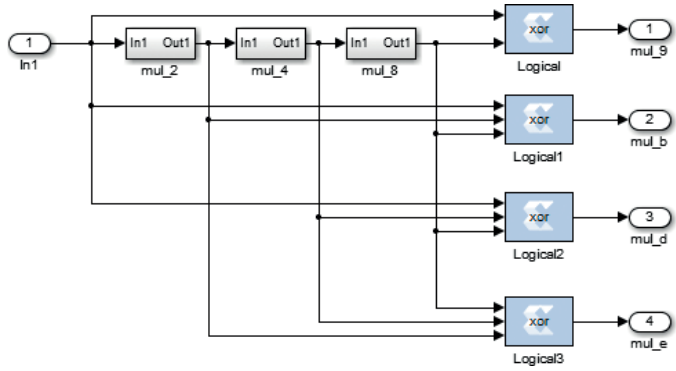

**Figure 21.**
*Implementation of multiplication block.*

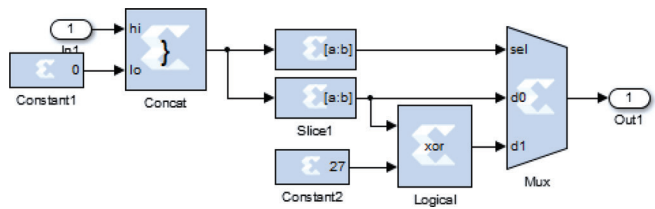

**Figure 22.**
*Implementation of multipliers.*

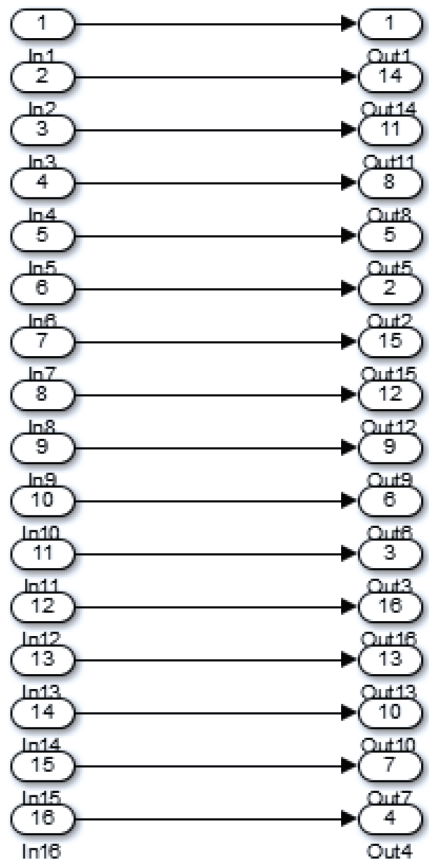

**Figure 23.**
*Implementation of inverse shift row.*

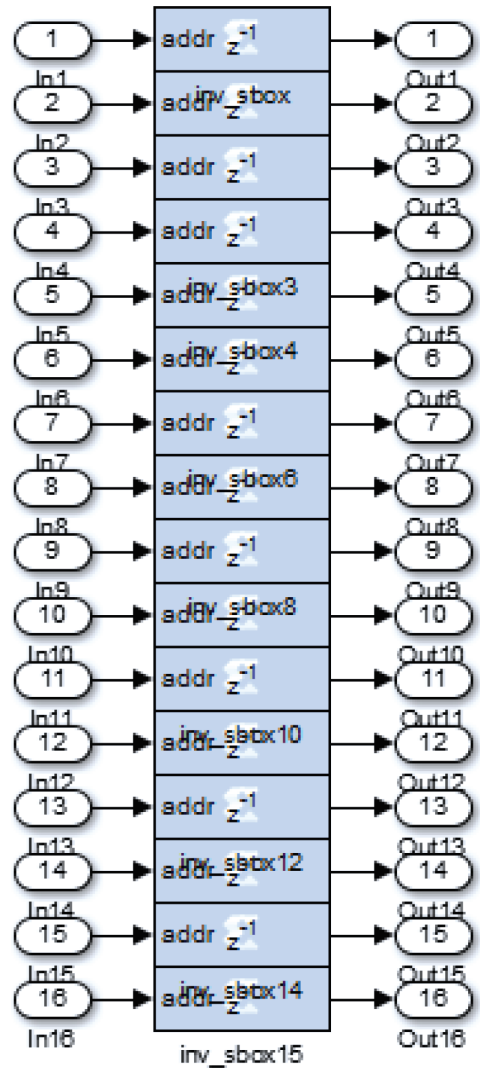

**Figure 24.**
*Implementation of inverse s-box.*

r. 16 user LEDs

s. Two tri-color LEDs

t. PDM microphone

u. Temperature sensor

v. Two 4-digit 7-segment displays

w. USB HID Host for mice, keyboards, and memory sticks

x. PMOD for XADC signals

y. 12-bit VGA output

z. Four PMOD ports

## 4. Experimental results

### 4.1 RTL schematic

**Figure 25** shows detailed RTL schematic of the proposed implementation of AES algorithm.

### 4.2 Synthesis result

The design is synthesized using Xilinx XST synthesizer. In the proposed design, an optimized and synthesizable very high speed integrated circuit (VHSIC) hardware description language (VHDL) code for the implementation of image as well as 128-bit data encryption is developed so as to utilize less area and increase the speed. **Table 1** shows design utilization summary of the proposed design.

From the synthesis results of the proposed design, it is clear that this system utilizes only 121 slice registers, and its maximum operating frequency is 1102.536 MHz. The throughput of the system is calculated using the following formula:

$$\text{(Throughput) of the system} = \frac{128 \text{ bits} \times \text{Clock frequency}}{\text{Cycles per Encrypted block}} \tag{1}$$

By substituting the values in Eq. (1), throughput of the systems is 14.1125 Gbps.

### 4.3 Simulation result

**Figure 26** shows simulation result when an image is applied as an input.

### 4.4 Performance analysis

Performance analysis is a must to compare the performance of the proposed implementation with existing methods. The performance is compared on the basis of area and operating frequency. Till date various researchers have worked on FPGA-based implementations of AES algorithm; some of them have optimized speed and some have optimized area. In the proposed system, both area and speed are optimized. **Table 2** shows performance comparison of the proposed system with previous work.

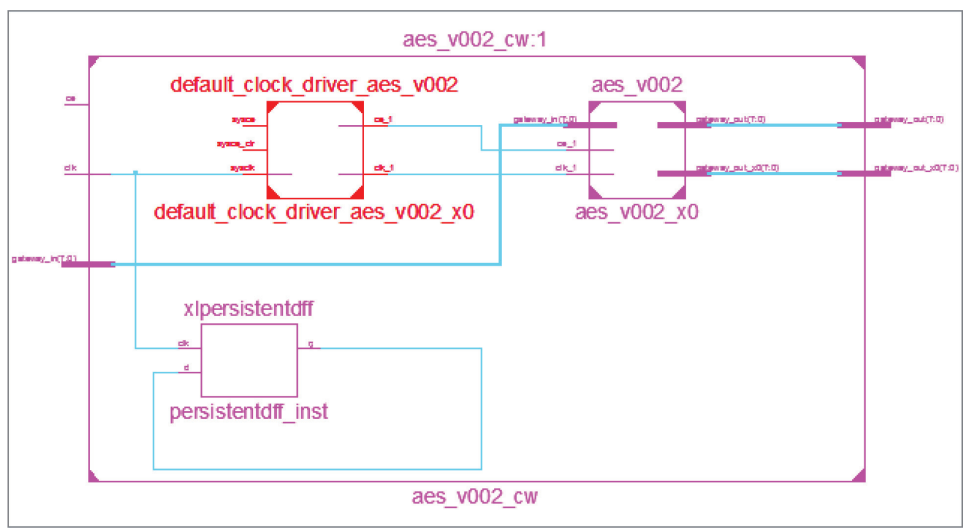

**Figure 25.**
*Detailed RTL schematic of AES algorithm.*

| Design utilization summary | | | |
|---|---|---|---|
| **Logic utilization** | **Used** | **Available** | **% utilization** |
| Number of slice registers | 121 | 126,800 | 0.00095 |
| Number of slice LUTs | 4782 | 63,400 | 7 |
| Number of bonded IOBs | 25 | 210 | 11 |

**Table 1.**
*Design utilization summary.*

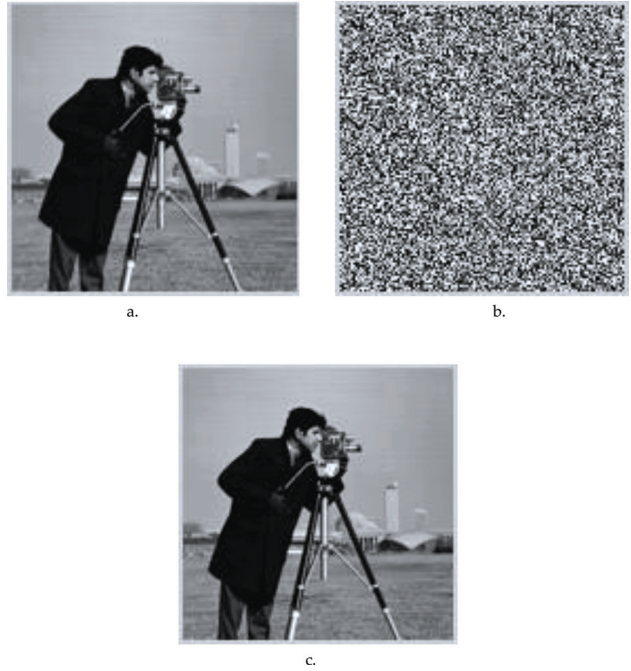

**Figure 26.**
*Simulation result (a) Original image, (b) Encrypted image, and (c) Decrypted image.*

| Sr. No. | Authors | Slices | Operating freq. (MHz) |
|---|---|---|---|
| 1 | Proposed work | 121 | 1102.536 |
| 2 | [3] | 1464 | — |
| 3 | [4] | 161 | 886.64 |
| 4 | [7] | 5493 | 277.4 |
| 5 | [8] | 3376 | — |
| 6 | [9] | 150 | 90 |
| 7 | [10] | 201 | 70 |
| 8 | [12] | 1403 | 160.875 |
| 9 | [15] | 1746 | — |
| 10 | [19] | 1853 | 140.390 |
| 11 | [21] | 6279 | 119.954 |

**Table 2.**
*Performance comparison of the proposed system with previous work.*

## 5. Conclusion

In this chapter, fast, area-efficient, and secure implementation of AES algorithm on FPGA is suggested. As per the literature survey, it is clear that Farooq and Faisal Aslam [4] achieved better performance in terms of speed, whereas Ibrahim [9] achieved better performance in terms of area. In this design, due to better Xilinx system generator-based design, the system is optimized, and it utilizes only 121 slice registers at maximum operating frequency of 1102.536 MHz. Also, throughput of the proposed system is 14.1125 Gbps.

## Conflict of interest

There is no conflict of interest.

## Acronyms and abbreviations

| AES | advanced encryption standard |
|-----|------------------------------|
| DDR | double data rate |
| DES | data encryption standard |
| FPGA | field-programmable gate array |
| Gbps | gigabits per second |
| MHz | megahertz |
| VHDL | VHSIC Hardware Description Language |
| VHSIC | very high speed integrated circuit |

**Author details**

Altaf O. Mulani* and Pradeep B. Mane
AISSMS Institute of Information Technology, Pune, Maharashtra, India

*Address all correspondence to: aksaltaaf@gmail.com

IntechOpen

# Hybrid Approaches to Block Cipher

*Roshan Chitrakar, Roshan Bhusal and Prajwol Maharjan*

## Abstract

This chapter introduces two new approaches to block cipher—one is DNA hybridization encryption scheme (DHES) and the other is hybrid graphical encryption algorithm (HGEA). DNA cryptography deals with the techniques of hiding messages in the form of a DNA sequence. The key size of data encryption standard (DES) can be increased by using DHES. In DHES, DNA cryptography algorithm is used for encryption and decryption, and one-time pad (OTP) scheme is used for key generation. The output of DES algorithm is passed as an input to DNA hybridization scheme to provide an added security. The second approach, HGEA, is based on graphical pattern recognition. By performing multiple transformations, shifting and logical operations, a block cipher is obtained. This algorithm is influenced by hybrid cubes encryption algorithm (HiSea). Features like graphical interpretation and computation of selected quadrant value are the unique features of HGEA. Moreover, multiple key generation scheme combined with graphical interpretation method provides an increased level of security.

**Keywords:** DNA hybridization encryption scheme (DHES), hybrid graphical encryption algorithm (HGEA), DNA cryptography, data encryption standard (DES), one-time pad (OTP), hybrid cube encryption algorithm (HiSea), block cipher

## 1. Introduction

There exist a number of cryptographic techniques for secure data communication [1], but many are vulnerable to attacks. With the failure of cryptographic algorithms like data encryption standard (DES), new approaches to cipher security are needed [2, 3]. A cryptographic scheme can be made more secure by combining it with relatively secure techniques. Theoretically, this hybridization method can be applied to any cryptographic scheme but block ciphers provide more rounds for working in terms of permutation and combination.

DNA-based method [4, 5] is one such approach that along with one-time pad (OTP) scheme can be applied to DES. OTP is the only unbreakable encryption that uses polyalphabetic randomness for the key [6]. So, OTP can be combined with DNA cryptography by taking longer message and key size ($\geq 64$ bit) so as to make brute force attack difficult and impractical [7].

As the first part of this chapter, DNA hybridization encryption scheme (DHES) is described, in which an improved algorithm named DDHO (that stands for DES and DNA-based hybridization with OTP) is proposed.

Another technique that can be combined with DES is "hybrid graphical encryption algorithms (HGEA), which is based on graphical interpretation by pattern recognition and transformation like hybrid cubes encryption algorithm (HiSea) [8, 9]. Most of the graphical encryption algorithms use mono-alphabetic or poly-alphabetic substitution and their range of input values is limited. But, HGEA uses a range of characters consisting of 256 possible values. It also produces output of 256 characters for single-input plaintext. Moreover, HGEA can be used by software as well as realized by implementing hardware devices.

The rest of this chapter is organized as follows. Section 2 describes DNA hybridization encryption scheme in detail. This section also presents performance analysis of algorithms and methods used by DHES. Similarly, Section 3 presents hybrid graphical encryption with illustrations. This section also presents performance analysis of encryption and decryption algorithms used by HGEA by comparing it with that of DES.

## 2. DNA hybridization encryption scheme (DHES)

### 2.1 DNA cryptography

DNA cryptography is an emerging field of cryptography that deals with hiding data in terms of DNA sequences [10, 11]. It can be implemented by using modern biological techniques as tools and DNA as information carrier to fully exert the inherent advantage of high storage density and high parallelism to achieve encryption [12]. The DNA cryptography uses the concept of molecular approach in traditional cryptographic technique to make the system more secure.

Some terminologies used in biochemical operations of DNA cryptography are annealing, melting, ligation, amplification, cutting, gel electrophoresis, oligonucleotides, etc. [13, 14].

### 2.2 DNA hybridization

DNA hybridization is a process in which two single-stranded DNAs (ssDNAs) are combined together to produce a single DNA sequence [15]. The two ssDNAs are complementary to each other and are of same length. That means: if one strand of DNA is 3′ to 5′, then the other strand must be 5″. If not, then the hybridization of the pairs fails and fragmentation occurs. To remove such fragmentation, fragment assembly has to be done [16].

#### 2.2.1 OTP scheme

OTP is the only potentially unbreakable encryption method. Plaintext encrypted using an OTP cannot be retrieved without the encryption key. The key generated by an OTP must be random and generated by a non-deterministic, non-repeatable process. The key also must be never re-used. In OTP scheme, the length of the key must be greater than or equal to the length of the plaintext.

#### 2.2.2 Key generation

OTP is used for key generation. The generated key is unique and only the sender and receiver know the key and that generated key is destroyed once it is used. The key generated by computers is not truly random; so, a pseudorandom number generator function is used for generating the key and the generated key is in the DNA. The size of ssDNA key is greater than the original size of the message, which results in a longer size of the encrypted message and that makes it difficult to break. The length of the key is the result of multiplication of the number of bits required to represent each character and total number of bits in the input message.

### 2.2.3 Encryption

The encryption process consists of the following steps (**Figure 1**):

1. Choose the plaintext to be sent.

2. Replace each letter in the text with its opposite alphabetical character excluding "A" and "a" (replacement algorithm), numerical values and symbols. Then, convert the plaintext into ASCII code and then into binary code.

3. The ssDNA OTP key is generated. (The length of the key depends on (i) the length of binary plaintext and (ii) the number of bits required to represent each nucleotide.)

4. Scan the binary sequence from left to right to find the occurrences of 0 and 1 s.

   • If the first digit of binary bit is 1, then this bit is compared with last n bases of OTP key and complementary data of DNA form are produced as the encrypted message where n is the number of bits required to represent the nucleotides.

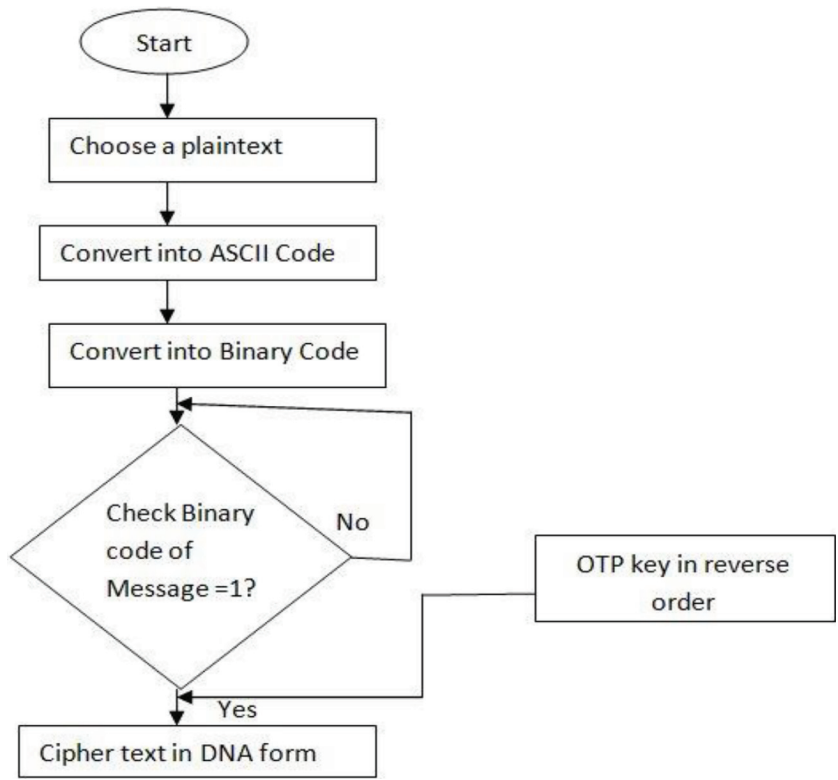

**Figure 1.**
*Flow chart of DNA hybridization encryption.*

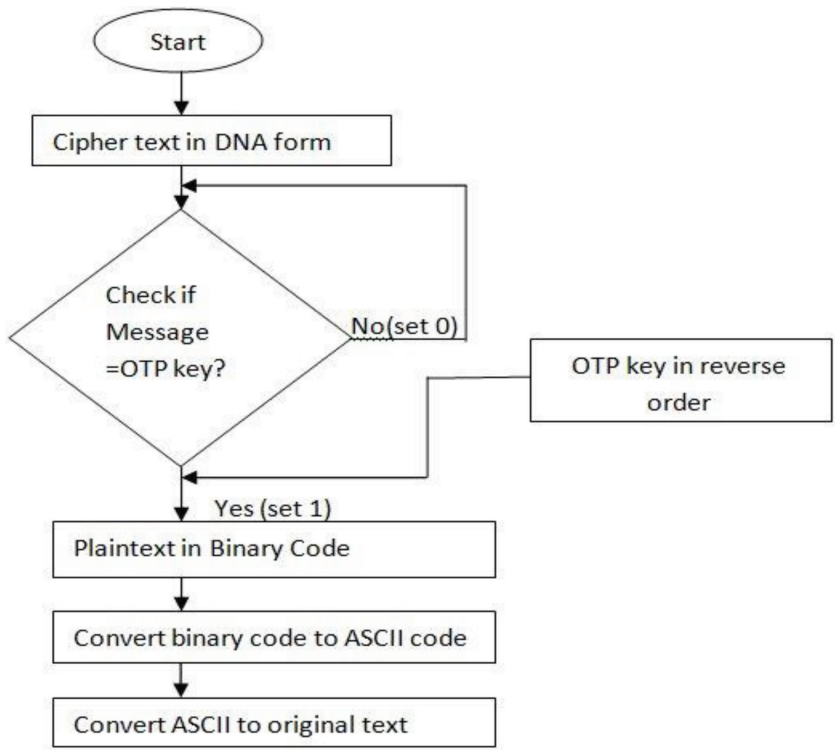

**Figure 2.**
*Flow chart of DNA hybridization decryption.*

- If the first digit of binary bit is 0, then no operation is carried out and the next n bases in the OTP key from reverse order are ignored where n is the number of bits required to represent nucleotides.

5. Repeat step 3 for all the occurrences of 1 and 0 s and put them all together to obtain the resulting ciphertext.

*2.2.4 Decryption*

The decryption process consists of the following steps (**Figure 2**):

1. Take n leftmost bits from the ciphertext and compare with the last n bits of the OTP key.

    - If they are found to be complementary, then binary bit "1" is formed; else, a binary "0" is formed.

2. Repeat step 1 for the subsequent n-bit sequence in the ciphertext till the end and put them all together to obtain the binary digit.

3. Apply reverse replacement algorithm.

4. Arrange the binary digits (in n bits form) and convert the value into ASCII code.

5. Convert the ASCII code to plaintext.

## 2.2.5 DDHO algorithm

Here, the proposed algorithm DDHO (combination of DES algorithm and DNA-based hybridization and OTP scheme) is explained in detail. First, the DES algorithm is performed on the given plaintext and key and the resultant ciphertext is taken as an input to the DNA hybridization and OTP scheme. Further, the algorithm proceeds as per the above illustrated DNA hybridization and OTP scheme.

The encryption algorithm for DDHO is as follows:

1. A block of 64 bits is permutated by an initial permutation called IP.

2. Resulting 64 bits are divided into two equal halves, each containing 32 bits, left and right halves.

3. The right half goes through a function F (Feistel function)

4. The left half is XOR-ed with output from the F function obtained in the above step.

5. The left and right halves are swapped (except the last round).

6. In the last round, apply an inverse permutation (IP-1) on both halves that is the last step which produces a ciphertext in binary form.

7. The ssDNA OTP key is generated and the length of this key depends upon the length of binary plaintext and the number of bits required to represent each nucleotide.

8. Start scanning the binary sequence, obtained in step 6, from left to right to find the occurrences of 0 and 1 s.

   • If first digit of binary bit is 1, then this bit is compared with last n bases of OTP key and complementary data of DNA form are produced as the encrypted message where n is the number of bits required to represent the nucleotides.

   • If the first digit of the binary bit is 0, then no operation is carried out and the next n bases in the OTP key from reverse order are ignored where n is the number of bits required to represent nucleotides.

9. Repeat step 8 for all the occurrences of 1 and 0 s and put them all together to obtain the resulting ciphertext.

The decryption algorithm of DDHO is the reverse process of DDHO encryption algorithm.

## 2.2.6 Analysis of DDHO

The plaintexts chosen for encryption and decryption using the DDHO algorithm are highly diverse. They include short and long texts, purely alphabetical text and text containing alphabets and many other characters. The plaintexts of diverse types are selected, so that they are very representative. With regard to difference of the lengths of the text, four plaintexts with increasing size are selected (**Table 1**).

The first plaintext contains only alphabetical and digital characters, the second plaintext contains only non-alphabetical and non-digital characters, and the third and the fourth plaintexts contain a combination of characters.

By applying the above test dataset to the DDHO algorithm, the original plaintext size, the resulting ciphertext size and the key size are examined, together with the encryption and decryption time. The encryption and decryption processes are performed five times for each plaintext and the average system time is obtained and listed to make the evaluation of time fair.

The results obtained are shown in the **Table 2**.

The number of bits needed to store the plaintext in ASCII format is eight times that of the length of the plaintext. For the output of DES in binary form, the number of nucleotides used to represent each bit is 10 so that the total size of key is 10 times the length of binary bits (i.e. output of DES). For instance, consider a plaintext of length 64 bits. The output of DES algorithm is also 64 bits, so the length of OTP key is equal to $64 \times 10$ bits = 640 bits. The length of ciphertext is 260 bits means that there are only 26 occurrences of number 1 in the output of DES algorithm (i.e. plaintext of DDHO) and the remaining $64 - 26 = 38$ bits are 0 s. Similarly, consider a plaintext of length 400 bits, the output of DES algorithm is equal to $400/64 = 6.25$ blocks of 64 bits. However, this length of plaintext is not an exact multiple of 64 bits. Therefore, it adds 48 bits 0 s at the end of plaintext and makes 7 blocks of 64 bit. The output of 7 blocks of 64 bits of DES is equal to $7 \times 64$ bits = 448 bits and hence the length of OTP key in DDHO algorithm is equal to $448 \times 10$ bits = 4480 bits. Similarly, there are only 102 occurrences of 1 s in the output of DES algorithm (i.e. plaintext of DDHO) and remaining bits have 0 s. The length of ciphertext depends upon how many 1 s (binary bit) are present in the plaintext. The more the number of 1 s, the more the length of the ciphertext.

As shown in **Table 2**, the ciphertext lengths are proportional to the corresponding plaintext lengths. The size of key increases hugely as the size of plaintext increases. The length of ciphertext is small as compared to the size of the key; this is because the length of ciphertext depends upon the number of 1 s present in the input plaintext.

| Dataset | Description |
| --- | --- |
| Test 1 | Only alphabetical and digital characters |
| Test 2 | Only non-alphabetical and non-digital characters |
| Test 3 | Combination of characters |
| Test 4 | Combination of characters |

**Table 1.**
*Plaintext of different contents for DDHO algorithm.*

| Dataset | Length of plaintext (bits) | Length of ciphertext (bits) | Size of key (bits) | Encryption time (milliseconds) | Decryption time (milliseconds) |
| --- | --- | --- | --- | --- | --- |
| Test 1 | 64 | 260 | 640 | 234 | 413 |
| Test 2 | 400 | 1020 | 4480 | 244 | 426 |
| Test 3 | 800 | 2330 | 8320 | 290 | 465 |
| Test 4 | 1600 | 3740 | 16,000 | 517 | 703 |

**Table 2.**
*Performance of DDHO with plaintexts of different lengths and contents.*

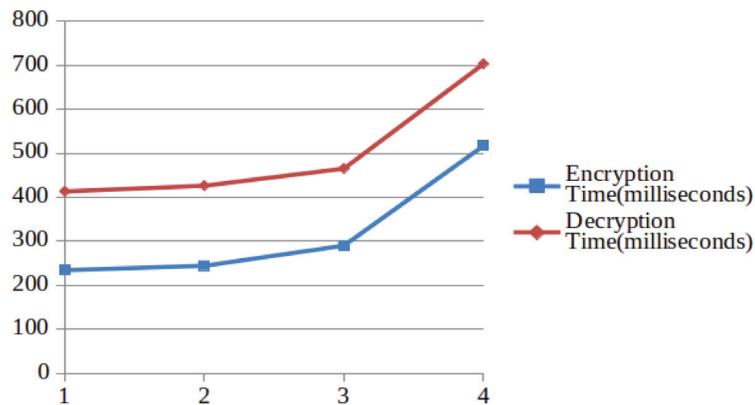

**Figure 3.**
*Analysis of encryption and decryption times of DDHO algorithm.*

The encryption and decryption times shown in **Table 2** and **Figure 3** show that the DDHO algorithm's encryption and decryption times for the different lengths of plaintext increase slower with the changes in the length of plaintext. This reveals that the processing time can be very fast even for relatively very long plaintext.

## 3. Hybrid graphical encryption

The hybrid graphical encryption algorithm (HGEA) is a unique graphical encryption algorithm based on mathematical transformations and graphical pattern realization. It is a symmetric key encryption in which a single 64-bit key is shared between two parties for encryption and decryption of data.

### 3.1 Hybrid cubes encryption algorithm (HiSea)

As the HGEA is inspired from hybrid cubes encryption algorithm (HiSea), it is explained here in brief.

Hybrid cubes encryption algorithm (HiSea) is the symmetric non-binary block cipher. The encryption and decryption keys, plaintext, ciphertext and internal operation in the encryption or decryption processes are based on the integer numbers. HiSea encryption algorithm was developed by Sapiee Jamel in 2011. The plaintext size for the encryption process is 64 bytes ASCII characters. Hybrid cube (HC) is generated based on the inner matrix multiplication of the layers between the two magic cubes (MCs). HC of order $4 \times 4$ is a matrix $H_{i,j}$, i {1, 2, 879} and j {1, 2, 3, 4}, defined as follows: $H_{i,j} = MC_{i,j} \times MC_{i + 1,j}$ where the $MC_{i,j}$ is a jth layer of ith magic cube [17–19].

### 3.2 Hybrid graphical encryption algorithm (HGEA)

HGEA performs the operation, like in HiSea, of generating $4 \times 4$ matrix, then mixing it with key and again mixing of rows and column. Further, the algorithm obtains decision parameters based on remainder value and exploit the correlation between distributions of two graphical patterns for manipulation of intermediate data.

The main concept of hybrid graphical encryption algorithm cipher is to realize the input data into 8 × 8 bit matrix pattern. Then, divide it into four 4 × 4 matrices by putting it up against XY axis quadrant graph. Again, each quadrant 4 × 4 matrix is expanded into four possible 16-bit 4 × 4 matrices by XOR operation with four 4 × 4 subkeys. Further quadrant selection operation selects one 4 × 4 bit matrix output for further processing. Finally, after the final XOR operation, each set of 4 × 4 bit matrix is plotted into XY axis plot (**Figure 4**).

### 3.3 The encryption illustrated

The binary conversion of data and its graphical representation are the key aspects of hybrid graphical encryption algorithm cipher. The five steps involved in the encryption process are: (1) conversion and arrangement, (2) transformation, (3) selection, (4) plotting, and (5) arrangement and conversion.

#### 3.3.1 Step 1: conversion and arrangement

This encryption algorithm can process any "n" plaintext ASCII characters from input file. The input string is split into 8 bytes of m parts. Then, the input ASCII message bit is put up against the standard ASCII table. The plaintext value is then replaced by its ASCII value according to the table. This encryption encompasses numbers, special characters and even spaces.

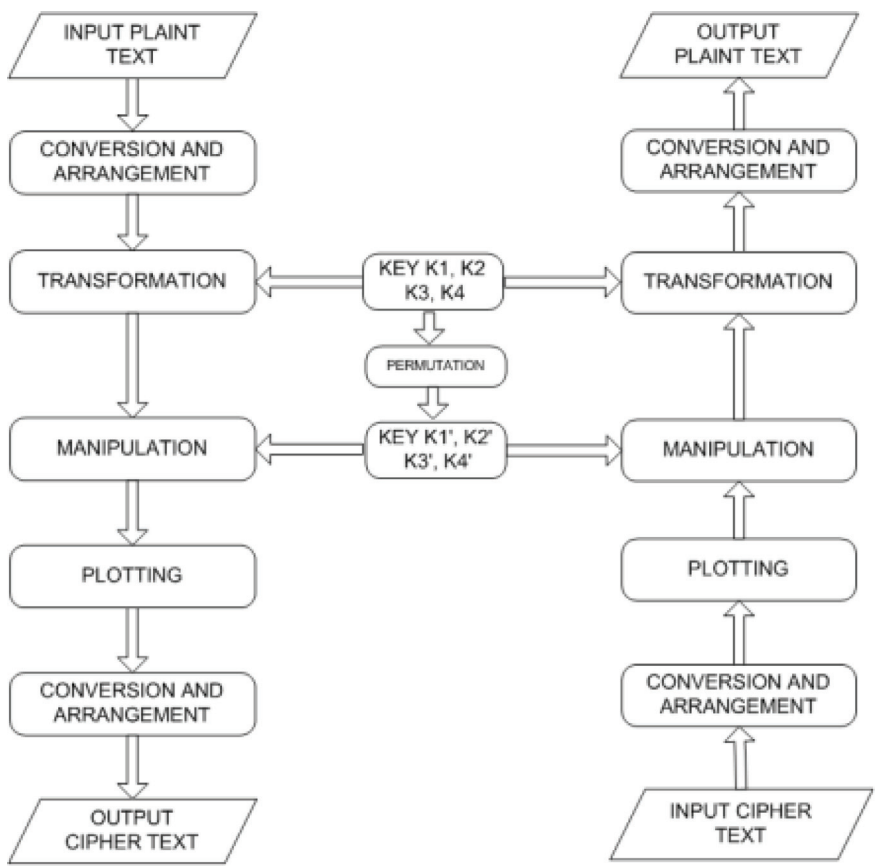

**Figure 4.**
*System architecture diagram of HGEA.*

Since this algorithm will be using the entire ASCII table for referencing, the case sensitivity of the message will play a very crucial part in output ciphertext.

After tabulating the plaintext in comparison with the ASCII table, ASCII and decimal values of the plaintext can be derived. Now, the decimal value has to be converted to binary value to move on to the next step. For binary values that do not reach the 8-bit mark, 0 s are added to the back. The obtained binary value is then tabulated in the form of an 8 × 8 matrix as shown in **Table 3**.

Since HGEA is a symmetric key encryption, a 64-bit binary key is shared for both encryption and decryption processes. These keys are also tabulated in the form of an 8 × 8 matrix of message bits.

Consider key bits as 64 random bits tabulated in an 8 × 8 matrix form, similar to message bits.

### 3.3.2 Step 2: transformation

In this step, initially the 8 × 8 matrix is divided into quadrant form as shown in **Table 4(a)**. The 8 × 8 matrix formed from plaintext is divided into four 4 × 4 matrices, that is, quarters named as $M_1$, $M_2$, $M_3$ and $M_4$ and generalized as $M_i$ Similarly, the 8 × 8 matrix form key is also divided into four quarters named as $K_1$, $K_2$, $K_3$ and $K_4$, generalized as $K_i$.

Again, $M_i$ which is a 4 × 4 matrix is converted into 8 × 8 matrix by performing XOR operation of Mi with $K_1$, $K_2$, $K_3$ and $K_4$. $M_1$ is XOR-ed with $K_1$, $K_2$, $K_3$ and $K_4$ and obtained value is populated to 1st, 2nd, 3rd and 4th quadrants, respectively, as shown in **Table 4(b)**.

| Plain Text | ASCII | Binary |
|:---:|:---:|:---:|
| M | 77 | 01001101 |
| E | 69 | 01000101 |
| S | 83 | 01010011 |
| S | 83 | 01010011 |
| A | 65 | 01000001 |
| G | 71 | 01000111 |
| E | 69 | 01000101 |
| 0 | 0 | 00000000 |

$$M_{Plaintext} = \begin{matrix} 0 & 1 & 1 & 0 & 0 & 0 & 1 & 1 \\ 0 & 1 & 1 & 0 & 0 & 0 & 1 & 1 \\ 0 & 1 & 1 & 0 & 0 & 0 & 1 & 1 \\ 0 & 1 & 1 & 0 & 0 & 0 & 1 & 1 \\ 0 & 1 & 1 & 0 & 0 & 0 & 1 & 1 \\ 0 & 1 & 1 & 0 & 0 & 0 & 1 & 1 \\ 0 & 1 & 1 & 0 & 0 & 0 & 1 & 1 \\ 0 & 1 & 1 & 0 & 0 & 0 & 1 & 1 \end{matrix}$$

**Table 3.**
*Conversion of plaintext to binary.*

|  | (a) |  | (b) |  |
|:---:|:---:|:---:|:---:|:---:|

|  | 01001101 | 0100 | 1101 |
|---|---|---|---|
| 1st    2nd | 01000101 | 0100 | 0101 |
|  | 01010011 | 0101 | 0011 |
|  | 01010011 | 0101 | 0011 |
| 4th    3rd | 01000001 | 0100 | 0001 |
|  | 01000111 | 0100 | 0111 |
|  | 01000101 | 0100 | 0101 |
|  | 00000000 | 0000 | 0000 |

**Table 4.**
*Reference XY axis and sample conversion of 8 × 8 matrix into 4 × 4 matrix.*

Then shifting operation is performed as follows:

For $M_1 \oplus K_1$, $M_2 \oplus K_1$, $M_3 \oplus K_1$ and $M_4 \oplus K_1$
        1st row, $R_1 \leftarrow$ no shift
        2nd row, $R_2 \leftarrow$ 1 bit
        3rd row, $R_3 \leftarrow$ 2 bits
        4th row, $R_4 \leftarrow$ 3 bits

For $M_1 \oplus K_2$, $M_2 \oplus K_2$, $M_3 \oplus K_2$ and $M_4 \oplus K_2$
        1st row, $R_1 \leftarrow$ 3 bit
        2nd row, $R_2 \leftarrow$ no shift
        3rd row, $R_3 \leftarrow$ 1 bit
        4th row, $R_4 \leftarrow$ 2 bits

For $M_1 \oplus K_3$, $M_2 \oplus K_3$, $M_3 \oplus K_3$ and $M_4 \oplus K_3$
        1st row, $R_1 \leftarrow$ 2 bits
        2nd row, $R_2 \leftarrow$ 3 bits
        3rd row, $R_3 \leftarrow$ no shift
        4th row, $R_4 \leftarrow$ 1 bit

For $M_1 \oplus K_4$, $M_2 \oplus K_4$, $M_3 \oplus K_4$ and $M_4 \oplus K_4$
        1st row, $R_1 \leftarrow$ 1 bit
        2nd row, $R_2 \leftarrow$ 2 bits
        3rd row, $R_3 \leftarrow$ 3 bits
        4th row, $R_4 \leftarrow$ no shift

### 3.3.3 Step 3: selection

As its name suggests, first quadrant selection operation is performed, which gives the selected quadrant value for further processing. In this step, only one $4 \times 4$ matrix value is selected for further processing for each Mi value. Thus, step. 2 and 3 via series of confusion and logical operations propose four possible $4 \times 4$ matrix values for further processing, and finally, the selection step selects only one $4 \times 4$ matrix value for further processing. For this purpose, counters are deployed, which will count the number of 1 s in each quadrant of subkeys $K_1$, $K_2$, $K_3$ and $K_4$ for $M_1$', $M_2$', $M_3$' and $M_4$', respectively. Then, the total number of 1 s in corresponding $K_i$ is divided by 4 and the remainder is found.

$$Remainder(R_i) = \frac{Total\_no\_of\_1s\_in\_K_i}{4} \tag{1}$$

Depending upon the total number of 1 s in $K_i$ for corresponding $M_i$, the selected quadrant value will be decided.

$$Q_s = R_i + 1 \tag{2}$$

For instance, let us consider that for $M_1$, the total number of 1 s in $K_1$ is calculated (consider 7, for example) and divided by 4. Now, considering the remainder which will be 3, hence $Q_s = 4$, the fourth quadrant is selected for further processing and denoted as $M_{is}$.

After this, we pass the $K_i$ via permutation box "P," which will shift the bit position of standard matrix as per bit position of randomly selected transposition matrix shown in the **Table 5**.

Finally, $M_i'$ XOR $K_i'$ is performed and $M_i''$ is generated. Thus, matrices $M_1^*$, $M_2^*$, $M_3^*$ and $M_4^*$ matrix are generated.

$$M_i^* \leftarrow M_{is} \oplus K_i' \qquad (3)$$

### 3.3.4 Step 4: plotting

Now, consider the standard matrix distribution of any $4 \times 4$ matrix as shown in **Table 6**. Each of the four $M_1^*$, $M_2^*$, $M_3^*$ and $M_4^*$ values will have different transformations when plotted in XY graph. Values of $M_1''$ will be populated to 1st quadrant as per graph position, which can be realized by permutation as shown in **Tables 7** and **8**.

In this way, the values of $M_1''$, $M_2''$, $M_3''$ and $M_4''$ are populated to the reference XY graph.

### 3.3.5 Step 5: arrangement and conversion

Finally, we have M1", M2", M3" and M4" plotted in $8 \times 8$ matrix form. Now, each row of the matrix is converted from binay to decimal and then to plaintext

| 4,3 | 1,4 | 3,1 | 4,2 |
|-----|-----|-----|-----|
| 1,2 | 1,1 | 4,4 | 2,1 |
| 2,3 | 3,2 | 3,4 | 2,2 |
| 2,4 | 4,1 | 3,3 | 1,3 |

**Table 5.**
*Reference standard $4 \times 4$ transposition matrix distribution.*

Standard matrix distribution of 4x4 matrix

| 1,1 | 2,1 | 3,1 | 4,1 |
|-----|-----|-----|-----|
| 1,2 | 2,2 | 3,2 | 4,2 |
| 1,3 | 2.3 | 3,3 | 4,3 |
| 1,4 | 2,4 | 3,4 | 4,4 |

Reference standard XY graph plot

**Table 6.**
*Plotting the values to standard XY axis graph.*

P1 =

| 4,4 | 3,4 | 2,4 | 1,4 |
|-----|-----|-----|-----|
| 4,3 | 3,3 | 2,3 | 1,3 |
| 4,2 | 2.2 | 2,2 | 1,2 |
| 4,1 | 3,1 | 2,1 | 1,1 |

P2 =

| 1,4 | 2,4 | 3,4 | 4,4 |
|-----|-----|-----|-----|
| 1,3 | 2,3 | 3,3 | 4,3 |
| 1,2 | 2.2 | 3,2 | 4,2 |
| 1,1 | 2,1 | 3,1 | 4,1 |

**Table 7.**
*Permutations $P_1$ and $P_2$ for $M_1$ and $M_2$.*

| P3 = | 1,1 | 2,1 | 3,1 | 4,1 |
|------|-----|-----|-----|-----|
|      | 1,2 | 2,2 | 3,2 | 4,2 |
|      | 1,3 | 2.3 | 3,3 | 4,3 |
|      | 1,4 | 2,4 | 3,4 | 4,4 |

| P4 = | 4,1 | 3,1 | 2,1 | 1,1 |
|------|-----|-----|-----|-----|
|      | 4,2 | 3,2 | 2,2 | 1,2 |
|      | 4,3 | 3,3 | 2,3 | 1,3 |
|      | 4,4 | 3,4 | 2,4 | 1,4 |

**Table 8.**
*Permutations $P_3$ and $P_4$ for $M_3$ and $M_4$.*

| Binary | ASCII | Plain Text |
|--------|-------|------------|
| 01001101 | 77 | M |
| 01000101 | 69 | E |
| 01010011 | 83 | S |
| 01010011 | 83 | S |
| 01000001 | 65 | A |
| 01000111 | 71 | G |
| 01000101 | 69 | E |
| 00000000 | 0 | 0 |

**Table 9.**
*Example conversion of ASCII cipher into binary.*

characters by referring the standard ASCII table, as illustrated in **Table 9**, and this is our ciphertext.

## 3.4 The decryption process

Decryption is also performed in the same manner as encryption but in reverse order. The steps involved in decryption are: (1) conversion and arrangement, (2) plotting, (3) selection, (4) transformation and (5) arrangement and conversion.

## 3.5 The algorithms

In this section, the algorithms of HGEA encryption and decryption are explained step by step. First, the encryption algorithm is given below.

---

**Algorithm 1: HGEA encryption**

---

***Input***: Plaintext p, which is divided into n 8-bit characters. Each 8-bit value ASCII character is converted into 64-bit binary values in 8 × 8 matrix Mp, which is the input for the algorithm. Similarly, 64-bit random key (K) is used.

---

***Output***: Ciphertext C, where C can be decrypted using key K to its corresponding plain-text P.

---

1: Mp ← a(i, j); where 0 < (i, j) < 7

2: Split 8 × 8 matrix into $M_1$, $M_2$, $M_3$, $M_4$, 4 × 4 matrix.

3: for, (i = 0; i < 4; i++);

---

| 4: | for, (j = 0; j < 4; j++); |
| 5: | $M_1$ [i] [j] ← Mp; |
| 6: for, (i = 0; i < 4; i++); |
| 7: | for, (j = 0; j < 4; j++); |
| 8: | l = j + 4; |
| 9: | $M_2$ [i] [j] ← Mp; |
| 10: for, (i = 0; i < 4; i++); |
| 11: | k = i + 4: |
| 12: | for, (j = 0; j < 4; j++); |
| 13: | $M_3$ [i] [j] ← Mp; |
| 14: for, (i = 0; i < 4; i++); |
| 15: | k = i + 4: |
| 16: for, (j = 0; j < 4; j++); |
| 17: | l = j + 4; |
| 18: | $M_4$ [i] [j] ← Mp; |
| 19: Generate $K_1$, $K_2$, $K_3$, $K_4$ from K similar to plaint text. |
| 20: Compute: |
| 21: | $M_1 \oplus K_1$, $M_1 \oplus K_2$, $M_1 \oplus K_3$, $M_1 \oplus K_4$; |
| 22: | $M_2 \oplus K_1$, $M_2 \oplus K_2$, $M_2 \oplus K_3$, $M_2 \oplus K_4$; |
| 23: | $M_3 \oplus K_1$, $M_3 \oplus K_2$, $M_3 \oplus K_3$, $M_3 \oplus K_4$; |
| 24: | $M_4 \oplus K_1$, $M_4 \oplus K_2$, $M_4 \oplus K_3$ and $M_4 \oplus K_4$; |
| 25: Perform Right Shift Operation circularly: |
| 26: With $M_1 \oplus K_1$, $M_2 \oplus K_1$, $M_3 \oplus K_1$ and $M_4 \oplus K_1$ |
| 27: | 1st row no shift |
| 28: | Right circular shift the 2nd row by 1 bit |
| 29: | Right circular shift the 3rd row by 2 bits |
| 30: | Right circular shift the 4th row by 3 bits |
| 31: With, $M_1 \oplus K_2$, $M_2 \oplus K_2$, $M_3 \oplus K_2$ and $M_4 \oplus K_2$ |
| 32: | Right circular shift the 1st row by 3 bits |
| 33: | 2nd row no shift |
| 34: | Right circular shift the 3rd row by 1 bit |
| 35: | Right circular shift the 4th row by 2 bits |
| 36: With $M_1 \oplus K_3$, $M_2 \oplus K_3$, $M_3 \oplus K_3$ and $M_4 \oplus K_3$ |
| 37: | Right circular shift the 1st row by 2 bits |
| 38: | Right circular shift the 2nd row by 3 bits |
| 39: | 3rd row no shift |
| 40: | Right circular shift the 4th row by 1 bit |
| 41: With $M_1 \oplus K_4$, $M_2 \oplus K_4$, $M_3 \oplus K_4$ and $M_4 \oplus K_4$ |
| 42: | Right circular shift the 1st row by 1 bit |
| 43: | Right circular shift the 2nd row by 2 bits |
| 44: | Right circular shift the 3rd row by 3 bits |
| 45: | 4th row no shift |
| 46: | Compute selected quadrant value ($Q_s$): |
| 47: | Remainder ($R_{is}$) ← $\sum$1's in $K_i$/4, for corresponding $M_i$ respectively. |
| 48: | For $Q_{is} = R_i + 1$. |
| 49: | Generate $K_1$', $K_2$', $K_3$' and $K_4$' |
| 50: | $K_i$' = $\pi_i(K_i)$,where $\pi$ = fixed permutation Table. |
| 51: | Compute: $M_i^* ← M_{is} \oplus K_i$' |
| 52: | $M_i$ cipher = $\pi$ ($Mi^*$) |
| 53: | M cipher ← ($M_1$cipher + $M_2$cipher) ($M_4$cipher + $M_3$cipher) |
| 54: | Convert each row of 8 × 8 matrix binary into ciphertext |

Next, the decryption algorithm is given below.

**Algorithm 2: HGEA decryption**

***Input***: Ciphertext C which is divided into n 8-bit characters. Each 8-bit value ASCII character is converted into 64-bit binary values in 8 × 8 matrix $M_c$, which is the input for the algorithm. Similarly, 64 bit random key (K) is used.

***Output***: Plaintext P, where P is decrypted to plaintext using key K.

| | |
|---|---|
| 1: | Mc ← a(i, j); where 0 < (i, j) < 7 |
| 2: | Split 8 × 8 matrix into $M_1$", $M_2$", $M_3$", $M_4$" 4 × 4 matrix. |
| 3: | for, (i = 0; i < 4; i++); |
| 4: | for, (j = 0; j < 4; j++); |
| 5: | $M_1$" [i] [j] ← Mc; |
| 6: | for, (i = 0; i < 4; i++); |
| 7: | for, (j = 0; j < 4; j++); |
| 8: | i = j + 4; |
| 9: | $M_2$" [i] [j] ← Mc; |
| 10: | for, (i = 0; i < 4; i++); |
| 11: | k = i + 4: |
| 12: | for, (j = 0; j < 4; j++); |
| 13: | $M_3$" [i] [j] ← Mc; |
| 14: | for, (i = 0; i < 4; i++); |
| 15: | k = i + 4: |
| 16: | for, (j = 0; j < 4; j++); |
| 17: | l = j + 4; |
| 18: | $M_4$" [i] [j] ← Mp; |
| 19: | Generate $K_1$, $K_2$, $K_3$, $K_4$ from K similar to plaint text. |
| 20: | $M_i$ cipher = $\pi$ ($M_i^*$) |
| 21: | Generate $K_1$', $K_2$', $K_3$' and $K_4$' |
| 22: | $K_i$' = $\pi_i(K_i)$,where $\pi$ = fixed permutation table. |
| 23: | Compute: $M_i^* ← M_{is} \oplus K_i$' |
| 24: | Compute selected quadrant value ($Q_u$): |

| 25: | Remainder(R) ← $\sum$1's in $K_i$ / 4, for corresponding $M_i$ respectively. |
| --- | --- |
| 26: | For $R_i$ = n, $Q_{is}$ = n + 1. |
| 27: Perform Right Shift Operation circularly: | |
| 28: If $Q_s$ = 1 | |
| 29: | 1st row no shift |
| 30: | Left circular shift the 2nd row by 1 bit |
| 31: | Left circular shift the 3rd row by 2 bits |
| 32: | Left circular shift the 4th row by 3 bits |
| 33: If $Q_s$ = 2 | |
| 34: | Left circular shift the 1st row by 3 bits |
| 35: | 2nd row no shift |
| 36: | Left circular shift the 3rd row by 1 bit |
| 35: | Left circular shift the 4th row by 2 bits |
| 37: If $Q_s$ = 3 | |
| 38: | Left circular shift the 1st row by 2 bits |
| 39: | Left circular shift the 2nd row by 3 bits |
| 40: | 3rd row no shift |
| 41: | Left circular shift the 4th row by 1 bit |
| 42: if $Q_s$ = 4 | |
| 43: | Left circular shift the 4th row by 1 bit |
| 44: | Left circular shift the 4th row by 2 bits |
| 45: | Left circular shift the 4th row by 3 bits |
| 46: | 4th row no shift |
| 47: Selected quadrant value ($K_{is}$) ← $Q_s$ | |
| 48: $M_i$ ← $M_i$' ⊕ $K_{is}$ | |
| 49: Generate $M_1$, $M_2$, $M_3$ and $M_4$. | |
| 50: Mp ← ($M_1$cipher + $M_2$cipher) ($M_4$cipher + $M_3$cipher) | |
| 51: Convert above result to string to get plaintext. | |

## 3.6 Performance evaluation

Performance measurement criterion is the time taken by the algorithms to perform encryption and decryption of the input text file, that is, the encryption computation time and decryption computation time. [20]

### 3.6.1 Computation time for encryption and decryption

The encryption computation time of the encryption algorithm is the time taken by the algorithm to produce the ciphertext from the plaintext. The encryption time can be used to calculate the encryption throughput of the algorithms.

The decryption computation time is the time taken by the algorithms to produce the plaintext from the ciphertext. The decryption time can be used to calculate the decryption throughput of the algorithms.

For evaluation, files of sizes 1, 2, 5, 10, 20, 40, 60, 100 and 200 KB were used as input data files. For the sake of comparison, same input files have been set as input to both HGEA and DES. [14] DES has also been implemented in the same environment, JAVA version 8 and simulated in IntelliJ IDEA v.2018.1.4 on the same machine (**Tables 10** and **11**).

The same data set was encrypted via both DES and HGEA algorithms. Each sample data set was encrypted with five randomly chosen passwords, and the encryption execution time and decryption execution time were listed as shown in **Table 12**.

The whole test was performed very carefully. During the test, it was observed that sometimes, when new input data are fed into the computation model, the test results returned false encryption and decryption time values; for example, the test model resulted in very high encryption and decryption execution time values or the test model's decryption execution time was very high compared to encryption computation time and the decrypted plaintext differed from input plaintext. So to resolve this issue, a new terminal was generated in Intellij whenever the input value was changed.

The average values of execution time for encryption and decryption were computed and are tabulated in **Tables 12** and **13**.

| Key<br>File Size | P@ssw0rD | | 11110000 | | a1b2c3d4 | | PRAJWOLM | | TESTHGEA | |
|---|---|---|---|---|---|---|---|---|---|---|
| | Enc Time | Dec Time | Enc Time | Dec Time | Enc Time | Dec Time | Enc Time | Dec Time | Enc Time | Dec Time |
| 1 KB | 4 | 1 | 5 | 2 | 5 | 2 | 6 | 3 | 5 | 2 |
| 2 KB | 6 | 4 | 4 | 3 | 5 | 3 | 6 | 4 | 4 | 2 |
| 5 KB | 7 | 6 | 6 | 5 | 5 | 4 | 6 | 5 | 6 | 5 |
| 10 KB | 12 | 9 | 9 | 7 | 10 | 9 | 10 | 8 | 9 | 8 |
| 20 KB | 17 | 15 | 19 | 16 | 16 | 12 | 15 | 12 | 18 | 15 |
| 40 KB | 33 | 26 | 35 | 27 | 30 | 25 | 29 | 22 | 33 | 25 |
| 60 KB | 49 | 34 | 46 | 30 | 48 | 32 | 47 | 32 | 45 | 29 |
| 100 KB | 68 | 38 | 70 | 39 | 67 | 36 | 69 | 37 | 66 | 35 |
| 200 KB | 94 | 65 | 89 | 58 | 91 | 62 | 90 | 61 | 91 | 59 |

Table: Test results DES

**Table 10.**
*Execution time of DES.*

| Key<br>File Size | P@ssw0rD | | 11110000 | | a1b2c3d4 | | PRAJWOLM | | TESTHGEA | |
|---|---|---|---|---|---|---|---|---|---|---|
| | Enc Time | Dec Time | Enc Time | Dec Time | Enc Time | Dec Time | Enc Time | Dec Time | Enc Time | Dec Time |
| 1 KB | 24 | 16 | 24 | 16 | 26 | 16 | 28 | 18 | 28 | 20 |
| 2 KB | 33 | 19 | 30 | 17 | 31 | 18 | 34 | 20 | 28 | 16 |
| 5 KB | 53 | 34 | 48 | 32 | 51 | 33 | 46 | 31 | 47 | 30 |
| 10 KB | 66 | 51 | 60 | 46 | 63 | 47 | 59 | 44 | 62 | 47 |
| 20 KB | 91 | 87 | 84 | 80 | 87 | 85 | 86 | 83 | 83 | 83 |
| 40 KB | 111 | 104 | 119 | 109 | 109 | 98 | 116 | 105 | 115 | 109 |
| 60 KB | 181 | 167 | 178 | 161 | 168 | 157 | 175 | 159 | 168 | 151 |
| 100 KB | 245 | 222 | 236 | 217 | 247 | 225 | 237 | 216 | 240 | 215 |
| 200 KB | 467 | 357 | 471 | 364 | 457 | 355 | 462 | 361 | 458 | 368 |

Table: Test results HGEA

**Table 11.**
*Execution time of HGES.*

| Input File Size in KB | DES Encryption Time in ms | HGEA Encryption Time ms |
| --- | --- | --- |
| 1 | 5 | 26 |
| 2 | 5 | 31 |
| 5 | 6 | 49 |
| 10 | 10 | 62 |
| 20 | 17 | 86 |
| 40 | 32 | 114 |
| 60 | 47 | 174 |
| 100 | 68 | 241 |
| 200 | 91 | 463 |

**Table 12.**
*Computation time of encryption for various input sizes.*

| Input File Size in KB | DES Decryption Time in ms | HGEA Decryption Time ms |
| --- | --- | --- |
| 1 | 2 | 17 |
| 2 | 3 | 18 |
| 5 | 5 | 32 |
| 10 | 8 | 47 |
| 20 | 14 | 84 |
| 40 | 25 | 105 |
| 60 | 31 | 159 |
| 100 | 37 | 219 |
| 200 | 64 | 361 |

**Table 13.**
*Computation time of decryption for various input size.*

The above tabulated values represent the computation time of various sizes of sample data sets being processed by DES and HGEA. Results show that for the proposed HGEA, execution time is 26, 49 and 62 for 1, 5 and 10 KB of data, respectively. HGEA has a much slower execution time compared to DES. The execution time of the proposed algorithm increases with an increase in data size.

Further, by using the values of **Tables 12** and **13**, various graphs showing the encryption execution and decryption execution times for DES and HGEA with different input sizes were generated as shown in **Figure 5(a)–(c)**.

The above graph shows that the throughput of DES is considerably better than HGEA for small-size input files and it increases with an increase in file size compared to HGEA. It also shows that encryption time for both algorithms is more than decryption time.

The variation in encryption time and decryption time in HGEA is due to steps followed during decryption. During encryption process in the transformation step, each Mi is XOR-ed with all of the possible subkeys and only one of four possible 4 × 4 matrices is chosen for further processing. However, during decryption in the transformation step, only one 4 × 4 matrix is directly generated depending upon selected quadrant value. So, the generation of all of the possible intermediate values is one of the reasons for variation of encryption and decryption times in HGEA.

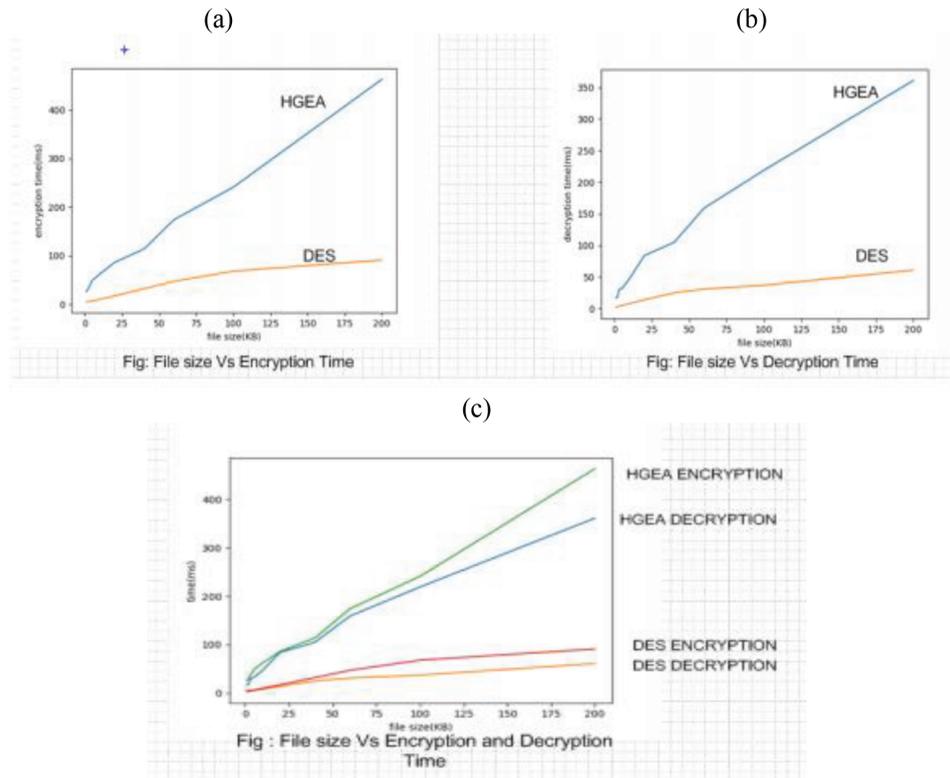

**Figure 5.**
*File size vs. execution time of DES and HGEA.*

## 4. Conclusion

The DNA cryptosystem containing DNA hybridization technique and secure key generation technique OTP are studied, explained and implemented by taking input from the DES algorithm output. The output of DES-based DNA cryptography algorithm has encrypted message in the form of DNA sequences and the decrypted message is the original plaintext.

The DDHO algorithm is tested on different types of plaintext; the encryption and decryption times are calculated; the analysis of length of plaintext, length of ciphertext and size of key is done and found that the length of ciphertext is proportional to the corresponding plaintext length. The encryption and decryption times increase slower with the changes in the length of plaintext.

From the process of analyzing various cryptographic algorithms, a unique encryption algorithm "hybrid graphical encryption algorithm" has been proposed. The algorithm was based on hybrid cubes encryption algorithm (HiSea). The features like graphical interpretation and computation of selected quadrant value are the unique features of this algorithm, which is different from existing standard encryption algorithms. Also, the multiple transformation, multiple key generation provides, combined with graphical interpretation provided added security to the algorithm. The realization of computation model proved the proposed encryption algorithm is practically realizable. Further comparative analysis of computation time of realized model was made.

Although the comparative analysis of proposed model is proven to be more secure, the model was slower than DES. However, the processing units of modern day computer system are extremely high and developing rapidly, the implementation of HGEA possible.

## Author details

Roshan Chitrakar[1,2]*, Roshan Bhusal[2] and Prajwol Maharjan[2]

1 Nepal Open University, Nepal

2 Nepal College of Information Technology, Nepal

*Address all correspondence to: roshanchi@gmail.com

IntechOpen

*Edited by Jaydip Sen*

In the era of Internet of Things (IoT), and with the explosive worldwide growth of electronic data volume and the associated needs of processing, analyzing, and storing this data, several new challenges have emerged. Particularly, there is a need for novel schemes of secure authentication, integrity protection, encryption, and non-repudiation to protect the privacy of sensitive data and to secure systems. Lightweight symmetric key cryptography and adaptive network security algorithms are in demand for mitigating these challenges. This book presents state-of-the-art research in the fields of cryptography and security in computing and communications. It covers a wide range of topics such as machine learning, intrusion detection, steganography, multi-factor authentication, and more. It is a valuable reference for researchers, engineers, practitioners, and graduate and doctoral students working in the fields of cryptography, network security, IoT, and machine learning.

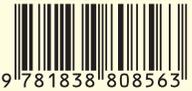